\documentclass[11pt,a4paper]{article}
\pdfoutput=1
\usepackage{jheppub}

\usepackage[T1]{fontenc}
\usepackage[utf8]{inputenc}

\usepackage{amsmath,amssymb,slashed,hyperref}
\usepackage{mathbbol}
\usepackage{amsfonts}
\usepackage{upgreek}
\usepackage{comment}

\numberwithin{equation}{section}

\usepackage[english]{babel}

\usepackage[mathscr]{eucal}
\usepackage{epsfig}
\usepackage{booktabs}
\usepackage{bbold}
\usepackage{slashed}

\DeclareMathAlphabet{\pxitfont}{OML}{pxmi}{m}{it}
\DeclareMathAlphabet{\pxitfontn}{U}{pxmia}{m}{it}
\def\Qb{\pxitfont Q}

\DeclareFontFamily{U}{euc}{}%
\DeclareFontShape{U}{euc}{m}{n}{<-6>eurm5<6-8>eurm7<8->eurm10}{}%
\DeclareSymbolFont{AMSc}{U}{euc}{m}{n} %
\DeclareMathSymbol{\psitt}{\mathord}{AMSc}{"20}    
\DeclareMathSymbol{\chitt}{\mathord}{AMSc}{"1F}    
\DeclareMathSymbol{\sigmatt}{\mathord}{AMSc}{"1B}    

\def\ang{\vartheta}

\newcommand{\bea}{\begin{array}}
\newcommand{\eea}{\end{array}}
\newcommand{\beq}{\begin{equation}}
\newcommand{\eeq}{\end{equation}}
\newcommand{\beqn}{\begin{eqnarray}}
\newcommand{\eeqn}{\end{eqnarray}}
\newcommand{\tr}{{\rm tr}}

\newcommand{\nnr}{\nonumber\\}
\newcommand{\ex}{{\rm e}}

\newcommand{\T}{\theta}
\newcommand{\eps}{\epsilon}

\newcommand{\CC}{{\mathbb C}}
\newcommand{\RR}{{\mathbb R}}
\newcommand{\ZZ}{{\mathbb Z}}

\newcommand{\fr}{\frac}

\newcommand{\cN}{\mathcal{N}}
\newcommand{\cA}{\mathcal{A}}

\newcommand{\dA}{{\dot{A}}}
\newcommand{\dB}{{\dot{B}}}

\newcommand{\da}{{\dot{a}}}
\newcommand{\db}{{\dot{b}}}
\newcommand{\CS}{{\mathrm {CS}}}

\newcommand{\bz}{{\bar{z}}}
\newcommand{\calK}{\mathcal{K}}

\newcommand{\slq}{\slashed{q}}

\def\SO{{\rm SO}}

\DeclareMathOperator{\re}{Re}
\DeclareMathOperator{\im}{Im}

\font\teneusm=eusm10 
\font\seveneusm=eusm7 
\font\fiveeusm=eusm5
\newfam\eusmfam
\textfont\eusmfam=\teneusm \scriptfont\eusmfam=\seveneusm
\scriptscriptfont\eusmfam=\fiveeusm

\font\tencmmib=cmmib10 \skewchar\tencmmib='177
\font\sevencmmib=cmmib7 \skewchar\sevencmmib='177
\font\fivecmmib=cmmib5 \skewchar\fivecmmib='177
\newfam\cmmibfam
\textfont\cmmibfam=\tencmmib \scriptfont\cmmibfam=\sevencmmib
\scriptscriptfont\cmmibfam=\fivecmmib
\def\cmmib#1{{\fam\cmmibfam\relax#1}}

\font\teneurm=eurm10 
\font\seveneurm=eurm7 
\font\fiveeurm=eurm5
\newfam\eurmfam
\textfont\eurmfam=\teneurm \scriptfont\eurmfam=\seveneurm
\scriptscriptfont\eurmfam=\fiveeurm

\font\teneufm=eufm10 
\font\seveneufm=eufm7 
\font\fiveeufm=eufm5
\newfam\eufmfam
\textfont\eufmfam=\teneufm \scriptfont\eufmfam=\seveneufm
\scriptscriptfont\eufmfam=\fiveeufm

\def\bar{\overline}

\def\hat{\widehat}

\def\p{{\cmmib p}}

\def\d{{\mathrm d}}

\def\YM{{\mathrm{YM}}}

\def\tilde{\widetilde}

\def\2{{\bf 2}}
\def\1{{\bf 1}}
\def\0{{\bf 0}}

\def\bar{\overline}

\def\V{{\mathcal V}}

\title{Teichm\"uller TQFT vs Chern-Simons theory}

\author{Victor Mikhaylov}

\affiliation{Simons Center for Geometry and Physics,\\State University of New York,\\ Stony Brook, NY 11794}

\emailAdd{victor.mikhaylov@gmail.com}

\abstract{Teichm\"uller TQFT is a unitary 3d topological theory whose Hilbert spaces are spanned by Liouville conformal blocks. It is related but not identical to PSL$(2,\RR)$ Chern-Simons theory. To physicists, it is known in particular in the context of 3d-3d correspondence and also in the holographic description of Virasoro conformal blocks. We propose that this theory can be defined by an analytically-continued Chern-Simons path-integral with an unusual integration cycle. On hyperbolic three-manifolds, this cycle is singled out by the requirement of invertible vielbein. Mathematically, our proposal translates a known conjecture by Andersen and Kashaev into a conjecture about the Kapustin-Witten equations. We further explain that Teichm\"uller TQFT is dual to complex SL$(2,\CC)$ Chern-Simons theory at integer level $k=1$, clarifying some puzzles previously encountered in the 3d-3d correspondence literature. We also present a new simple derivation of complex Chern-Simons theories from the 6d (2,0) theory on a lens space with a transversely-holomorphic foliation.}

\keywords{Teichmuller TQFT, complex Chern-Simons theory, 3d-3d correspondence, Kapustin-Witten equations}

\arxivnumber{1710.04354}

\begin{document}

\maketitle

\section{Introduction and summary}
Chern-Simons theory in three dimensions was one of the first ever constructed topological quantum field theories in dimension greater than one \cite{WittenJones}. Its version with a compact gauge group $G$ is also among the best understood interacting quantum field theories. To a closed oriented two-manifold $C$ it associates a finite-dimensional Hilbert space which is the space of conformal blocks of the WZW model with gauge group $G$. To a closed oriented three-manifold $W$ it associates a partition function known as the Witten-Reshetikhin-Turaev invariant, which admits a rigorous mathematical definition \cite{RT}. Chern-Simons theories with non-compact gauge groups are less well understood. For the case of a complex gauge group, some early work includes \cite{WittenCCS,WittenGrav}. Rather spectacular recent progress is summarized in \cite{DimofteRev2}.

Often an extremely useful way to study a quantum field theory is to embed it into string or M-theory. Chern-Simons theory can be engineered \cite{WittenTopStr} by open topological strings, and its large-N limit
is described by closed topological strings in the geometry, related to the original
one by a geometric transition \cite{GVtrans}. Via the relation between topological
strings and string/M-theory, Chern-Simons invariants become related to counting BPS states of membranes in M-theory \cite{GV1,GV2,OV}. This description is naturally amenable to categorification~\cite{GSV}. In a closely related more recent development, analytically-continued Chern-Simons theory \cite{WittenCS} and theories with complex gauge groups were constructed by compactification of the M5-brane theory, and thus appeared on one of the two sides of the 3d-3d correspondence. For a recent review and references, see \cite{Dimofte-rev}.

Among gauge groups that are not compact or complex, the first example that one would like to understand is SL$(2,\RR)$. In the first years of Chern-Simons theory it was realized \cite{VerlindeVerlinde,Verlinde} that SL$(2,\RR)$ Chern-Simons theory must be related to quantization of Teichm\"uller spaces, and at least part of its Hilbert space is the space of Liouville conformal blocks. To avoid some unnecessary details, we will actually consider PSL$(2,\RR)$ rather than SL$(2,\RR)$.

Teichm\"uller space $\mathcal{T}$ of a two-manifold\footnote{All our manifolds are assumed to be oriented, so we will mostly omit mentioning this explicitly.} $C$ is the space of conformal or complex structures on $C$, modulo diffeomorphisms isotopic to the identity. For a closed surface of genus $g\ge 2$, $\mathcal{T}$ is diffeomorphic to $\RR^{6g-6}$. The reason that it makes appearance in PSL$(2,\RR)$ Chern-Simons theory  is simple. The classical phase space of this Chern-Simons theory is the moduli space $\mathcal{M}^\RR$ of flat PSL$(2,\RR)$ connections on $C$, which includes $\mathcal{T}$ as one of its connected components. Flat connections in $\mathcal{T}$ are precisely the ones that come from uniformization, that is, from representing $C$ as a quotient of the hyperbolic upper half-plane by a subgroup of its group of isometries PSL$(2,\RR)$. In a convenient real structure, an $\mathfrak{sl}(2)_\RR$ gauge field can be written as
\beq
\cA=\left(\bea{cc}ia& e\\e^\dagger&-ia\eea\right)\,.\label{sl2rA}
\eeq
For the connections that belong to the component $\mathcal{T}\subset\mathcal{M}^\RR$, there exists a gauge in which the one-form $e$ is everywhere non-degenerate. Then it can be taken as a vielbein on $C$ and defines a hyperbolic metric $\d s^2=e^\dagger e$, in which $a$ is the Levi-Civita connection.

Chern-Simons action endows $\mathcal{T}$ with a symplectic form which is a multiple of the Weil-Petersson form $\omega_{\rm WP}$. Quantization with this symplectic structure and with some real choice of polarization produces a Hilbert space $\mathcal{H}\simeq L^2(\RR^{3g-3})$. It has been argued in \cite{VerlindeVerlinde,Verlinde}, based on a Ward identity, that $\mathcal{H}$ can be identified with the space of Liouville conformal blocks on $C$. For a quick check, let us count the parameters. A trivalent graph from which $C$ can be obtained by thickening has $3g-3$ legs. A Liouville conformal block is labeled by the corresponding $3g-3$ real Liouville momenta, which are the variables in a wavefunction, belonging to $L^2(\RR^{3g-3})$. 

More recently, quantization of Teichm\"uller spaces was put on rigorous mathematical footing \cite{KashaevQ,ChekhovFock,TeschnerLanglands}. The Hilbert space $\mathcal{H}$ comes with a Hermitian product, an action of the mapping class group of $C$ and an action of an algebra of Verlinde operators \cite{TeschnerLanglands,DGS}. This structure depends on the coupling constant $b^2>0$, or equivalently on the central charge 
\beq
c=13+6(b^2+b^{-2})\ge 25\,.\nonumber
\eeq

$\mathcal{H}$ is a part of the Hilbert space of PSL$(2,\RR)$ Chern-Simons theory, but it is not the whole Hilbert space. Indeed, the moduli space $\mathcal{M}^\RR$ has $4g-3$ components, of which $\mathcal{T}$ is only one. The other components include gauge fields, for which the vielbein $e$ in (\ref{sl2rA}) is not invertible.\footnote{More precisely, besides $\mathcal{T}$ there is one more component with invertible vielbein, isomorphic to $\bar{\mathcal T}$.} Then a TQFT with Hilbert spaces $\mathcal{H}$, if it exists, is distinct from PSL$(2,\RR)$ Chern-Simons theory. This theory is sometimes called Teichm\"uller TQFT, and will be the main subject of this paper.

The first question is, why would Teichm\"uller TQFT exist? It is not true in general that one can throw out components from the phase space, and still get a quantum field theory that respects factorization. The reason one expects that Teichm\"uller TQFT exists is that the Hilbert spaces $\mathcal{H}$ define an analog of an extended modular functor \cite{TeschnerMod}. It means that they carry representations of (central extensions of) the mapping class groups, which moreover agree with cutting and gluing punctured two-manifolds. It is known that having an extended modular functor is enough to construct a TQFT \cite{BK}. For Teichm\"uller TQFT, an important complication is that the Hilbert spaces are infinite-dimensional, and, to our knowledge, a mathematically rigorous construction of a TQFT from the modular functor of Liouville conformal blocks has not yet been completed. 

There exists another mathematical approach \cite{AK,AKnew} to Teichm\"uller TQFT, in which partition functions for a class of triangulated three-manifolds are constructed  by gluing together wavefunctions associated to tetrahedra. The answer is then shown to be independent of the triangulation. A closely related construction appears in the physics literature \cite{DimofteQuantum,labelled,Dimofte-k}. For three-manifolds with boundaries, the partition function takes values in function spaces which can be naturally identified with $\mathcal{H}$.

The first objective of our paper is to give a physicist's definition of Teichm\"uller TQFT on a three-manifold, which would clearly show that it exists as a unitary 3d TQFT, distinct from PSL$(2,\RR)$ Chern-Simons theory. We will propose that it is equivalent to an analytically-continued \cite{WittenQM,WittenCS} Chern-Simons theory with a particular unusual integration cycle. Our definition will depend on some conjectures about the Kapustin-Witten equations.

Recently, Teichm\"uller TQFT has reappeared in physics in the context of the 3d-3d correspondence \cite{Yamazaki,DGS,Dimofte-rev}. Consider the 6d $(2,0)$ theory of type $A_1$, put on a product manifold $S^3_b\times W$. To preserve some supersymmetry, one turns on a suitable supergravity background, and the partition function of the theory depends on the squashing parameter $b^2$ of $S^3_b$. It is expected that the reduction on $S^3_b$ produces some topological Chern-Simons theory on the three-manifold $W$. Consider the case that $W$ is a product $\RR\times C$, and view the direction $\RR$ as time. The Hilbert space of the 6d $(2,0)$ theory in this geometry is what the Chern-Simons theory is supposed to associate to the two-manifold $C$. On the other hand, reversing the order of compactification, we obtain a 4d $\cN=2$ $\mathcal{S}$-class theory associated to $C$, and its Hilbert space on $S^3_b$ is expected to be the space of Liouville conformal blocks, according to the AGT correspondence \cite{AGT,NW}. Thus, one expects that the Chern-Simons theory obtained by a three-sphere reduction from six dimensions is precisely Teichm\"uller TQFT. In \cite{Yamazaki}, this proposal was tested for $W$ being a mapping cylinder. This did not require a definition of Teichm\"uller TQFT on a general three-manifold, so it was sufficient to think of it as a subsector of PSL$(2,\RR)$ Chern-Simons theory. In the present paper, we would like to understand the general situation.

After this initial work on the 3d-3d correspondence, three-sphere reduction from six dimensions was preformed explicitly in\footnote{For similar reductions on $S^1\times S^2$, see \cite{YagiM5,YamazakiM5}.} \cite{CJ}. The result was perhaps a little surprising. It was found that the Chern-Simons theory on $W$ is not SL$(2,\RR)$, but rather SL$(2,\CC)$ at integer level $k=1$ and complex level, determined by the parameter $b^2$ of the geometry. 

The second objective of our paper is to explain the relation between Teichm\"uller TQFT and complex SL$(2,\CC)$ Chern-Simons theory as a duality. In fact, we will show that there exist two different complex Chern-Simons theories, to which Teichm\"uller TQFT is equivalent. 

Via the 3d-3d correspondence, partition function of Teichm\"uller TQFT on a three-manifold $W$ is related to the $S^3_b$ supersymmetric partition function of a 3d superconformal $\cN=2$ theory ${\rm T}[W]$, obtained from the 6d $(2,0)$ theory by reduction on $W$ \cite{labelled,Dimofte-rev}. The definition of this supersymmetric partition function involves a supergravity background on the $S^3$, designed so as to preserve some Killing spinor. Recently, a systematic understanding of such backgrounds was achieved in \cite{Komar3d1,Komar3d2,Komar3d3}. It was shown that the partition function does not depend on much of the details of the background, but only depends on a geometric structure known as the transversely holomorphic foliation.

The third objective of our paper is to apply this machinery to derive complex Chern-Simons theories from six dimensions. This can be done uniformly for supersymmetric backgrounds on $S^1\times S^2$, $S^3$ or higher lens spaces, leading to Chern-Simons theories with $k=0$, $k=\pm 1$ or $|k|>1$. In the special case of $S^3$, different orders of reduction give rise to two different complex Chern-Simons theories or to Teichm\"uller TQFT. We expect that this approach will eventually lead to a systematic understanding of integration cycles in Chern-Simons theories obtained from six dimensions, and will help to better understand the holomorphic blocks \cite{HolomBlocks}. But in this paper, we only make first steps in this direction.

In the rest of this section, we provide a short summary of the paper.

In section~\ref{tquant}, we make the first step towards defining Teichm\"uller TQFT. Using the approach to quantization via branes in the A-model \cite{BQ}, for a two-manifold $C$ we construct a quantum mechanics, whose Hilbert space is obtained from quantization of the Teichm\"uller space of $C$. (This setup was previously considered in \cite{NW,DGS} in the same AGT context, and we add just a few minor details to this story.) Specifically, we consider an A-model on $\RR\times\mathcal{I}$ with the target being the SO$(3)$ Hitchin moduli space $\mathcal{M}_H$ for $C$, with the boundary conditions at the two ends of the interval $\mathcal{I}$ set by a Lagrangian and a coisotropic brane. The Lagrangian brane is supported on the Hitchin section in $\mathcal{M}_H$, which is well-known to be isomorphic to the Teichm\"uller component $\mathcal{T}\subset\mathcal{M}^\RR$.

In section~\ref{TeichCS}, we observe that the Hitchin sigma-model on $\RR\times\mathcal{I}$ with these branes can in fact be naturally lifted to a theory with three-dimensional covariance. This theory is the Kapustin-Witten twisted $\cN=4$ super Yang-Mills on the geometry $W\times\mathcal{I}$, where $W$ is an arbitrary three-manifold and $\mathcal{I}$ is an interval, which we take to be parameterized by $y\in[0,y_0]$. The boundary condition at $y=0$ is a Nahm pole, or equivalently D5-brane type, while the boundary condition at $y=y_0$ is NS5-type, as defined in \cite{GWbc,5branes}. For $W\simeq \RR\times C$, this theory reduces to the same Hitchin sigma-model setup that is defined in section~\ref{tquant}, as follows from some well-known results \cite{KW,GWJones}.

The $\cN=4$ super Yang-Mills theory on $W\times \mathcal{I}$ with these boundary conditions is the unique natural 3d-covariant lift of the sigma-model of section~\ref{tquant}. As we explain later, it can also be obtained by reduction from six dimensions. Hence we are led to propose that this setup gives precisely the definition of Teichm\"uller TQFT on a three-manifold $W$. To understand its meaning, we localize the path-integral onto solutions to the Kapustin-Witten equations. This allows to interpret the Teichm\"uller TQFT partition function as a path-integral in analytically-continued Chern-Simons theory \cite{WittenCS},
\beq
Z_{\rm Teichm}=\int_{\mathcal S}D\cA\,\exp\left(\fr{ib^2}{4\pi}\int\tr\left(\cA\d\cA+\fr{2}{3}\cA^3\right)\right)\,,
\eeq
where the integration cycle $\mathcal{S}$ is the space of fields $\cA=A+i\phi$ at $y=y_0$ that can be reached by Kapustin-Witten flows on $W\times\mathcal{I}$ that start with the Nahm pole at $y=0$. (It is a non-trivial conjecture that the space $\mathcal{S}$, for a suitable class of three-manifolds, is a meaningful integration cycle for the Chern-Simons path-integral.) We also explain how this integration cycle, provided that it makes sense, can be expanded in Lefschetz thimbles.

In section~\ref{hyperb}, we turn specifically to the case that the three-manifold $W$ is hyperbolic. It is a known conjecture by Andersen and Kashaev \cite{AK} that in the semiclassical limit $b^2\rightarrow\infty$, the Teichm\"uller partition function on a hyperbolic three-manifold decays as
\beq
|Z_{\rm Teichm}|\sim \exp\left(-b^2{\rm Vol}(W)/2\pi\right)\,,\label{introas}
\eeq
where ${\rm vol}(W)$ is the hyperbolic volume. Some further support to this conjecture comes from holography \cite{Hologr}. On a hyperbolic three-manifold, there exist two special flat PSL$(2,\CC)$ connections, the geometric $\cA_{\rm geom}=\omega+ie$ and the conjugate geometric $\cA_{\bar{\rm geom}}=\omega-ie$, constructed from the Levi-Civita connection $\omega$ and the vielbein $e$ of the hyperbolic metric. The behavior (\ref{introas}) is characteristic of the flat connection $\cA_{\bar{\rm geom}}$. Moreover, it is known that the contribution of this critical point is the most subleading \cite{Reznikov}. Therefore, one may conclude that the integration cycle $\mathcal{S}$ is precisely the Lefschetz thimble for the conjugate geometric flat connection $\cA_{\bar{\rm geom}}$. (This implication of the Andersen-Kashaev conjecture for the Chern-Simons integration cycle has been previously proposed in \cite{Hologr,BGL}.) Interestingly, we see that it is possible to have a consistent Chern-Simons theory where one restricts to vielbeins that are invertible, at least semiclassically. One may wonder if this teaches us something about 3d gravity and its relation to Chern-Simons theory.

Since we have a description of the integration cycle $\mathcal{S}$ in terms of the Kapustin-Witten equations, we are able to formulate a conjecture about counting their solutions on a four-manifold $W\times \RR^+$, which would imply that $\mathcal{S}$ is indeed the Lefschetz thimble for $\cA_{\bar{\rm geom}}$. So far we could not prove this statement, but we do perform some  non-trivial tests that it successfully passes. 

In section~5, we attempt to understand how S-duality of $\cN=4$ super Yang-Mills acts on Teichm\"uller TQFT. First we discuss an issue that could potentially invalidate our proposal for the integration cycle. It is sometimes assumed that Chern-Simons path-integrals over Lefschetz thimbles are related by S-duality to half-space partition functions with a Nahm pole boundary condition. If it were true, it would imply that Teichm\"uller TQFT partition function is a Laurent series (possibly with some non-integer powers) in variable $q=\exp(2\pi i/b^2)$, which is known not to be the case experimentally. We explain that Lefschetz thimble integrals are not related by S-duality to anything with obvious definition. The reason is some peculiarity of the NS5-type boundary condition used in defining these path-integrals. On the way, we make contact with the recent work \cite{GMP} that explains how to extract $q$-series from Chern-Simons partition functions.

The same issues with the NS5-type boundary condition make it difficult to apply S-duality to the Teichm\"uller TQFT partition function, and we have to make some assumptions about BRST-exact terms in the action before we can act with S-duality. With these assumptions made, we see that Teichm\"uller TQFT is invariant under the exchange $b\rightarrow b^{-1}$, which of course is a known fact \cite{DGS,AK}. Two other elements of the SL$(2,\ZZ)$ S-duality group relate Teichm\"uller TQFT to complex SL$(2,\CC)$ Chern-Simons theories. The most general form of the path-integral in a complex Chern-Simons theory is
\beq
\int_{\mathcal{C}}D\cA_\ell D\cA_r\exp\left(\fr{v+ik}{2}\CS(\cA_{\ell})-\fr{v-ik}{2}\CS(\cA_r)\right)\,,\nonumber
\eeq
where $k\in\ZZ$ and $v\in\CC$ are the coupling constants, and $\mathcal{C}$ is some suitable integration cycle. We find that Teichm\"uller TQFT is equivalent to two versions of complex Chern-Simons theory, with the coupling constants
\beqn
&{\rm CS-I}:&\quad k=-1\,,\quad v=i\fr{b^2+1}{b^2-1}\,,\nnr
&{\rm CS-II}:&\quad k=1\,,\quad v=-i\fr{b^2-1}{b^2+1}\,.\nonumber
\eeqn
For CS-I, we make a proposal for the integration cycle, and it certainly is not the usual one. For CS-II, understanding the integration cycle seems to be more difficult. Note that CS-II is the theory obtained in \cite{CJ} from six dimensions. We now understand the formula\footnote{Apart from a factor of $i$ which seems to be missing in \cite{CJ}.} for the coupling constant obtained in that paper as a consequence of S-duality.

In section~\ref{redu}, we derive complex Chern-Simons theories from six dimensions. We put the 6d $(2,0)$ theory on a geometry $W\times L_k$, where $W$ is some three-manifold and $L_k$ is a lens space $S^3/\ZZ_k$. First, for a moment replace $L_k$ with $\RR^3$. To preserve some supersymmetry, we twist the theory along $W$, which leaves four unbroken supercharges that together with translations on $\RR^3$ form a 3d $\cN=2$ superalgebra. Viewing the 6d theory on $W\times \RR^3$ as a 3d $\cN=2$ supersymmetric theory on $\RR^3$, we propose that it is possible to use the backgrounds of \cite{Komar3d1} to supersymmetrically compactify on $L_k$. Using the results of \cite{Komar3d2} we prove that it is possible to deform the metric on $L_k$ without affecting the partition function, so that the geometry becomes an arbitrarily small two-torus fibered over a finite interval $\mathcal{I}$. Then we reduce on the torus fiber and obtain topologically twisted 4d $\cN=4$ super Yang-Mills theory on $W\times\mathcal{I}$. We explicitly identify the boundary conditions and the coupling constants in terms of the geometry of the background and verify that the effective theory on $W$ is a complex Chern-Simons theory. We make some comments on the integration cycle in this theory, but we cannot yet derive it from the first principles. 

Our approach takes as an input directly the background that can be used for localizing the 3d theory ${\rm T}[W]$, and gives as an output a Chern-Simons theory engineered in terms of the twisted 4d super Yang-Mills, which is the natural framework to define analytically continued Chern-Simons theories. For these reasons, we hope optimistically that it will be possible to completely understand the question of integration cycles in the future.

In section~\ref{cocol}, we return to Teichm\"uller TQFT and address some natural questions related to conformal blocks and the usual 2d-4d AGT correspondence. A useful representation for the wavefunctions in the Hilbert space $\mathcal{H}$ is by holomorphic wavefunctions in the K\"ahler quantization of $\mathcal{T}$. We explain how to construct the basis of states $\langle {\bf q}|$ of K\"ahler quantization in terms of brane corners. This essentially repeats the argument of Nekrasov and Witten \cite{NW} on the derivation of AGT, with some minor variations. As a natural extension of our conjectures of section~\ref{hyperb}, we propose that the wavefunction of Teichm\"uller TQFT on a three-manifold $W$ with boundary, presented in the basis $\langle \mathbf{q}|$, is given by the path-integral with the dominant contribution coming from the complete hyperbolic metric on $W$, with the conformal structure at infinity fixed by ${\bf q}\in\mathcal{T}$. For the special case that the three-manifold $W$ is a handlebody (or just a ball) with a network of defects, we make contact with the usual holographic setup for the Virasoro conformal blocks \cite{Hijano,Perlmutter}. We also explain, mostly following \cite{Harlow}, how a Liouville partition function on $C$ can be obtained from Teichm\"uller TQFT on a hyperbolic three-manifold $C\times\mathcal{I}$ with two asymptotic boundaries.

Section~\ref{SecRemarks} contains an informal discussion of Teichm\"uller TQFT, PSL$(2,\RR)$ Chern-Simons theory and modular functors. It aims to add a few technical details to the facts mentioned in this introduction, and is not meant to contain any new results.

In section~\ref{out}, we list some open questions.

Appendix~\ref{hypap} contains some computations for the Kapustin-Witten equations that support our conjectures of section~\ref{hyperb}. In appendix~\ref{THFs}, we review the transversely holomorphic foliations and the backgrounds of \cite{Komar3d1,Komar3d2}, and prove that for lens spaces, they can be deformed into a bundle of a small torus over a long interval. In appendix~\ref{5branesusy}, following \cite{GWbc}, we consider $\cN=4$ super Yang-Mills half-BPS boundary conditions that do not preserve Lorentz symmetry. We extend to finite values of parameters some results for the NS5-type boundary condition that in \cite{GWbc} were obtained to the first order in the Lorentz symmetry violating parameter. Then we specialize to the case of boundary conditions that preserve the 3d Lorentz symmetry of the twisted theory.

\section{Branes and quantization of the Teichm\"uller space}\label{tquant}
The Teichm\"uller space $\mathcal{T}$ of a two-manifold $C$ has a natural symplectic structure. Teichm\"uller TQFT, by its very definition, associates  to $C$ a quantum mechanics, whose Hilbert space is the quantization of the Teichm\"uller space of $C$. A natural way to construct such a quantum mechanics is to use the approach via branes in the A-model \cite{BQ}. The target of the A-model should be a complexification of the phase space $\mathcal{T}$, and for that one can naturally choose the SO$(3)$ Hitchin moduli space. The corresponding setup has been explored in section 4.6 of \cite{NW} (see also \cite{TeschnerLanglands}) in relation to the AGT correspondence. The goal of this section is to briefly recall this story. Throughout the paper, we follow the standard notations \cite{Hitchin,KW} for the Hitchin moduli space, its hyperk\"ahler structure, the Hitchin topological sigma-model and branes. 

Let $\mathcal{M}_H$ be the moduli space of solutions to the Hitchin equations on a hyperbolic Riemann surface $C$ of genus $g$. For simplicity, we mostly assume that there are no marked points on $C$ (and therefore $g\ge 2$), although at times we will lift this restriction. The gauge group $G$ is chosen to be SO$(3)$. The space $\mathcal{M}_H$ is hyperk\"ahler, with the Hitchin metric $g_H$, a triplet of complex structures $I$, $J$ and $K$, which generate the algebra of quaternions, and with symplectic forms $\omega_I$, $\omega_J$ and $\omega_K$, which are K\"ahler forms in the respective complex structures. The cohomology class of $\omega_I$ is non-trivial, while $\omega_J$ and $\omega_K$ are exact forms, in the absence of punctures.

\begin{figure}
 \begin{center}
   \includegraphics[width=4cm]{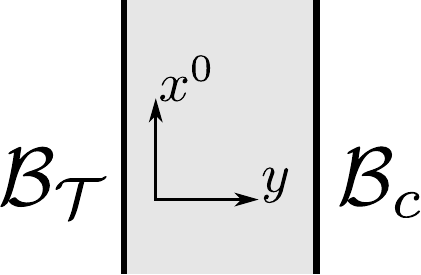}
 \end{center}
\caption{\small The worldsheet of a string, whose space of states is the quantization of the Teichm\"uller space. The brane $\mathcal{B}_c$ on the right is coisotropic, and the brane $\mathcal{B}_{\mathcal{T}}$ on the left is Lagrangian, supported on the Hitchin section.}
 \label{BLBc}
\end{figure}

Consider the two-dimensional sigma-model of maps into $\mathcal{M}_H$. We take the metric in the target to be $b^2\, g_H$, where $b^2>0$ is a real parameter. We also turn on a B-field $B=-b^2\,\omega_I$. For the topological BRST operator we take the supercharge of the A-model in complex structure $K$, or, equivalently, we set the Kapustin-Witten parameter $t$ to be $-1$. The K\"ahler form is then $\omega=b^2\,\omega_K$.  The sigma-model is put on the worldsheet $\RR\times \mathcal{I}$, where ${\mathcal I}=[0,y_0]$ is an interval, parameterized by a variable $y$. The setup is shown on figure~\ref{BLBc}.
At $y=y_0$, we choose the boundary condition to be determined by a coisotropic brane $\mathcal{B}_c$ of type $(B,A,A)$. Its support is the whole of $\mathcal{M}_H$, the Chan-Paton bundle $\mathcal{L}_c$ is trivial, and the connection has zero curvature $F$, but $F+B$ is non-zero. The physics of this brane is governed by the complex structure $\omega^{-1}B=J$. At $y=0$ we put a Lagrangian brane $\mathcal{B}_\mathcal{T}$, also of type $(B,A,A)$. Its support is a section of the Hitchin fibration, which is defined as a component of the fixed point set of an involution of the Hitchin equations $\sigma:\,(A,\phi)\rightarrow(A,-\phi)$, isomorphic \cite{Hitchin} to the Teichm\"uller space $\mathcal{T}$ of $C$. The isomorphism depends on a choice of a reference complex structure on $C$. The involution $\sigma$ is holomorphic in complex structure $I$, which therefore induces a complex structure on the fixed point set. The submanifold $\mathcal{T}\subset\mathcal{M}_H$ is isomorphic to $\CC^{3g-3}$ in this complex structure. It is Lagrangian with respect to the holomorphic symplectic form $\Omega_I=\omega_J+i\omega_K$. The brane $\mathcal{B}_\mathcal{T}$ is therefore the usual Lagrangian brane in our A-model.  The connection on its Chan-Paton line bundle $\mathcal{L}_\mathcal{T}$ is chosen to have curvature $F=b^2\omega_I|_{\mathcal{T}}$, so that the condition $F+B=0$ for the Lagrangian brane is satisfied. We remark that often in the A-model approach to quantization one takes the B-field to be zero, the Chan-Paton bundle on the Lagrangian brane to be flat, and moves the curvature to the coisotropic brane. In our case that would not work. The reason is that for $b^2\notin\ZZ$, the cohomology class of the form $b^2\omega_I$ is not integral, and there can be no line bundle on $\mathcal{M}_H$ with a connection of such curvature. The restriction of $b^2\omega_I$ to the Teichm\"uller brane, however, is trivial in the cohomology, as $\mathcal{T}$ is contractible. Thus, unless $b^2$ is an integer, we have to keep  non-zero $B$-field and non-zero curvature on the Lagrangian brane. 

Let us also remark that our gauge group SO$(3)$ is not simply-connected, which brings in the possibility of some discrete theta-angles in the sigma-model, as explained in section~7 of \cite{KW}. However, for our purposes they all are irrelevant, because one end of the string is always bound to the Lagrangian brane $\mathcal{B}_{\mathcal{T}}$, which is contractible.

The vector space $\mathcal{H}$ of $(\mathcal{B}_\mathcal{T},\mathcal{B}_c)$ string states in this A-model can be naturally regarded \cite{BQ} as a quantization of the support of the Lagrangian brane, that is, the Teichm\"uller space $\mathcal{T}$, in the symplectic structure $F=b^2\,\omega_I|_{\mathcal{T}}$. The reason for that is simple: the effective theory of zero-modes of the string in this setting reduces to a quantum mechanics for the phase space $\mathcal{T}$, the support of the Lagrangian brane, with prequantum line bundle $\mathcal{L}_\mathcal{T}\otimes \mathcal{L}^{-1}_c$ with a unitary connection of curvature $F$. The Hilbert space of this quantum mechanics is what one calls the quantization of $\mathcal{T}$. Note that the symplectic form $\omega_I$ reduces \cite{Hitchin} on $\mathcal{T}$ to the Weil-Petersson form $\omega_{WP}$. Thus, the Hilbert space of our theory $\mathcal{H}$ is the quantization of the Teichm\"uller space of $C$ with the symplectic form $b^2\,\omega_{WP}$. By itself, this claim is rather vacuous, because $\mathcal{T}$ is symplectomorphic to $\RR^{6g-6}$ with its standard symplectic form. But there exists a lot of extra structure that makes the statement interesting. 

Generally in the A-model, the space of $(\mathcal{B}_1,\mathcal{B}_2)$ open string states need not have a hermitian inner product. It only has a natural pairing with the space of $(\mathcal{B}_2,\mathcal{B}_1)$ strings, which comes from the two-point disc amplitude. However, in our case $\mathcal{H}$ does have a hermitian product, and thus is indeed a Hilbert space. As explained in \cite{BQ}, this follows from the fact that the support of the Lagrangian brane is a component of the fixed point set of the involution $\sigma$, which is antiholomorphic in complex structure $K$, used in the definition of the A-model. 

The space $\mathcal{H}$ is naturally a module for the algebra of $(\mathcal{B}_c,\mathcal{B}_c)$ strings, which act by adjoining from the right. This algebra is a quantum deformation of the algebra of holomorphic functions on $\mathcal{M}_H$ in the complex structure which appears in the definition of the coisotropic brane $\mathcal{B}_c$\,, that is, $J$. These holomorphic functions are generated by traces of holonomies of flat SL$(2,\CC)$ connections around a basis of one-cycles on $C$. The quantum deformation is governed to the first order by the holomorphic symplectic form $\omega-iB=b^2\Omega_J$. The quantum algebra, to be denoted ${\bf A}_q$, actually depends\footnote{This fact is known in the explicit constructions of these algebras, see {e.g.} \cite{DimofteQuantum}. The following argument suggests an explanation. In the reduction from the $\cN=4$ super Yang-Mills theory, the coisotropic brane appears from the NS5-brane. It is invariant under the S-duality transformation $S^{-1}TS$, which shifts $1/b^2$ by one. It would be interesting to prove this more carefully by T-duality in the Hitchin sigma-model.} on the exponentiated parameter $q=\exp(2\pi i/b^2)$.
Similarly, there is an action of $(\mathcal{B}_\mathcal{T},\mathcal{B}_\mathcal{T})$ strings on $\mathcal{H}$ from the left. As one can see from the mirror model, the algebra of such strings is similarly\footnote{To be precise, there is a slight difference between the two algebras, related to the fact that S-duality exchanges SO$(3)$ with SU$(3)$. For us, this will be unimportant. Some discussion of these matters can be found in section~4 of \cite{DGS}.} the quantum-deformed algebra of traces of holonomies, now with the deformation parameter $\tilde{q}=\exp(2\pi i b^2)$.  One expects the space $\mathcal{H}$ to be an irreducible module of the algebra ${\bf A}_q\otimes{\bf A}_{\tilde q}$\,. Another piece of structure comes from the fact that the Teichm\"uller space has an action of the mapping class group of $C$. Upon quantization, this gives rise to a projective-unitary representation of the mapping class group on $\mathcal{H}$. The corresponding operators act on the algebra ${\bf A}_{q}\otimes {\bf A}_{\tilde q}$ by natural automorphisms. A mathematically-rigorous quantization of the Teichm\"uller space, exhibiting this rich structure, was first constructed in \cite{ChekhovFock}, \cite{KashaevQ}, where the reader can find details and precise statements. (See also \cite{TeschnerLanglands}, \cite{DGS} and references therein.)

The vectors in the Hilbert space $\mathcal{H}$ can be identified with Liouville conformal blocks. This works as follows \cite{Teschner03b,TeschnerLanglands}. Choose a pants decomposition of $C$, to be labeled by $\sigmatt$. Cutting into pairs of pants goes along $3g-3$ circles $c_a$ which form a maximal set of non-isotopic simple circles on $C$. For a circle $c_a$, $a=1,\dots,3g-3$, one defines a function $l_a$ of a complex flat connection as twice the logarithm of an eigenvalue of the holonomy of the connection around $c_a$. On a point in $\mathcal{T}\subset\mathcal{M}_H$, that is, for a PSL$(2,\RR)$ flat connection that comes from uniformization, $l_a$ is equal to the hyperbolic length of the geodesic, isotopic to the circle $c_a$. The functions $l_a$, $a=1,\dots,3g-3$, all Poisson-commute in symplectic structure $\Omega_J$, and upon quantization give a maximal set of commuting hermitian operators $\hat l_a$ on $\mathcal{H}$. We denote by $|\sigmatt,{\bf l}\rangle$ the common eigenvector of these operators\footnote{One might think that the operators $\hat l_a$ can be defined purely in terms of quantized traces of holonomies, that is, elements of the algebra $\mathbf{A}_q$, by somehow taking logarithms. In this case they would commute with the whole $\mathbf{A}_{\tilde q}$. These statements would be false. The model situation for this phenomenon is the modular double, see \cite{Faddeev} or section 6 of \cite{DimofteQuantum}. We recall some of these matters in section~\ref{SecRemarks}.} with real non-negative eigenvalues ${\bf l}=(l_1,\dots,l_{3g-3})$. Next, take the natural complex structure on $\mathcal{T}$, with some choice of holomorphic coordinates ${\bf q}$. The symplectic form $\omega_{WP}$ in this complex structure is the K\"ahler form for the Weil-Petersson metric. With this data, one can quantize $\mathcal{T}$ with symplectic form $b^2\omega_{WP}$ in K\"ahler polarization. As a result, the state vector $|\sigmatt,{\bf l}\rangle$ gives a holomorphic function on $\mathcal{T}$,
\beq
\Psi^{\sigmatt,{\bf l}}({\bf q})\equiv\langle{\bf q}|\sigmatt,{\bf l}\rangle\,.\label{conf1}
\eeq
It can be shown that this wavefunction is equal to the Liouville conformal block, associated to the pants decomposition $\sigmatt$, with the intermediate momenta in the corresponding $3g-3$ channels equal to $l_a$. The parameter $b^2$ that appears as a coefficient in the symplectic form becomes the usual Liouville parameter with the same name. In particular, the central charge is equal to
\beq
c=6(b^2+b^{-2})+13\,.\label{Llec}
\eeq
The modular-invariant hermitian product on $\mathcal{H}$ is identified with the Liouville correlation function. More details as well as references can be found in section 6 of \cite{TeschnerLanglands}.

In physics literature, the idea that Virasoro conformal blocks arise from quantizing the Teichm\"uller space was first proposed by \cite{VerlindeVerlinde,Verlinde}. In those papers, $\mathcal{T}$ was also viewed as a component of the space of flat PSL$(2,\RR)$ connections. The quantization was performed in a K\"ahler polarization. Wavefunctions were constructed as functionals on the infinite-dimensional space of all PSL$(2,\RR)$ connections, and the Gauss law constraint was imposed only after the quantization. The identification of the quantum wavefunction with the conformal block was based on the observation that the aforementioned constraint is identical to the Virasoro Ward identity.

In this paper, we mainly concentrate on the case of real positive $b^2$, but, of course, Virasoro conformal blocks can be defined for complex $b^2$. In the Hitchin sigma-model, one can consider a natural generalization of the setup that we used. On the right, we still put a coisotropic brane, whose physics is governed by the holomorphic symplectic form $\omega-iB=b^2\Omega_J$, except that $b^2$ now is not real, so
\beqn
\omega&=&|b^2|\,(\omega_K\cos\alpha-\omega_I\sin\alpha)\,,\nnr
B&=&-|b^2|\,(\omega_I\cos\alpha+\omega_K\sin\alpha)\,,
\eeqn
where $\alpha$ is the phase of $b^2$. On the left, we put a Lagrangian brane. It is supported on a submanifold which is holomorphic in the complex structure, in which the K\"ahler form is $B$. It is Lagrangian with respect to the forms $\omega$ and $\omega_J$. This submanifold is related by a Hitchin diffeomorphism $\CC^*$ to the submanifold of opers. The Lagrangian brane also supports a Chan-Paton bundle of curvature\footnote{This bundle indeed exists, because the cohomology class $[B]$ is trivial, when restricted to the brane. Indeed, for $\alpha\ne 0,\pi$, it is proportional to $[\omega]$, which vanishes on the Lagrangian brane, while for $\alpha=0$ or $\pi$ we reduce to our usual Teichm\"uller case.} $-B$, to cancel the $B$-field. The space of states is still a module for the algebra $\mathbf{A}_q\otimes\mathbf{A}_{\tilde q}$, and therefore one expects that it can be naturally identified with the space of conformal blocks. 

By way of a digression, we point out that the case of $b^2\in i\RR$ is special. The $B$-field is trivial in the cohomology, and can be traded for a curvature on the coisotropic brane. We then have a simple setup with zero $B$-field, and Lagrangian and coisotropic branes both of type $(A,A,B)$. They are a hyperk\"ahler rotation of the usual brane of opers and the canonical coisotropic brane. It is easy to apply S-duality (or Hitchin mirror symmetry) to this configuration. The result is essentially the same configuration, with the two branes exchanged. This gives a simple realization of the $b\rightarrow b^{-1}$ symmetry of the Liouville theory. Understanding this symmetry for the brane setup with the general $B$-field is more tricky and lies beyond the scope of this paper.

Finally we point out that our setup for quantizing the Teichm\"uller space in the A-model is not the only possible one. One can choose different branes and different symplectic structures in the A-model, but, as long as the Hilbert space is an irreducible module for ${\bf A}_q\otimes {\bf A}_{\tilde q}$ with $q=\exp(2\pi i/b^2)$ and $\tilde q=\exp(2\pi i b^2)$, one expects it to be isomorphic to the space of the conformal blocks. For example, in \cite{NW} it was found that the brane setup in $\mathcal{M}_H$ which arises by compactification from a four-dimensional Omega-deformed gauge theory in the context of the AGT correspondence \cite{AGT} is different from what we have used here. Namely, the configuration in section 4 of that paper is almost what we have just described for $b^2\in i\RR$, but with the Lagrangian brane being of type $(A,B,A)$, which is the brane of opers itself and not its hyperk\"ahler rotation.\footnote{Interestingly, this setup does not fit into the paradigm of quantization via branes, because the would-be symplectic form vanishes, when restricted to the Lagrangian brane. Nevertheless, as explained in \cite{NW}, the theory makes sense and is expected to produce the same Liouville conformal blocks.}

We shall encounter these variations of the setup later in section~\ref{cocol}. However, our main interest in this paper lies in three-dimensional quantum field theories. It turns out that of all obvious brane configurations for quantizing the Teichm\"uller space, with a coisotropic brane on the right and a Lagrangian brane on the left, the only setup (up to trivial equivalences) that can be lifted to a theory with three-dimensional covariance is the one that we considered initially, namely, the  A-model in symplectic structure $\omega_K$ with the branes $\mathcal{B}_{\mathcal T}$ and $\mathcal{B}_c$, with real $b^2$. (We later return to this fact in section~\ref{d5sec}.) This is the setup that we will mostly focus on.

\section{Teichm\"uller TQFT as a Chern-Simons theory}\label{TeichCS}
So far we understand our tentative TQFT for three-manifolds of the form $\RR\times C$. To define the theory on a general three-manifold $W$, we recall that the Hitchin sigma-model can be obtained \cite{KW} by dimensional reduction from the four-dimensional $\cN=4$ super Yang-Mills theory. Namely, taking the four-manifold to be a product $\Sigma\times C$ and compactifying on $C$, one can see that the Yang-Mills theory reduces to a sigma-model on the worldsheet $\Sigma$ with the target space being the Hitchin moduli space for $C$. In our construction, the worldsheet was a product $\Sigma\simeq \RR\times {\mathcal I}$, and therefore the four-manifold was $\mathcal{I}\times\RR\times C$. Then to put the quantum Teichm\"uller theory on a general three-manifold $W$, it should be enough to replace $\RR\times C$ by $W$. Now we will explain this in more detail.

\subsection{The setup}
The $\cN=4$ Yang-Mills theory can be twisted and put on an arbitrary four-manifold. The twist that we need is the one used by Kapustin and Witten. On a general four-manifold, the twisted Lorentz group in this case preserves two supercharges, which square to zero on gauge-invariant quantities. A linear combination of these supercharges, parameterized by a $\mathbb{P}^1$-valued parameter $t$, can be taken as a BRST operator $\Qb$ of the topological field theory. The unbroken subgroup of the R-symmetry group is U$(1)_{\rm gh}$, the charge under which defines the ghost number. Bosonic fields of the theory include a gauge field $A$, an adjoint-valued one-form $\phi$ of ghost number zero, and an adjoint-valued complex field $\sigma$ of ghost number two. The bosonic part of the twisted super Yang-Mills action on a four-manifold $V$ is\footnote{We always work in Euclidean signature. The argument in the path-integral is $\exp(-I_\YM)$.}
\beqn
I_\YM&=&-\fr{1}{g_\YM^2}\int_V{\rm Vol}_V\,\tr\left(\fr{1}{2}F_{\mu\nu}F^{\mu\nu}+\nabla_\mu\phi^\nu\nabla^\mu\phi^\nu+{\mathcal R}_{\mu\nu}\phi^\mu\phi^\nu+\fr{1}{2}[\phi_\mu,\phi_\nu]^2\right)\nnr
&+&\fr{i\theta_\YM}{8\pi^2}\int_V\tr(F\wedge F)\,,
\eeqn
where ${\mathcal R}_{\mu\nu}$ is the Ricci tensor of $V$. We have omitted terms involving the field $\sigma$, since their form will not be important for our purposes. To study the topological theory, it is useful to rewrite the action in a form, in which it is manifestly $\Qb$-invariant,
\beqn
I_\YM&=&\fr{i\calK}{4\pi}\int_V\tr(F\wedge F)+\{\Qb,\dots\}\,.\label{N4act}
\eeqn
This equality is correct up to some total derivative terms. We introduced the notation
\beq
\calK=\fr{\theta_\YM}{2\pi}+\fr{4\pi i}{g_\YM^2}\fr{t-t^{-1}}{t+t^{-1}}\label{canonical}
\eeq
for the so-called canonical parameter, determined by the Yang-Mills couplings $g_\YM$, $\theta_\YM$ and the twisting parameter $t$. (The points $t=\pm i$ are special. For them, the formula above is not valid.) The $\Qb$-exact terms in the action can be expressed as a linear combination of gauge-invariant squares of the following expressions,
\beqn
\V^+&=&(F-\phi\wedge \phi+t\d_A\phi)^+\,,\label{lhss1}\\
\V^-&=&(F-\phi\wedge\phi-t^{-1}\d_A\phi)^-\,,\label{lhss2}\\
\V^0&=&\d_A\star\phi\,.\label{lhss3}
\eeqn
For real $t$, the super Yang-Mills path-integral can be localized onto the subspace of fields on which $\V^+$, $\V^-$ and $\V^0$ vanish. The BPS equations
\beq
\V^+=0\,,\quad \V^-=0\,,\quad \V^0=0\label{KWeqs}
\eeq
are known as the Kapustin-Witten equations.

\begin{figure}
 \begin{center}
   \includegraphics[width=6.5cm]{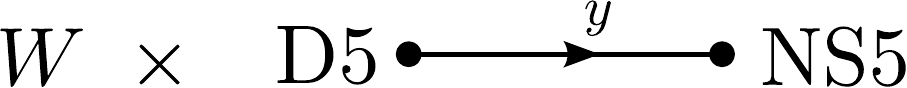}
 \end{center}
\caption{\small Teichm\"uller TQFT on a three-manifold $W$ can be engineered by the four-dimensional $\cN=4$ super Yang-Mills theory on $W\times\mathcal{I}$ with the D5-type and the NS5-type boundary conditions at the two ends of the interval.}
 \label{branes3d}
\end{figure}
For our application, we take the $\cN=4$ super Yang-Mills theory with gauge group SO$(3)$ and put it on a four-manifold $V\simeq W\times\mathcal{I}$ with the product metric $\d s_V^2=\d s_W^2+\d y^2$, see figure~\ref{branes3d}. We choose the twisting parameter to be $t=-1$, and the canonical parameter to be $\calK=b^2>0$. Then $\theta_\YM/2\pi = b^2$. The gauge coupling will also be set to $4\pi/g_\YM^2=b^2$, although nothing really depends on it, since it appears in $\Qb$-exact terms only. It is necessary to specify the boundary conditions at the two ends of the interval $\mathcal{I}$.
At $y=y_0$, we put an NS5-type boundary condition. It is called so, because it can be obtained \cite{GWbc,5branes} in a brane construction, where the Yang-Mills theory is engineered in the type IIB string theory on a stack of semi-infinite D3-branes which end on an NS5-brane.\footnote{To be completely precise, such brane construction would give a boundary condition which differs from ours by some $\Qb$-exact terms and field redefinitions, see section~\ref{secblocks}. } The scalar field $\sigma$ and the component $\phi_y$ of the adjoint one-form, normal to the boundary, are required to vanish at $y=y_0$. The three components $\phi_i$, tangent to the boundary $W$, as well as the gauge field $A$ satisfy mixed boundary conditions which can be determined by setting to zero the boundary variation of the action.\footnote{This is true in the physical super Yang-Mills theory. In topological theory, one introduces a set of auxiliary fields in order to have $\Qb^2=0$ off-shell. With these fields in the action, the fields $\phi_i$ and $A$ have free boundary conditions at $y=y_0$.} The action, with boundary terms at $y=y_0$ included, is equal to
\beq
I_{\rm YM}=\fr{ib^2}{4\pi}\int_V\tr(F\wedge F)+\fr{b^2}{4\pi}\int_W\tr\left(2\phi F+i\phi \d_A \phi-\fr{2}{3}\phi^3\right)+\{\Qb,\dots\}\,.\label{ginv}
\eeq
(The wedges between differential forms are omitted, when this does not lead to ambiguity.) The fields in the integral over $W$ are the restrictions of the bulk fields to $y=y_0$.  The same action can be rewritten intuitively as
\beq
I_{\YM}=-ib^2\CS(\cA)+({\rm something\,at}\,y=0)+\{\Qb,\dots\}\,,\label{YMCS}
\eeq
where we introduced a complexified gauge field $\cA=A+i\phi$ and also the notation
\beq
\CS(\cA)=\fr{1}{4\pi}\int_W\tr\left( \cA\d\cA+\fr{2}{3}\cA^3\right)\label{csa}
\eeq
for the Chern-Simons functional. Unless $b^2\in \ZZ$, this expression is not invariant under large gauge transformations, and so should be understood as a formal rewriting of the manifestly gauge-invariant action (\ref{ginv}). 

To complete the definition of the boundary condition at $y=y_0$, one has to specify, what happens to the fermions. This is described {e.g.} in \cite{WittenQM}, but for our purposes no details are really needed. We will just mention that the fermion $\{\Qb,\cA\}$ should certainly vanish at the boundary, to ensure that the action is $\Qb$-invariant. 

The boundary condition at $y=0$ is known as the D5-type, or the Nahm pole \cite{Np1,Np2}. The fields $\phi_y$, $\sigma$ and the gauge field satisfy the Dirichlet boundary condition, while the tangential components $\phi_i$ are singular and diverge at $y\rightarrow 0$ according to a model Nahm pole singularity. This boundary condition is perhaps familiar, when $W$ is the flat space. The generalization to curved $W$ was described in section 3.4 of \cite{5branes}. On $W$, we choose a vielbein $e$ for the Riemannian metric, and denote by $\omega$ the corresponding Riemannian connection. It is defined by the property
\beq
\d_\omega e=0\,.
\eeq
The boundary condition identifies the SO$(3)$ gauge bundle at $y=0$ with the frame bundle of $W$. Then it makes sense to require the fields at $y\rightarrow 0$ to have the following asymptotics,
\beqn
\phi&=&-\fr{e}{y}+\dots\,,\label{bc1}\\
A&=&\omega+\dots\,,\label{bc2}
\eeqn
where dots stay for less singular terms. (In fact, one can require the subleading terms to vanish at $y\rightarrow 0$.) To verify that this is a good boundary condition, one needs at least to check that the singular terms in the fields do not make the action divergent. Using that $\d_\omega e=0$, $\d_\omega\star e=0$ and $e\wedge e=\star e$, it is easy to see that there are no divergent terms in $\V^+$, $\V^-$ and $\V^0$, and therefore the $\Qb$-exact part of the action is well-defined. The non-$\Qb$-exact part of the action, apart from the terms supported at the other end of the interval, contains only the topological term, which is finite, since the gauge field in (\ref{bc2}) has no singularity. 

An important subtlety in the definition of the boundary condition at $y=0$ is explained in section 3.5 of \cite{5branes}. With the behavior of the gauge field as in (\ref{bc2}), the topological term in the action (\ref{ginv}) depends on the metric on $W$. Indeed, its variation under changes of the metric is the same as that of the gravitational Chern-Simons term
\beq
ib^2\CS(\omega)=\fr{i b^2}{4\pi}\int_V\tr \left(\omega\d\omega+\fr{2}{3}\omega^3\right)\,.\label{counter}
\eeq
(Here the trace is taken over the fundamental representation of $\mathfrak{su}(2)$, as  everywhere else in our formulas.) To cancel this metric dependence, one adds a counterterm $-ib^2\CS(\omega)$ to the action. However, this functional is not gauge invariant, and its value depends on a choice of the trivialization of the tangent bundle of $W$. This introduces a framing dependence in the topological field theory. Explicitly, under a unit change of framing, the counterterm produces the factor
\beq
\exp\left(\fr{i\pi b^2}{2}\right)\label{fram}
\eeq
in the partition function.\footnote{This is true if we allow arbitrary changes of framing for the tangent bundle, as for an abstract SO$(3)$ bundle. We could also define the gravitational Chern-Simons term using the four-manifold signature, in which case the ambiguity would be a cube of (\ref{fram}). And using the choice of the spin structure, the elementary framing factor could be made the 48-th power of (\ref{fram}). }

The gravitational counterterm at $y=0$ cannot be the only source of framing dependence. The D5 and the NS5 boundary conditions are related by an S-duality transformation. Therefore, one expects that the boundary condition at $y=y_0$ should also contribute to the framing dependence.\footnote{The correspondent gravitational counterterm will be understood as a part of the definition of the Chern-Simons path-integral, and will not be written out explicitly.} In this case, it is a quantum effect which should appear upon regularizing the path-integral. The S-duality acts on the canonical parameter $b^2$ in the same way as on the gauge coupling, that is, changes it into $-b^{-2}$. The elementary framing factor from the boundary at $y=y_0$ should then be
\beq
\exp\left(\fr{i \pi b^{-2}}{2}\right)\,.\label{framns}
\eeq
(The minus sign in the coupling $-b^{-2}$ was canceled by another sign which comes from the fact that, in order to map the D5NS5 system to itself, one needs to accompany the S-duality by a reflection.) In a three-dimensional topological field theory which is associated to a two-dimensional CFT of central charge $c$ the factor produced by an elementary change of framing is \cite{WittenJones}
\beq
\exp\left(\fr{\pi ic}{12}\right)\,.
\eeq
From (\ref{fram}) and (\ref{framns}) we expect that our TQFT is related to a CFT with central charge
\beq
c\stackrel{?}{=}6\,(b^2+b^{-2})\,.
\eeq
This indeed is almost the Liouville central charge (\ref{Llec}). What is missing is a $b$-independent constant $13$, which one may expect to appear from the one-loop determinant. (For more details on such constant shifts in the context of analytically-continued Chern-Simons theory, see section 3.5.3 of \cite{5branes}.)

For $W\simeq \RR\times C$, our setup should reduce back to the sigma-model on $\Sigma\simeq \RR\times\mathcal{I}$, considered in section \ref{tquant}. How this works, has been explained in \cite{KW} and \cite{GWJones}. Since the four-dimensional theory is topological, we are free to rescale the metric to make the typical size of the Riemann surface $C$ much smaller than the length of the interval $\mathcal{I}$. The theory is then expected to reduce to a sigma-model on $\Sigma$, for which the target is the moduli space of vacuum field configurations. It can be shown that such configurations should solve the Kapustin-Witten equations (\ref{KWeqs}) simultaneously for all values of $t$. On $\RR\times C\times{\mathcal{I}}$, such solutions are given by pullbacks of solutions to the Hitchin equations on $C$. Thus, the low energy theory is indeed the Hitchin sigma-model.\footnote{The Hitchin moduli space has singularities, corresponding to reducible solutions. Near the singular loci, dimensional reduction from the super Yang-Mills theory produces some extra massless fields, which will play a role in section \ref{lefsec}.}

Since we have set $t=-1$, the topological field theory in two dimensions is the A-model in complex structure $K$. It is easy to see this directly. Indeed, the action (\ref{ginv}) contains a coupling $\fr{b^2}{2\pi}\int_W\tr(\phi F)$ at $y=y_0$. It can be rewritten as a bulk integral $-\fr{b^2}{2\pi}\int_V\tr(\d_A\phi\wedge F)$, which in two dimensions reduces to the integral of the pullback of the symplectic form $b^2\omega_K$,
\beq
-\fr{b^2}{2\pi}\int_V\tr(\d_A\phi\wedge F)=b^2\int_\Sigma X^*\hspace{-1mm}\left(\fr{1}{2\pi}\int_C\tr(\delta \phi\wedge\delta A)\right)\,.
\eeq
(The disappearance of the minus sign in this formula is explained in footnote 15 in \cite{KW}.) Similarly, the topological term in (\ref{ginv}) together with the coupling $\fr{ib^2}{4\pi}\int_W\tr(\phi\d_A\phi)$ reduce in two dimensions to the B-field $B=-b^2\omega_I$. The boundary condition at $y=y_0$ puts no restriction on the fields $\phi_i$ and $A_i$, and in the sigma-model gives rise to the coisotropic brane $\mathcal{B}_c$, whose support is the whole of $\mathcal{M}_H$. The BRST-invariant operators on $\mathcal{B}_c$ are gauge-invariant functions of the complexified gauge field $\cA$ which is annihilated by $\Qb$ at $y=y_0$. These give rise to algebraic functions on $\mathcal{M}_H$, holomorphic in complex structure $J$.

It is a bit more tricky to see that the D5-type boundary condition at $y=0$ reduces in the sigma-model to the Lagrangian brane $\mathcal{B}_\mathcal{T}$.
To do this, one considers the Kapustin-Witten equations on $\RR\times C\times \RR^+$, where the infinity in $\RR^+$ means just that we are very far from the D5-brane on the interval $\mathcal{I}$, in the scale set by the size of the Riemann surface $C$. For very large $y$, the KW solutions are required to become $y$-independent and therefore approach solutions to the Hitchin equations. It is not hard to see that any solution that has the Nahm pole at $y\rightarrow 0$, at large $y$  will approach a point of the Hitchin section $\mathcal{T}\subset\mathcal{M}_H$. Moreover, for each point of $\mathcal{T}$ there exists precisely one solution. This was explained in \cite{GWJones}, with the rigorous mathematical proof completed in \cite{MazzeoNew}.

\subsection{Chern-Simons theory with an exotic integration cycle}\label{exocont}
We have argued that Teichm\"uller TQFT is the four-dimensional topologically twisted $\cN=4$ super Yang-Mills theory, put on an interval with particular boundary conditions. Following \cite{WittenCS}, we would like to explain, how to understand this setup as an analytically-continued Chern-Simons theory.

The Yang-Mills path-integral can be localized onto the space of solutions to the Kapustin-Witten equations on $W\times\mathcal{I}$. A slight simplification comes from the fact that the boundary conditions at both ends of the interval set $\phi_y=0$. Then it is possible to show \cite{WittenCS} by a simple integration-by-parts argument that $\phi_y$ vanishes everywhere for any solution.

Let us denote by $\mathcal{G}$ and $\mathcal{G}_\CC$ the groups of SO$(3)$ and PSL$(2,\CC)$ gauge transformations on $W$, respectively, and by $\mathcal{G}^0$ and $\mathcal{G}^0_\CC$ the subgroups of gauge transformations, connected to the identity. We would like to gauge away the component $A_y$ of the gauge field, and to do so by a gauge transformation on $W\times\mathcal{I}$ that is trivial at $y=y_0$, the location of the NS5-type boundary. There is a price to pay for that. The boundary condition at $y=0$ will now say that the fields may not approach the Nahm pole precisely in the form (\ref{bc1})-(\ref{bc2}), but rather be \emph{conjugate to it} by a $\mathcal{G}^0$ gauge transformation. 
With $\phi_y=A_y=0$ and in the product metric on $W\times\mathcal{I}$, the Kapustin-Witten equations become
\beqn
\partial_y\phi&=&-\star\left(F-\phi\wedge\phi\right)\,,\label{flow1}\\
\partial_yA&=&-\star\left(\d_A\phi\right)\,,\label{flow2}\\
\d_A^*\phi&=&0\,.\label{mmap}
\eeqn
Here $A$ and $\phi$ should be viewed as fields on $W$, which depend on a parameter $y$. The Hodge star and the differential are those on $W$. 

Let $\Omega_{\cA}$ be the space of PSL$(2,\CC)$ gauge fields $\cA$ on $W$, or equivalently, the space of pairs $(A,\phi)$. 
The first two equations (\ref{flow1}) and (\ref{flow2}) can be rewritten as
\beq
\partial_y\cA=-i\star\bar{\mathcal{F}}\,,
\eeq
where $\bar{\mathcal{F}}$ is the field strength for $\bar\cA=A-i\phi$. They describe the downward gradient flow on $\Omega_\cA$ for the functional
\beq
h=-4\pi\im\left(\CS(\cA)\right)\,.\label{h}
\eeq
The flow is defined using the natural K\"ahler metric on $\Omega_\cA$. 
The moment map for the action of $\mathcal{G}_\CC^0$ on $\Omega_\cA$ is
\beq
\mu=\d^*_A\phi\,.\label{moment}
\eeq
The equation (\ref{mmap}) then is a sort of gauge-fixing condition. It is preserved by the flow and can be thought of as a part of the boundary condition.

Consider the space $\mathcal{S}_{y_0}\subset\Omega_\cA$ of fields, obtained by restricting to $y=y_0$ solutions to the localization equations (\ref{flow1})-(\ref{mmap}) which for $y\rightarrow 0$ are conjugate to the Nahm pole (\ref{bc1})-(\ref{bc2}) by a $\mathcal{G}^0$ gauge transformation.  One expects $\mathcal{S}_{y_0}$ to be, informally, a middle-dimensional real submanifold in the infinite-dimensional space $\Omega_\cA$. 
It is middle-dimensional, because the Nahm pole boundary condition, supplemented with a suitable gauge-fixing, is elliptic \cite{MazzeoWitten} and leaves free half of the modes. To make this argument completely explicit, one would need to compute the index of the linearization of the KW equations around the Nahm pole. Alternatively, just for illustration, we can perturbatively solve the KW equations for small $y$, and count the modes. The perturbative analysis was done in full generality in \cite{MazzeoWitten}, and is reviewed in a special case in our appendix~\ref{AppEx}. The space of perturbative solutions that approach the Nahm pole at $y\rightarrow 0$ is parameterized by six real functions on $W$. This is a middle-dimensional subspace in $\Omega_\cA/\mathcal{G}^0_\CC$. Our definition of $\mathcal{S}_{y_0}$ requires the fields to approach the Nahm pole only up to a $\mathcal{G}^0$ gauge transformation. This adds three more functions and gives nine modes, which is indeed one-half the dimension of $\Omega_\cA$. It is also easy to see that $\mathcal{S}_{y_0}$ is a real subspace, that is, no non-zero tangent vector over any point of $\mathcal{S}_{y_0}$ is mapped to another tangent vector by the complex structure operator $J$: $\delta\cA\rightarrow i\delta\cA$. Indeed, if $\delta \cA$ is such a tangent vector at the solution $\cA_0$, then it follows from the flow equations that $\partial_y\delta\cA=0$ and $\d_{\cA_0}\delta\cA=0$, which is easily seen to be incompatible with the Nahm pole boundary condition. 

Let us return to the Teichm\"uller TQFT partition function. The general construction \cite{WittenQM,WittenCS} of analytically-continued Chern-Simons theory implies that our path-integral can be rewritten as follows,
\beq
Z_{\rm Teichm}(W)=\int_{\mathcal{S}_{y_0}}D\cA\,\exp\left(ib^2\CS(\cA)\right)\,.\label{main1}
\eeq
It is an integral over ${\mathcal S}_{y_0}$ with the measure induced by the nowhere-vanishing holomorphic $\mathcal{G}_\CC^0$-invariant top-degree form $D\cA$ on $\Omega_\cA$. To understand the origin of this formula, one can look at the non-exact terms in the action (\ref{YMCS}). The choice of framing of $TW$ together with the boundary condition at $y=0$ defines a choice of framing of the gauge bundle on $W\times\mathcal{I}$. Then we can rewrite the bulk topological term as
\beq
\fr{ib^2}{4\pi}\int_V\tr(F\wedge F)=-ib^2\CS(A)|_{y=y_0}+ib^2\CS(\omega)\,.\label{rewr}
\eeq
The second term is canceled by the gravitational counterterm, introduced after (\ref{counter}). The second term combines with the boundary action at $y=y_0$ to produce precisely the Chern-Simons action in (\ref{main1}).
(For a detailed derivation of the relation between 4d super Yang-Mills and analytically-continued Chern-Simons, see \cite{WittenQM,WittenCS}.)

 The effective 3d theory (\ref{main1}) sitting at $y=y_0$ has gauge symmetry $\mathcal{G}^0$, which is a symmetry of $\mathcal{S}_{y_0}$ and the only part of the 4d gauge symmetry that we haven't used in fixing $A_y=0$.

The formula (\ref{main1}) may be slightly imprecise. The 4d path-integral is localized onto the space of solutions to the KW equations, which is fibered over the space of the boundary data $\mathcal{S}_{y_0}$. For a given point in $\mathcal{S}_{y_0}$, the bulk path-integral computes the Euler characteristic of the fiber over it, and this factor should be included into the path-integral, so we really have
\beq
Z_{\rm Teichm}(W)=\int_{\mathcal{S}_{y_0}}D\cA\,\chi_{\rm fiber}(\cA)\exp\left(ib^2\CS(\cA)\right)\,.\label{main2}
\eeq
We hope that with a generic choice of the metric, each fiber is simply a point, so that the extra factor is not needed. (This, for example, is the case for the analogous problem in the 2d sigma-model: for every point on the Hitchin section $\mathcal{T}$, there is one and precisely one time-independent solution to the KW equations on $C\times\RR\times\RR^+$ with the Nahm pole at $y=0$, approaching the given point at large $y$.) For the rest of this subsection, we simply ignore this factor, and soon we will obtain eq. (\ref{lefexp}), which automatically takes it into account.

For the path-integral (\ref{main1}) to make sense, the space of fields $\mathcal{S}_{y_0}$ should be a suitable integration cycle. This means first of all that the functional $h$ should be bounded from above on $\mathcal{S}_{y_0}$, so that the integral doesn't obviously diverge. To get some flavor of what this bound can be, we can evaluate $h$ on a given solution for small $y_0$, where the solution is well approximated by the Nahm pole (\ref{bc1})-(\ref{bc2}). In this case,
\beq
h\sim\fr{1}{y_0^3}{\rm Vol}(W)\,,
\eeq
where ${\rm Vol}(W)=-\fr{2}{3}\int_W\tr(e^3)$ is the volume of $W$. We may hope optimistically that 
$h$ is bounded on $\mathcal{S}_{y_0}$ by a constant $C(y_0)$, which for small $y_0$ is of the order of ${\rm Vol}(W)/y_0^3$.

It is also important that we are able to integrate by parts in the functional integral, which means that $\mathcal{S}_{y_0}$ should not have boundaries, except possibly for an asymptotic end in the directions, in which $h$ goes to minus infinity. If we denote by $\Omega_\cA^T$ the subspace of $\Omega_\cA$, on which $h<-T$, then $\mathcal{S}_{y_0}$ being a good integration cycle is equivalent to it being an element of the middle-dimensional relative homology $H(\Omega_\cA,\Omega_\cA^T)$, for $T\rightarrow\infty$. 

A slightly uncustomary property of Teichm\"uller TQFT is that its Hilbert space is of infinite dimension. Then, in computing the partition function on a three-manifold by cutting and gluing, one may encounter problems with the convergence of wave function integrals.  To some three-manifolds the theory simply does not assign a finite partition function. For example, if $W$ is a product $S^1\times C$ for a two-manifold $C$, then the partition function is infinite, since it is equal to the dimension of the Hilbert space on $C$. At the same time, for a class of three-manifolds a finite partition function has been constructed\footnote{See also a recent proposal in \cite{BGL}.} \cite{AK,AKnew}. Consistency of our proposed construction of the Teichm\"uller TQFT partition function requires

\noindent{\bf Conjecture 1.} {\it 
	Let $W$ be a three-manifold to which Teichm\"uller TQFT assigns a finite partition function. Then for any $y_0>0$ the subspace $\mathcal{S}_{y_0}\subset\Omega_\cA$ defines an element of $H(\Omega_\cA,\Omega_\cA^T)$, for arbitrarily large $T>0$. In particular, there exists a function $C(y_0)$ which provides a uniform bound on the functional $h$ for all flows on $W\times\RR^+_y$, defined by the equations (\ref{flow1})-(\ref{mmap}) and the boundary condition (\ref{bc1})-(\ref{bc2}).
}

(In fact, it would be sufficient to demand these properties to be true for some particular $y_0$. The bound would then also hold for any larger values of $y_0$.)

The properties that this conjecture claims for the Kapustin-Witten equations are highly non-trivial. Let us again summarize the arguments which make us think that this is worth considering. There is ample mathematical evidence that Teichm\"uller TQFT does exist and assigns finite partition functions to some three-manifolds. The path-integral (\ref{main1}) is the unique 3d covariant lift of the two-dimensional construction of section~\ref{tquant}. If Teichm\"uller TQFT possesses a path-integral definition, it must be (\ref{main1}). Furthermore, as we explain in section~\ref{redu}, the picture with $\cN=4$ super Yang-Mills on $W\times\mathcal{I}$, with a D5- and an NS5-type boundary conditions, can be obtained from six dimensions. The fact that this setup computes Teichm\"uller partition function is known from localization computations in the 3d-3d correspondence. Finally, in section~\ref{hyperb}, we will start directly with the definition of the integration cycle $\mathcal{S}_{y_0}$, and see that this definition matches nicely with some facts that are expected to hold for Teichm\"uller TQFT. In particular, we will propose and test an explicit bound $C(y_0)$ for $h$ on $\mathcal{S}_{y_0}$ for a hyperbolic three-manifold $W$. 

In our expectation, the most pessimistic scenario is that Conjecture~1 is false, but the functional integral (\ref{main2}) can still be somehow defined. Perhaps a slightly more robust definition is eq.~(\ref{vacs}), which will be explained later. The optimistic scenario is that Conjecture~1 is true, and we will assume this throughout the text, unless explicitly indicated otherwise.

Since one expects the integration cycles $\mathcal{S}_{y_0}$ to be homologous for any $y_0>0$, we will often omit the subscript.

As the last remark, let us show how unitarity of the theory manifests itself. For two closed oriented three-manifolds $W$ and $-W$ which differ by a choice of the orientation, partition functions of a unitary TQFT should differ by complex conjugation. One consequence of a change of the orientation is the change of sign of the Hodge star in the localization equations (\ref{flow1})-(\ref{flow2}). The Nahm pole boundary condition was defined with the vielbein such that $e\wedge e=\star e$. Then, because of the Hodge star, the sign of $e$ on $-W$ should also be chosen to be the opposite. These two sign differences are equivalent to $\phi\rightarrow-\phi$, that is, a complex conjugation of the integration cycle $\mathcal{S}\rightarrow\bar{\mathcal{S}}$. The orientation flip also causes the Chern-Simons action in the path-integral to change sign. As a result, we have
\beq
Z_{\rm Teichm}(-W)=\int_{\bar{\mathcal{S}}}D\cA\,\exp\left(-\fr{ib^2}{4\pi}\int_W\tr\left(\cA\d\cA+\fr{2}{3}\cA^3\right)\right)\,,
\eeq
which is indeed the complex conjugate of $Z_{\rm Teichm}(W)$, as long as $b^2$ is real.

\subsection{Lefschetz thimbles}\label{lefsec}
We recall that in the context of analytic continuation of complex integrals, the so-called Lefschetz thimbles are particular integration cycles which are indexed,  roughly speaking, by critical points of the action, and provide a basis for the space of admissible integration cycles. For an introduction to this machinery and for its application to analytically-continued Chern-Simons theory, see \cite{WittenQM} and \cite{WittenCS}.
We will not review these matters here.

Let $\Omega^{\rm flat}_\cA\subset\Omega_\cA$ be the space of flat PSL$(2,\CC)$ connections on $W$. By $\mathbf{a}$, $\mathbf{b},\dots$ we denote elements of $\Omega^{\rm flat}_\cA/\mathcal{G}_\CC$, the moduli space of flat bundles. The critical points of the Chern-Simons action are labeled by the elements of $\Omega^{\rm flat}_\cA/\mathcal{G}_\CC^0$, since we are working modulo  gauge transformations, homotopic to the identity. Such a critical point  is labeled by a pair $({\bf a},{s})$ of a flat PSL$(2,\CC)$ bundle ${\bf a}$ and an integer ${s}$ which parametrizes a lift of ${\bf a}$ to $\Omega_\cA/\mathcal{G}_\CC^0$. We often denote pairs $(\mathbf{a},{s})$, $(\mathbf{b},{s}'),\dots$ by $\mathbb{a}$, $\mathbb{b},\dots$. Let $\CS({\mathbf a})\in \CC/2\pi\ZZ$ be the Chern-Simons invariant of a flat bundle ${\bf a}$ and $\CS(\mathbb{a})\in\CC$ be the invariant of its lift. It is convenient to define 
\beq
s=-\fr{1}{2\pi}\re(\CS(\mathbb{a}))\,,
\eeq
so that it is not quite an integer, but takes values in\footnote{Despite the fact that the gauge group is SO$(3)$, $s$ takes values in a $\ZZ$-torsor and not a $\fr{1}{4}\ZZ$-torsor. The reason is that at the D5 boundary, the gauge bundle is identified with the tangent bundle, which has a trivial Stiefel-Whitney class. Because of that, all flat bundles $\mathbf{a}$ that may appear also need to have a trivial Stiefel-Whitney class.} $-\fr{1}{2\pi}\re\CS(\mathbf{a})+\ZZ$.

A critical point $\mathbb{a}$ defines a $\mathcal{G}^0_\CC$-orbit  $\mathcal{O}^\CC_{\mathbb a}$ in $\Omega_\cA^{\rm flat}$. This orbit has points with zero moment map~$\mu$ (\ref{moment}), if and only if the flat bundle $\mathbf{a}$ is semistable \cite{Corlette}. Since the requirement of zero moment map is a part of our localization equations, we will always restrict to semistable bundles. Then the subspace of $\mathcal{O}^\CC_{\mathbb a}$ with $\mu=0$ is a $\mathcal{G}^0$-orbit, to be denoted $\mathcal{O}_{\mathbb a}$, and in fact, $\mathcal{O}^\CC_{\mathbb a}=T^*\mathcal{O}_{\mathbb a}$. We sometimes denote by $\cA_{\mathbb{a}}$ some flat connection, belonging to $\mathcal{O}_{\mathbb a}$.

A Lefschetz thimble $\mathcal{C}_{\mathbb a}$, by definition, is obtained by downward gradient flows that originate from $\mathcal{O}_{\mathbb a}$. To be precise, this definition needs a small modification, when flat bundle $\mathbf{a}$ is a part of a moduli space. To simplify matters, throughout the paper we will restrict to the case that all flat connections are isolated. Removing this restriction should not change anything essential in our arguments.

Assuming that ${\mathcal S}$ is an admissible integration cycle, it can be expanded in Lefschetz thimbles,
\beq
{\mathcal S}=\sum_{\mathbb{a}} n_{\mathbb{a}}^{\mathcal{S}}\,\mathcal{C}_{\mathbb{a}}\,. 
\eeq
The coefficients in this expansion are computed by intersecting $\mathcal{S}$ with a system of dual cycles $\mathcal{C}^\vee_{\mathbb a}$, such that $\mathcal{C}_{\mathbb a}\cap\mathcal{C}^\vee_{\mathbb b}=\delta_{\mathbb{a},\mathbb{b}}$. As explained in \cite{WittenCS}, the cycle $\mathcal{C}^\vee_{\mathbb a}$ is constructed by upward gradient flows from an arbitrary fixed representative $\cA_{\mathbb a}$ of the orbit $\mathcal{O}_{\mathbb a}$. Then $n_{\mathbb a}^{\mathcal S}$ is the signed count of flows in $y\in[y_0,\infty)$ that begin at $\mathcal{S}$ and approach the flat connection $\cA_{\mathbb a}$ at infinity. 

But $\mathcal{S}$ itself is obtained by flows on $(0,y_0]$. Then we can as well count flows for $y\in(0,+\infty)$ that are conjugate to the Nahm pole at $y\rightarrow 0$ and approach the flat connection $\cA_{\mathbb a}$ at infinity. The signed count of such flows will be called $n_{\mathbb a}'$. We  pedantically introduce this new notation just for the unlikely case that $n_{\mathbb a}'\ne n_{\mathbb a}^{\mathcal S}$, which may happen if the factor $\chi_{\rm fiber}$ in the path-integral (\ref{main2}) is non-trivial. It is then obvious that the path-integral (\ref{main2}) is equal to
\beq
Z_{\rm Teichm}=\sum_{\mathbb{a}} n_{\mathbb{a}}'\,Z^{{\rm CS}}_{\mathbb{a}}(b^2)\,,\label{lefexp}
\eeq
which is true whether $\chi_{\rm fiber}$ is equal to one or not, so we no longer need to worry about this factor. The partition function $Z^{\rm CS}_{\mathbb a}(b^2)$ here is the Chern-Simons path integral with coupling constant $b^2\in\RR^+$ over the Lefschetz thimble $\mathcal{C}_{\mathbb a}$. 

Let $n_{\mathbb a}$ be the signed count of flows for $y\in(0,+\infty)$ that start with the Nahm pole (\ref{bc1})-(\ref{bc2}) at $y\rightarrow 0$ and approach the orbit $\mathcal{O}_{\mathbb a}$ for $y\rightarrow +\infty$. Superficially, this counting problem looks very similar to how we defined $n_{\mathbb a}'$, but there is a difference in how the gauge invariance is treated. For $n_{\mathbb a}'$, we required the flows to be conjugate to the Nahm pole at $y\rightarrow 0$ and to approach precisely a fixed representative $\cA_{\mathbb a}$ of the orbit $\mathcal{O}_{\mathbb a}$ for $y\rightarrow 0$. When $\mathbb{a}$ is irreducible, a global $\mathcal{G}_0$-gauge transformation maps one counting problem into another, so $n_{\mathbb a}'=n_{\mathbb a}$. But when $\mathbb{a}$ is reducible, with an isotropy group $H_{\mathbb a}\subset{\rm SO}(3)$, any flow that contributes to $n_{\mathbb a}$ becomes a moduli space which is a copy of $H_{\mathbb a}$, when we count flows for $n_{\mathbb a}'$. Since the Euler characteristic of a Lie group is zero, $n_{\mathbb a}'$ is zero for reducible critical points. (This is a direct analog of the fact that Lefschetz thimbles for irreducible critical points at Stokes walls cannot jump by thimbles, labeled by reducible critical points, see section 3.3.4 of \cite{WittenCS}.) Then we can rewrite eq.~(\ref{lefexp}) as
\beq
Z_{\rm Teichm}=\sum_{{\rm stable}\,\mathbb{a}} n_{\mathbb{a}}\,Z^{{\rm CS}}_{\mathbb{a}}(b^2)\,,\label{lefexp1}
\eeq
where the sum goes over stable critical points only. (Unstable critical points were excluded by the moment map condition, while strictly semistable ones are precisely the reducibles.)

Note that if the three-manifold $W$ has finite volume, then a critical point being reducible is equivalent to the corresponding flat bundle having holonomies in an abelian subgroup of PSL$(2,\CC)$. If the volume of the three-manifold is infinite, such abelian connections should not be considered as reducible, but rather are parts of moduli spaces, generated by would-be-gauge transformations at infinity. In such a case, our argument above would not prove that such connections do not contribute to the Teichm\"uller TQFT partition function. In this paper, we mostly focus on three-manifolds of finite volume, with the exception of some examples in section~\ref{cocol}. In those examples, however, the three-manifolds are such that there are no abelian flat connections. Thus, for the partition functions considered in the present paper, abelian connections never contribute.

There is a further minor point that has to be mentioned. Generically, a Lefschetz thimble $\mathcal{C}_{\mathbb a}$ is a nice integration cycle,  but it may fail to be compact (even modulo its part that goes to infinity in the field space), when there are downward flows from $\mathbb{a}$ to another critical point. Usually, the possibility of such flows is prevented by the existence of an integral which is conserved by the flow equations and is generically different for all critical points. However, we have specialized the Kapustin-Witten parameter to $t=-1$, in which case the conserved integral is $\re(\CS(\cA))$, and its value is the same for any two complex conjugate flat connections. Thus, any critical point with positive Chern-Simons volume $-4\pi\im\CS(\cA)$ has potentially non-compact Lefschetz thimble. Then in the expansion (\ref{lefexp1}), we have to perturb slightly $t$ from $-1$ and $b^2$ from being in $\RR^+$, for the Lefschetz thimbles and the counts $n_{\mathbb a}$ to make sense. We will argue however that in the case of hyperbolic three-manifolds, which is the most important example for the present paper, flat connections with positive Chern-Simons volume do not contribute, and we do not need to worry about these complications.

The fact that reducible connections do not contribute to the partition function is very important, so let us look at it from several different points of view. 

\begin{figure}
 \begin{center}
   \includegraphics[width=12cm]{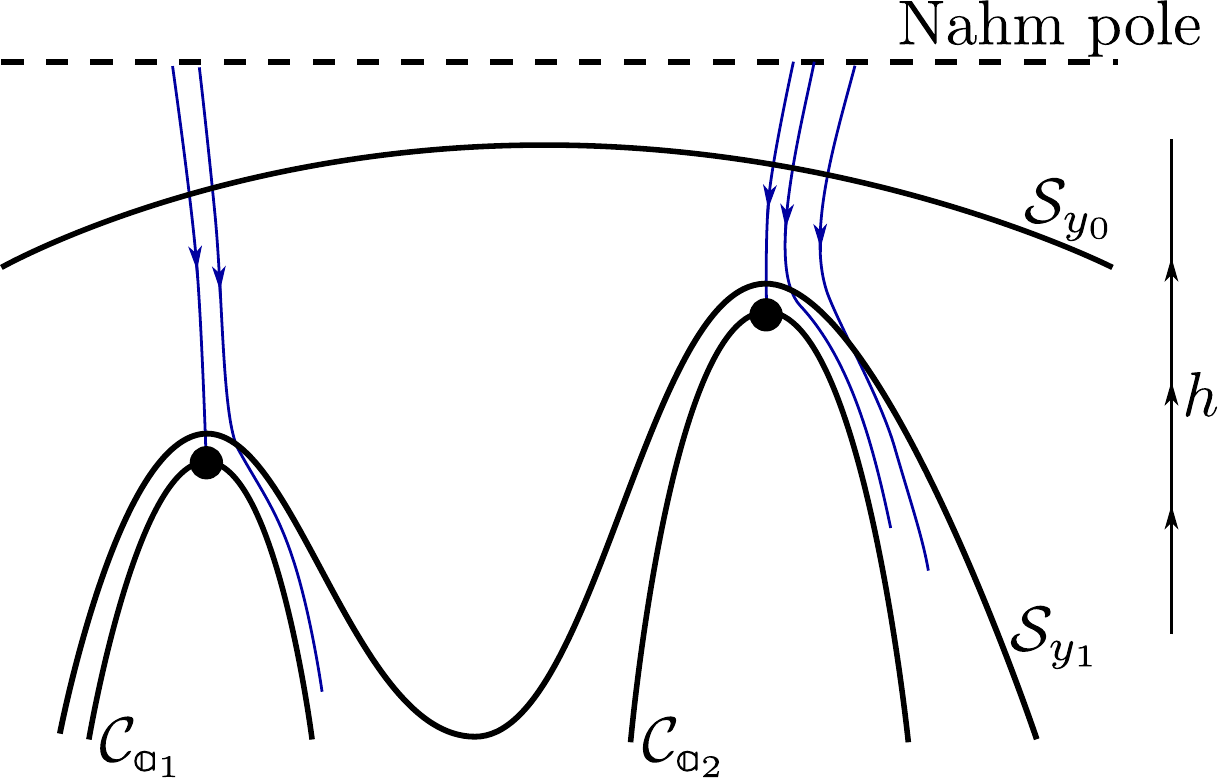}
 \end{center}
\caption{\small The functional $h$ grows in the upward vertical direction. The integration cycles $\mathcal{S}_{y_0}$ and $\mathcal{S}_{y_1}$, $y_0<y_1$, are obtained by flowing downward from the Nahm pole. If $y_1$ is large, the flows (shown in blue) that end up at finite points on $\mathcal{S}_{y_1}$ tend to spend most of their time near critical points $\mathbb{a}_1$, $\mathbb{a}_2$. This defines the decomposition of $\mathcal{S}_{y_1}$ into the Lefschetz thimbles $\mathcal{C}_{{\mathbb a}_1}$ and $\mathcal{C}_{{\mathbb a}_2}$.}
 \label{flowpic}
\end{figure}
First, we can stretch the interval $\mathcal{I}$ to make it very long, compared to any scale set by the three-manifold $W$. Then, intuitively, any flow that starts at the Nahm pole at $y=0$ and ends at a finite point in the field space at $y=y_0$ is expected to spend most of the time $y$ near some critical point. The space of flows, and therefore the integration cycle $\mathcal{S}$, can accordingly be decomposed into subspaces, labeled by the critical points near which the flows tend to stay. This is shown on figure~\ref{flowpic}, and, of course, is just another way to view the decomposition coefficients $n_{\mathbb a}'$. If the intermediate flat connection $\mathbb a$ has a non-trivial isotropy subgroup $H_{\mathbb a}$, then there is a symmetry on the space of flows that rotates the solution in the region $y<y_0/2$ by elements of $H_{\mathbb a}$. (As usual, it is important that we are counting flows that are \emph{conjugate} to the Nahm pole at $y\rightarrow 0$.) This creates an $H_{\mathbb a}$-worth of solutions and sets the corresponding algebraic count $n_{\mathbb a}'$ to zero.

Let us introduce the following objects,
\beqn
Z^{\rm D5}_{\bf a}(\tilde{q})&=&\sum_{s} n_{({\bf a},{s})}\exp\left(-2\pi ib^2s\right)\label{Dblock}\,,\\
Z^{\rm NS5,top}_{\bf a}(b^2)&=&Z_{({\bf a},{s})}^{\rm CS}(b^2)\,\exp\left(2\pi ib^2s\right)\,.\label{NSblock}
\eeqn
Recall that $\mathbf{a}$ is a flat bundle modulo all gauge transformations $\mathcal{G}_\CC$, not necessarily topologically trivial. A pair $(\mathbf{a},{s})$ is the same thing as what was denoted by $\mathbb{a}$. 
In the first line (\ref{Dblock}), we explicitly sum over ${s}\in-\fr{1}{2\pi}\re\CS(\mathbf{a})+\ZZ$, while keeping $\mathbf{a}$ fixed. This is the reason that we had to revert back to the notations with $\mathbf{a}$ and ${s}$ written our explicitly. The variable $\tilde q=\exp(2\pi i b^2)$ has already been introduced in section~\ref{tquant}. The right hand side of (\ref{Dblock}) is a Laurent series in $\tilde q$, up to an overall prefactor with a non-integer power of $\tilde{q}$. 

In the second line (\ref{NSblock}), the function $Z^{\rm NS5,top}_{\bf a}(b^2)$ is essentially the integral over a Lefschetz thimble $\mathcal{C}_{(\mathbf{a},s)}$, except that we have added a $c$-number $ib^2\re\CS(\cA_{({\bf a},{s})})$ to the action. What it gains us is that $Z^{\rm NS5,top}_{\bf a}(b^2)$, unlike the Lefschetz thimble integral, depends on the flat bundle $\mathbf{a}\in\Omega^{\rm flat}_\cA/\mathcal{G}_\CC$ and not on the lift $s$.

\begin{figure}
 \begin{center}
   \includegraphics[width=13cm]{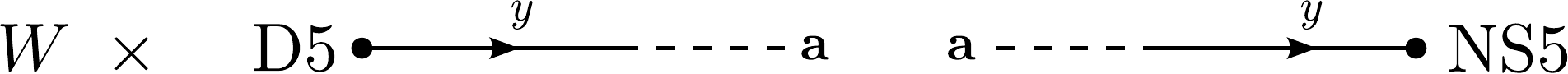}
 \end{center}
\caption{\small We can stretch and cut the interval $\mathcal{I}$ to factorize the partition function on $W\times\mathcal{I}$ into products of partition functions in two half-spaces $W\times\RR^+$. The partition functions in the left and the right half-space, respectively, are $Z^{\rm D5}_{\bf a}(\tilde{q})$ and $Z^{\rm NS5,top}_{\bf a}(b^2)$.}
 \label{vacua}
\end{figure}
The objects $Z^{\rm D5}_{\bf a}(\tilde{q})$ and $Z^{\rm NS5,top}_{\bf a}(b^2)$ are nothing but the partition functions of the $\cN=4$ super Yang-Mills in the half-spaces, as shown on figure~(\ref{vacua}). We will sometimes call them blocks. Note that they are labeled by flat bundles $\mathbf{a}$ modulo all gauge transformations, as it should be for the vacua of a quantum theory. (The theory in question is the quantum mechanics, obtained by compactifying the $\cN=4$ super Yang-Mills on $W$.) In the D5-block, the quantity $-2\pi ib^2s$ is the contribution of the bulk topological term. The sum in (\ref{Dblock}) goes over the instanton number. In the NS5-block $Z^{\rm NS5,top}_{\bf a}(b^2)$, the shift $2\pi ib^2s$ in the action comes from the fact that the bulk topological term, written in some gauge as a total derivative, gives a contribution not only at the location of the NS5 brane, but also at the infinite end. More details about these objects are given in section~\ref{secblocks}, where we also explain the notation ``top'' in $Z^{\rm NS5,top}_{\bf a}(b^2)$.

Using these half-space partition functions, we can rewrite the formula (\ref{lefexp1}) for the Teichm\"uller TQFT partition function as
\beq
Z_{\rm Teichm}=\sum_{{\rm stable}\,{\bf a}} Z^{\rm D5}_{\bf a}(\tilde q) Z^{\rm NS5,top}_{\bf a}(b^2) \,.\label{vacs}
\eeq
In good circumstances, the sum here goes over a finite set. This formula may be taken as our most clear proposal for the Teichm\"uller partition function. Note that $Z^{\rm D5}_{\bf a}(\tilde q)$ is generically a Laurent series. A substitute for {Conjecture~1} could then be the requirement that $Z^{\rm D5}_{\bf a}(\tilde q)$ is a finite polynomial, or at least can be naturally regularized, when $\mathbf{a}$ is stable and the three-manifold is such that the Teichm\"uller TQFT partition function is finite.

It may seem that there is something strange about eq.~(\ref{vacs}). The physical meaning of this formula is that we factorize the amplitude on $W\times \mathcal{I}$ by cutting the interval. The vacua of the effective quantum mechanics on $\RR_y$, obtained by compactifying the $\cN=4$ super Yang-Mills on $W$, are labeled by semistable flat bundles $\mathbf{a}$. Yet among the intermediate states in (\ref{vacs}) we only see stable flat bundles, which makes it look that something is wrong with the factorization. The explanation of this puzzle is the following.
The twisted super Yang-Mills theory has a so-far neglected complex field $\sigma$, which together with its conjugate $\bar\sigma$ is a part of a BRST multiplet
\beq
\Qb\sigma=0\,
,\quad \Qb\bar\sigma=\eta\,,\quad\Qb\eta=[\sigma,\bar\sigma]\,.\label{Qsigma}
\eeq
The localization equations for $\sigma$ say that it has to be zero, if the solution for $A$ and $\phi$ is irreducible. Equivalently, an irreducible complex connection $\cA$ induces a mass for all components of $\sigma$ in the Lagrangian. Irreducible flat connections ${\bf a}$ correspond to massive supersymmetric vacua of the quantum mechanics. But if ${\bf a}$ is invariant under a subgroup $H$ of the gauge group, the field $\sigma$ has zero modes, valued in the complexified Cartan $\mathfrak{t}_\CC^H$ of $H$. It means that classically ${\bf a}$ corresponds not to a single vacuum, but to a moduli space, labeled by a $\mathfrak{t}_\CC^H$-valued expectation value of $\sigma$. With our assumption that $W$ has finite volume, the corresponding zero modes are normalizable. Then they should be quantized, and, in particular, the blocks $Z^{\rm NS5, top}_{\bf a}$ and $Z^{\rm D5}_{\bf a}$ are valued not in numbers, but in wavefunctions. This situation has been explored in great detail section 5 of \cite{WittenQM}. In our case, let us denote by $\sigma_0$, $\bar\sigma_0$ and $\eta_0$ components of the fields, valued in $\mathfrak{t}_\CC^H\simeq\CC$. Then the BRST operator (\ref{Qsigma}) becomes the Dolbeault differential $\partial_{\bar\sigma_0}$ on $\CC$. Since $\CC$ is non-compact, there exist two natural sorts of cohomology of $\Qb$. The first is the usual $H_{\bar\partial}$, with a basis given by holomorphic polynomials $\sigma_0^k$, $k\ge 0$. The second is the cohomology with compact support $H^{c}_{\bar\partial}$, spanned by classes $\partial_{\sigma_0}^k\delta^{(2)}(\sigma_0,\bar\sigma_0)\eta_0$, $k\ge 0$. It is easy to see that the blocks $Z^{\rm D5}_{\bf a}$ and $Z^{\rm NS5}_{\bf a}$ both take values in $H^c_{\bar\partial}$. Indeed, we can act on them by inserting an operator $\tr\,\sigma^2$ at $y\rightarrow\infty$. This is a $\Qb$-invariant scalar operator, and the insertion point does not really matter. Thus, it can be moved to $y\rightarrow 0$, where $\sigma$ vanishes via the boundary condition. Therefore, the blocks take values in wavefunctions which are annihilated by $\tr\,\sigma^2$, and these must be elements of the cohomology with compact support. Whenever we encounter a reducible vacuum $\mathbf{a}$ in the decomposition of $Z_{\rm Teichm}$, we are instructed to pair two such wavefunctions. For two compactly-supported cohomology classes, the pairing is zero because of too many fermion insertions. 

It is instructive to compare the spaces of solutions to the localization equations on a long stretched interval $\mathcal{I}$ and on the two intervals of figure~{\ref{vacua}}. Let us call these setup I and setup II, respectively. Assume that we cut the interval at $y=y_0/2$ and obtain two disjoint intervals $\mathcal{I}_\ell\cup\mathcal{I}_r$, and the connection at $y=y_0/2$ belongs to the orbit $\mathcal{O}_{\mathbf a}$. The difference between setups I and II is that in the second case, we have two flat connections $\cA_\ell$ and $\cA_r$ in the orbit $\mathcal{O}_{\mathbf a}$ and two copies of the gauge group $\mathcal{G}$, sitting at the point $y=y_0/2$. If $\mathbf{a}$ is irreducible, then one copy of the gauge group $\mathcal{G}$ is eaten, when we align $\cA_\ell$ and $\cA_r$, and the problem becomes the same as for the single interval $\mathcal{I}$. When $\mathbf{a}$ is reducible, setup II has more gauge symmetry than setup I precisely by a factor of  the isotropy group $H_{\mathbf a}$. It means that the moduli space in setup I is larger by the same factor. This is the same factor of $H_{\mathbf{a}}$ that we saw in the proof that $n_{\mathbb a}'=0$ for abelian $\mathbb a$. The role of the multiplet (\ref{Qsigma}) is precisely to ensure that factorization in quantum gauge theory is consistent with this ``mismatch'' in the classical configuration spaces.

The reason we went into so much detail with these explanations is that later we will use the same arguments to show, why reducible flat connections can be discarded in complex Chern-Simons theory.

\section{Teichm\"uller TQFT on a hyperbolic three-manifold}\label{hyperb}
Our definition of the integration cycle $\mathcal{S}$ was so far quite abstract. Now assume that $W$ is a complete finite-volume hyperbolic three-manifold. It may either be closed or have cusps. 

We would like to conjecture that, when $W$ has a complete hyperbolic metric, the integration cycle $\mathcal{S}$ has a simple universal description. At first this may seem like a surprise. Indeed, the expansion coefficients $n_{\mathbb{a}}$ or, equivalently, the blocks $Z_{\bf a}^{\rm D5}$ are expected to be non-trivial invariants of the three-manifold. They should contain much of the information about Chern-Simons partition functions for rank one real and complex gauge groups. However, we are interested not in all blocks, but only in the ones labeled by irreducible flat connections,\footnote{Another restriction is that we fix the Kapustin-Witten parameter $t=-1$. As is explained in section~\ref{secblocks}, one expects that the blocks, most easily related to Chern-Simons partition functions, are the ones computed at $t=0$ or $t=\infty$. } and those, we conjecture, are very simple for a complete hyperbolic $W$. 

We recall a few basic facts about flat connections on a hyperbolic three-manifold. 
Let $e$ and $\omega$ be the vielbein and the Levi-Civita connection for the hyperbolic metric. From them, we can construct two flat PSL$(2,\CC)$ connections, the geometric $\cA_{\rm geom}=\omega+ie$ and the conjugate geometric $\cA_{\rm \bar{geom}}=\omega-ie$. Their flatness is equivalent to the defining equation of the Levi-Civita connection $\d_\omega e=0$ and the constant negative curvature condition $\d\omega+\omega\wedge\omega=e\wedge e$. The flat bundle $\mathbf{a}_{\rm geom}$ corresponding to $\cA_{\rm geom}$ comes from the geometric structure of $W$, that is, from the isomorphism with a quotient of the hyperbolic space $H_3$ by a subgroup of its isometry group PSL$(2,\CC)$. According to the Mostow theorem, the geometric flat bundle is rigid. Note that\footnote{To avoid possible confusion, we stress that by the Chern-Simons invariant of a complex flat connection we mean the functional (\ref{csa}) and not just its real part.}
\beq
\CS(\cA_{\rm \bar{geom}})=\CS(\omega)+\fr{i}{2\pi}{\rm Vol}(W)    \,.\nonumber
\eeq

It is known \cite{Reznikov} that for any flat bundle ${\bf a}$ on a hyperbolic three-manifold,
\beq
|\im\CS({\bf a})|\le\fr{1}{2\pi}{\rm Vol}(W)\,.\label{bound}
\eeq
We will assume that our three-manifold is generic, so that equality holds for ${\mathbf a}_{\rm geom}$ and ${\mathbf a}_{\rm \bar{geom}}$ only.

The key to identifying the integration cycle $\mathcal{S}$ on a hyperbolic $W$ is a conjecture by Andersen and Kashaev \cite{AK}. By studying examples of knot complements, they proposed that in the semiclassical limit, the Teichm\"uller TQFT partition function behaves as
\beq
b^2\rightarrow\infty:\quad |Z_{\rm Teichm}|\sim\exp\left(-\fr{b^2}{2\pi}{\rm Vol}(W)\right)\,.\label{akassymp}
\eeq

Although in this paper we only consider gauge group SO$(3)$, it should be possible to construct analogs of Teichm\"uller TQFT for any compact simple Lie group. The corresponding higher Teichm\"uller spaces \cite{FG} are defined as particular Lagrangian subspaces in Hitchin moduli spaces for these gauge groups. We expect that all our statements should carry over to those cases, with the analog of the geometric flat connection defined using the principal embedding of $\mathfrak{su}(2)$. In particular, it should be possible to play this game for PSU$(N)$, taking the limit $N\rightarrow\infty$. The Teichm\"uller TQFT partition functions can alternatively be computed in 3d $\cN=2$ superconformal theories, via the 3d-3d correspondence. In \cite{Hologr}, these partition functions were analyzed in holography, and it was found that the leading large $N$ asymptotics is precisely (\ref{akassymp}), provided that one uses the volume for the principally-embedded geometric flat connection. (The corresponding group-theoretical factor introduces an $N^3$ scaling.) What is nice about this argument is that it is valid universally for any hyperbolic $W$. In the papers of Andersen and Kashaev, the formula (\ref{akassymp}) was tested for $N=2$, but only for a few knot complements.

In light of these experimental facts, we propose that the asymptotics (\ref{akassymp}) hold for the Teichm\"uller TQFT partition function on an arbitrary complete finite-volume hyperbolic three-manifold.

In the decomposition (\ref{lefexp1}), the contribution of a critical point $\cA_{\mathbb{a}}$ has leading semiclassical asymptotics
\beq
b^2\rightarrow\infty:\quad Z_{{\mathbb a}}^{\rm CS}\sim \exp\left(ib^2\CS(\cA_{\mathbb{a}})\right)\,.
\eeq
Clearly, this agrees with (\ref{akassymp}) for $\cA_{\mathbb a}$ being the conjugate geometric flat connection $\cA_{\rm \bar{geom}}$. Moreover, the inequality (\ref{bound}) implies that the term for $\cA_{\rm \bar{geom}}$ is the most subleading in the expansion (\ref{lefexp1}) over the Lefschetz thimbles. Since $Z_{\rm Teichm}$ decays as fast as this term, no other flat connections can contribute in the expansion.

To be precise, the same flat bundle ${\bf a}_{\bar{\rm geom}}$ corresponds to many Lefschetz thimbles, labeled by different lifts $s$. Chern-Simons partition functions on these thimbles differ by powers of $\tilde q=\exp(2\pi i b^2)$, up to a framing factor. The integration cycle $\mathcal{S}$ must be the Lefschetz thimble for one particular lift of ${\bf a}_{\bar{\rm geom}}$, which we will continue to call $\cA_{\rm \bar{geom}}$. Indeed, if it were a combination of Lefschetz thimbles for different lifts, the asymptotics (\ref{akassymp}) would involve the absolute value of a Laurent polynomial in $\tilde{q}$ as a prefactor. This does not seem to agree with the experimental data. As will be clear from the following discussion, precisely which lift of ${\mathbf a}_{\rm \bar{geom}}$ to $\Omega_\cA/\mathcal{G}^0_\CC$ to take is determined by the choice of framing.

These implications of the Andersen-Kashaev conjecture for the Chern-Simons integration cycle have been previously pointed out in \cite{Hologr} and \cite{BGL}. We hope that our paper adds something to the understanding of the origin and the nature of this Chern-Simons theory, as well as the possible origin of these phenomena from the properties of the Kapustin-Witten equations, as will be pointed out soon.

Note that (\ref{akassymp}) is similar to the celebrated volume conjecture, except for the minus sign. But this sign makes the statement very powerful. The volume conjecture says that the Chern-Simons integration cycle, in appropriate setting, must contain the Lefschetz thimble for $\cA_{\rm geom}$. The formula (\ref{akassymp}) says that the Teichm\"uller TQFT integration cycle must contain the Lefschetz thimble for $\cA_{\rm \bar{geom}}$, and nothing else. 

We can now propose that the Andersen-Kashaev conjecture is equivalent to the following

{\bf Conjecture 2.} {\it Let $W$ be a complete hyperbolic three-manifold. Consider the Kapustin-Witten equations (\ref{KWeqs}) for gauge group SO$(3)$ and $t=-1$ on $W\times \RR^+_y$. At $y=0$, we impose the Nahm pole boundary condition (\ref{bc1})-(\ref{bc2}). At $y\rightarrow\infty$, we require the solution to approach an irreducible flat PSL$(2,\CC)$ bundle ${\bf a}$. Let $n_{{\bf a}, s}$ be the signed count of solutions of instanton number ${s}$. Then $n_{{\bf a},{s}}$ is zero for all ${\bf a}$, except for ${\bf a}_{\bar{\rm geom}}$, and for all instanton numbers, except for one particular $s_0$. The number $n_{{\bar{\rm geom}},{s_0}}$ is non-zero.}

A remark may be in order. The statement could have been equivalently formulated for the flow equations (\ref{flow1})-(\ref{mmap}) which differ from the Kapustin-Witten equations by setting $\phi_y$ to zero. Indeed, when the flat bundle ${\bf a}$ is irreducible, $\phi_y$ has to vanish\footnote{The proof is a simple exercise in integration by parts, which we leave to the reader.} for $y\rightarrow\infty$. Since it also vanishes for $y\rightarrow 0$, it will be automatically zero everywhere for any solution to the Kapustin-Witten equations.

Also, we did not require the hyperbolic manifold to be of finite volume. This condition was necessarily for all the examples that we used to motivate Conjecture~2, but some experimentation in section~\ref{cocol} suggests that the conjecture may hold as well for more general complete hyperbolic three-manifolds.

Our conjecture may seem to be too strong a statement. Perhaps the mildest question that arises is, why $n_{\bar{\rm geom},s_0}$ is non-zero for any hyperbolic three-manifold. The following consideration gives a hint. We are free to choose any metric to define the D5 boundary condition (\ref{bc1})-(\ref{bc2}). On a hyperbolic three-manifold, we may of course choose the hyperbolic metric\footnote{This statement may not look so innocent, if we want to replace {e.g.} an $S^3$ with a monodromy operator of parabolic conjugacy class along a knot $K$ by a complete hyperbolic three-manifold $S^3\setminus K$ with a cusp. From looking at the classical flat connections and the allowed gauge transformations, one can convince oneself that equivalence of these two setups in Chern-Simons theory at least is not implausible.} for that purpose. Then it is easy to see that the flow equations always have a model solution
\beqn
\phi=-e\coth y\,, \quad A=\omega\,, \label{model}
\eeqn
which satisfies the correct boundary condition at $y=0$. At $y\rightarrow +\infty$, the complexified gauge field $A+i\phi$ approaches precisely the conjugate geometric connection! This is the first non-trivial sign that our definition of the integration cycle $\mathcal{S}$ may be on the right track. (Still, this does not quite prove even that $n_{\bar{\rm geom},s_0}\ne 0$, since in the signed count of flows there may be other flows that cancel the contribution of the model solution above.) The most vexing question, of course, is why all the other coefficients $n_{{\bf a},{s}}$ are zero. We now propose and test a statement which, if true, would explain the mechanism behind Conjecture~2. 

\subsection{A stronger conjecture}\label{stronger}
The equations (\ref{flow1})-(\ref{flow2}) describe the downward gradient flow for the functional
\beq
h=-4\pi\im\CS(\cA)\,.
\eeq
The value of $h$ for the model solution (\ref{model}), evaluated at flow time $y$, is equal to
\beq
h_0(y)=(\coth^3 y-3\coth y){\rm Vol}(W)\,.
\eeq
The following is a pure guess, for which we will try to build some mathematical evidence.

{\bf Conjecture 3.} {\it Consider a solution to the flow equations (\ref{flow1})-(\ref{mmap}) with the boundary condition (\ref{bc1})-(\ref{bc2}), where the data $e$ and $\omega$ are those for the hyperbolic metric. Then for any $y>0$, the value of the functional $h$, evaluated on the solution at flow time $y$, is no greater than $h_0(y)$, with the equality holding only for the model solution (\ref{model}).}

This statement provides a bound $C(y)$, whose existence was a big part of  Conjecture 1. $C(y)$ is equal to $h_0(y)$. It also implies Conjecture 2. Indeed, if any solution other than (\ref{model}) would approach some flat connection at $y\rightarrow\infty$, this flat connection would violate the bound (\ref{bound}), which is impossible. We also stress that Conjecture 3 concerns the flow equations (\ref{flow1})-(\ref{flow2}) and not the Kapustin-Witten equations (\ref{KWeqs}). Equivalently, it prohibits solutions to the Kapustin-Witten equations that would for $y\rightarrow\infty$
\renewcommand{\theenumi}{\alph{enumi}}
\begin{enumerate}
\item approach an irreducible connection;
\item approach a reducible connection and $\phi_y=0$\,.
\end{enumerate}
Thus, it does not imply that the blocks $Z^{\rm D5}_{\bf a}$ are zero for all reducible flat connections ${\bf a}$. The corresponding solutions of the Kapustin-Witten equations just need to have $\phi_y\ne 0$ at infinity. The experimental evidence for Conjecture 3 that we are about to present would fail, if we tried to apply it to the Kapustin-Witten equations instead of the flow equations.

\subsubsection{A symmetric ansatz}
For an illustration and a simple test, let us make the maximally-symmetric ansatz (inspired by \cite{Hen})
\beq
A=\omega+eg(y)\,,\quad \phi=ef(y)\,,\label{syman}
\eeq
where $f$ and $g$ are some functions of $y$. The functional $h$ reduces to
\beq
{h}=\left(1+g^2-\fr{1}{3}f^2\right)f\,,
\eeq
where we set ${\rm Vol}(W)=1/3$ for convenience. The flow equations are the downward gradient flow for $h$, defined using the metric $\int\d y((\delta f)^2+(\delta g)^2)$. The boundary condition requires $f$ to approach $-1/y$ and $g$ to vanish for $y\rightarrow 0$.

\begin{figure}
 \begin{center}
   \includegraphics[width=10cm]{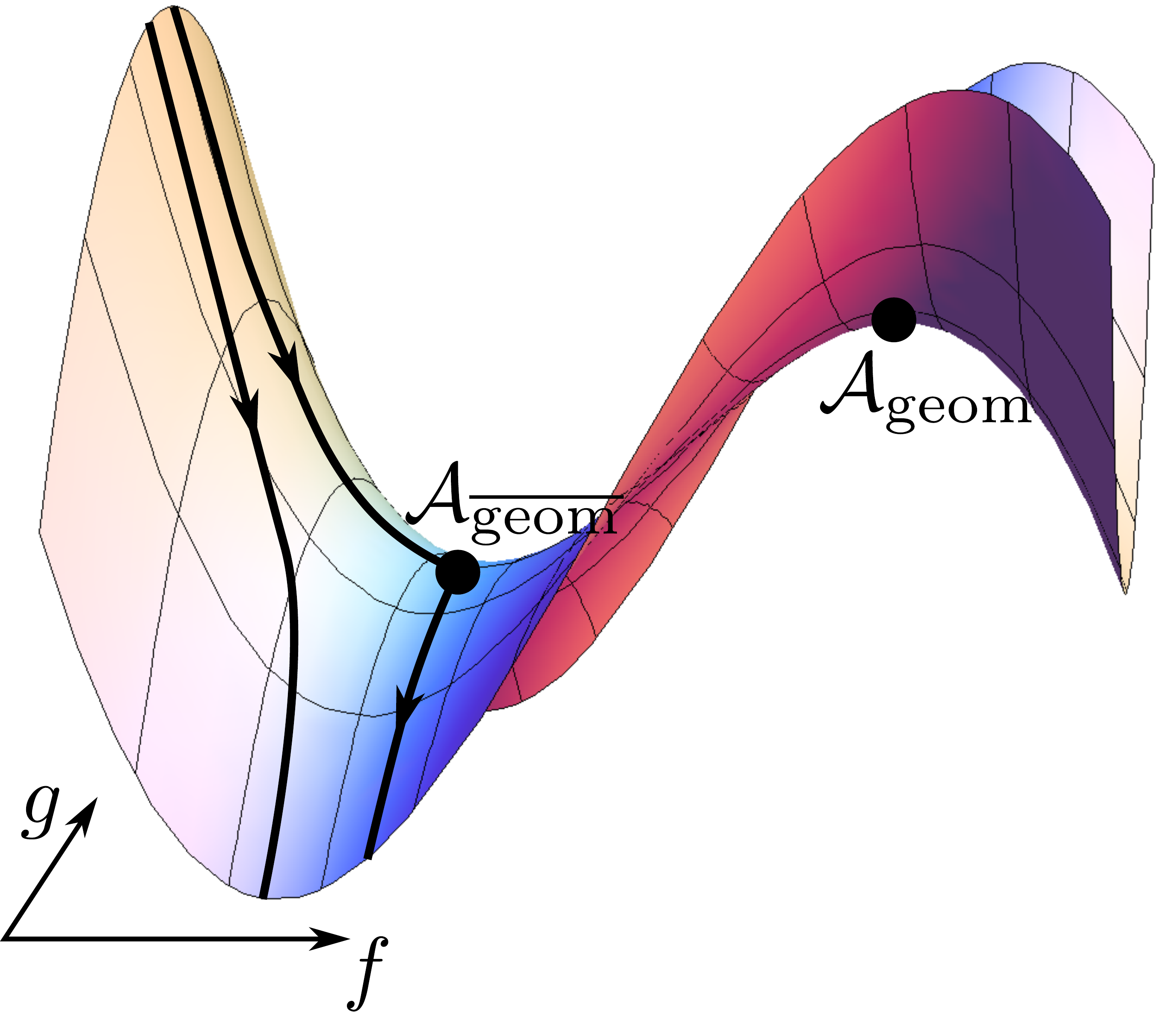}
 \end{center}
\caption{\small Profile of the function $h(f,g)$. The upper and the lower critical points $(1,0)$ and $(-1,0)$ correspond to the geometric and the conjugate geometric flat connections, respectively. The model solution descends from the hill on the left and ends at $\cA_{\bar{\rm geom}}$.}
 \label{profile}
\end{figure}
The profile of $h$ is shown on figure~\ref{profile}. There are two critical points, $\cA_{\rm \bar{geom}}$ at $(f,g)=(-1,0)$ and $\cA_{\rm geom}$ at $(f,g)=(1,0)$.
The flows that we are interested in descend from the hill on the left of the figure. The model solution goes along the ridge of the hill and ends up at the critical point $\cA_{\rm \bar{geom}}$. Conjecture 3 says essentially that any other flow slides down faster than the flow along the ridge. This claim may not be entirely obvious, but it is true, as we prove in appendix\ref{apsyman}.

\subsubsection{Perturbative expansion}\label{pertexp}
We can make a more stringent test of Conjecture~3 by solving the flow equations perturbatively for small $y$, and computing the functional $h$ on the solution. A big part of the computation has already been done in \cite{Hen} (see also \cite{MazzeoWitten}).

The Nahm pole boundary condition identifies the gauge bundle on $W$ with the tangent bundle, and preserves the diagonal subgroup of SO$(3)_{\rm gauge}\times{\rm SO}(3)_{\rm spin}$. Under this diagonal subgroup, the space of adjoint-valued one-forms $\Omega^1_{\rm adj}$ decomposes into subspaces of tensors of spin zero, one and two
to be denoted by $V_-$, $V_0$ and $V_+$, respectively. We will denote the corresponding projections of an adjoint-valued one-form by subscripts, {e.g.} $c=c_-+c_0+c_+$. In the orthonormal basis given by the vielbein, $c_0$ is the antisymmetric part of the three-by-three matrix $c$, $c_-$ is the trace part and $c_+$ is the symmetric traceless part. Let us parametrize the fields as
\beq
A=\omega+a\,,\quad \phi=-e\coth y+\varphi\,,\label{ansatz}
\eeq
where $a$ and $\varphi$ are adjoint-valued one-forms. For small $y$, the solution to the equations (\ref{flow1}-\ref{mmap}) can be expanded as
\beqn
\varphi&=&\varphi_1 y+\varphi_2 y^3+\dots\,,\nnr
a&=&a_1 y^2+a_2 y^4+\dots\,.
\eeqn
One finds for the first non-trivial order, 
\beqn
\varphi_1&=&c_+\,,\label{phi1}\\
a_1&=&c_--(\star\d_\omega c_+)_0-\fr{1}{3}(\star\d_\omega c_+)_+\,,
\eeqn
where $c_-\in V_-$ and $c_+\in V_+$ are the zero modes that are not constrained by the equations. All higher-order terms can be expressed in terms of $c_-$ and $c_+$ via the equations. (In particular, perturbative solutions are indeed parameterized by 1+5=6 real functions on $W$, as was claimed in section~\ref{exocont}.)

We would like to compute the functional $h(y)-h_0(y)$ for small $y$. It can be expanded as
\beq
\fr{1}{2}(h(y)-h_0(y))=\delta_1 y+\delta_2 y^3+\dots\,.
\eeq
We introduce a notation
\beq
(a,b)=-\int\tr(a\wedge\star b)
\eeq
for the symmetric positive-definite bilinear form on $\Omega^1_{\rm adj}$. We also denote $|a|^2\equiv(a,a)$. In appendix~\ref{AppEx}, we perturbatively solve the equations to the order which is necessary to compute $\delta_1$ and $\delta_2$. This gives the result
\beqn
\delta_1&=&-\fr{2}{5}|c_+|^2\,\label{delta1}\\
\delta_2&=&-\fr{37}{105}|c_+|^2-\fr{6}{7}|c_-|^2-\fr{13}{7}|(\star\d_\omega c_+)_0|^2-\fr{3}{7}|(\star\d_\omega c_+)_+|^2+\fr{16}{21}\int\tr(c_+^3)\label{delta2}\,.
\eeqn
Miraculously, all terms here are negative-definite, except for one. The last term in $\delta_2$ is proportional to the integral of the determinant of the $3\times 3$ matrix $c_+$. It is a symmetric traceless matrix (in an orthonormal basis), and its determinant is generically non-zero and can have either sign. However, its contribution is beaten by $-\fr{2}{5}|c_+|^2$ in $\delta_1$, as long as $y\delta_1+y^3\delta_2$ gives a reasonable approximation. We conclude that Conjecture 3 has passed the test. In appendix~\ref{AppEx}, we also consider the general case of Kapustin-Witten equations, that is, the case when $\phi_y\ne 0$. Then there exists an extra mode that, it seems, violates the negativity of $h-h_0$, although we do not quite prove this.

\section{Dualities}\label{dual}
It has been observed in the literature that Teichm\"uller TQFT should be somehow related to SL$(2,\CC)$ complex Chern-Simons theory at integer level $k=1$. We propose that there in fact exists a duality between the two theories, which is generated by a particular element of the SL$(2,\ZZ)$ S-duality group of the $\cN=4$ super Yang-Mills.

One encounters a serious difficulty in applying dualities to the brane configuration that we have used for Teichm\"uler TQFT. As a result, the relation with complex Chern-Simons theory will be partly conjectural, and the integration cycle in the dual theory will be fixed by an indirect argument.

We start with a discussion of simple half-BPS boundary conditions in $\cN=4$ super Yang-Mills and their S-duality. The motivation is two-fold. First, this will help to appreciate the problems in properly establishing the duality with complex Chern-Simons theory. Second, we would like to gain better understanding of the blocks that appear in the decomposition (\ref{vacs}) of the Teichm\"uller TQFT partition function. More specifically, one might think that there is a clash between S-duality and our conjecture about the integration cycle for Teichm\"uller TQFT. We proposed that on a hyperbolic three-manifold, the whole partition function is equal to the path-integral on a particular Lefschetz thimble. Naively, S-duality suggests that this integral is a holomorphic block, that is, a power series in $q=\exp(2\pi i/b^2)$. However, the Teichm\"uller TQFT partition function in known examples  does not seem to admit such a presentation. (Instead, it can be factorized into a sum of products of series in $q$ and $\tilde{q}$.) We explain what is wrong with this argument, and on the way make contact with the results of \cite{GMP} on resurgence in Chern-Simons theory.

\subsection{Half-BPS boundary conditions, blocks and Lefschetz thimbles}\label{secblocks}
In flat space, the physical 4d $\cN=4$ theory has SO$(4)$ Lorentz and SO$(6)_R$ R-symmetry group. The supercharges transform under these groups as $({\bf 2},{\bf 1})\otimes {\bf 4}\oplus ({\bf 1},{\bf 2})\otimes {\bf \bar 4}$. To put the theory on a product four-manifold $W\times\RR^+$, one does a twist by a subgroup SO$(3)_X\subset {\rm SO}(6)_R$, embedded in the obvious way. This leaves unbroken another subgroup of the R-symmetry, which we denote SO$(3)_Y$. Its Cartan subgroup is the U$(1)_{\rm gh}$ ghost number symmetry. Let SO$(3)_W'\simeq {\rm diag}({\rm SO}(3)_W\times {\rm SO}(3)_X)$ be the Lorentz group in the twisted theory on $W\times\RR^+$. Under the product SO$(3)_W'\times {\rm SO}(3)_Y$, the supersymmetries transform as 
\beq
({\bf 1}\otimes {\bf 2}\oplus {\bf 3}\otimes {\bf 2})\otimes V_2\,,\label{V2}
\eeq
where $V_2$ is an invariant two-dimensional vector space.

We choose a 1/2-BPS boundary condition which preserves the bosonic symmetries SO$(3)_W'\times {\rm SO}(3)_Y$ of the twisted theory. Then it also preserves an SO$(3)_Y$-doublet of scalar supercharges, of which we pick the element of ghost charge plus one and declare it to be the topological supercharge $\Qb$. The doublet is tensored with a fixed vector in $V_2$, determined by the boundary condition. This vector can be parameterized projectively by a possibly infinite complex number $t$, which is precisely the Kapustin-Witten parameter. Similarly, the boundary condition preserves an SO$(3)_Y$-doublet of vector supercharges, times a vector in $V_2$, which we parameterize by some $t'$. If the boundary condition preserves the Lorentz symmetry SO$(3)_W$ of the untwisted theory, then $t'=t$, but in general this need not be so. 

A class of half-BPS boundary conditions can be obtained (at least for classical gauge groups) from a brane construction, where a stack of D3-branes in the type IIB string theory ends on a $(p,q)$-fivebrane \cite{GWbc,Janus}. One can also turn on a flux of the fivebrane U$(1)$ gauge field. In order to preserve the twisted Lorentz group, the field strength should be proportional to the symplectic form on $T^*W$, and therefore depends on a single parameter.  It is not hard to show, as we do in appendix~\ref{5branesusy}, that such $(p,q)$-fivebrane boundary condition preserves supersymmetries with
\beqn
t&=&-\ex^{-i\ang_{p,q}}\tan\left(\fr{\pi}{4}-\fr{3\beta}{2}\right)\,,\label{t}\\
t'&=&-\ex^{-i\ang_{p,q}}\tan\left(\fr{\pi}{4}+\fr{\beta}{2}\right)\,,\label{tp}
\eeqn
where $\ang_{p,q}=\arg(p\tau+q)$, $\tau$ is the complexified string or Yang-Mills coupling, and $\beta$ is a $2\pi$-periodic angle, related to the magnitude of the fivebrane field strength. A $(-p,-q)$ boundary condition with angle $\beta$ preserves the same supersymmetry as a $(p,q)$ boundary condition with angle $\pi-\beta$, and the two are in fact equivalent. The boundary conditions with $\beta=\pm\pi/2$ are expected to be pathological, as we will see in some examples. Then it is enough to restrict to $\beta\in(-\pi/2,\pi/2)$.

The S-duality group SL$(2,\ZZ)$ acts on the 1/2-BPS boundary conditions \cite{GWS}. In our conventions, the generator $T$ transforms a boundary condition of type $(p,q)$ into the one of type $(p,q-p)$, and shifts the gauge coupling $\tau$ and the canonical parameter $\calK$ by one. It leaves the supersymmetry parameters $t$ and $t'$ invariant. The generator $S$ transforms type $(p,q)$ into type $(q,-p)$. It changes the gauge coupling $\tau$ to $-1/\tau$ and the canonical parameter $\calK$ to $-1/\calK$, and multiplies the supersymmetry parameters $t$ and $t'$ by a phase $-\tau/|\tau|$. The angle $\beta$ is duality-invariant. These transformation rules are consistent with the formulas (\ref{t}) and (\ref{tp}) and with the definition  (\ref{canonical}) of the canonical parameter.

\subsubsection{D5-type boundary condition}\label{d5sec}
In our notations, $(p,q)=(0,1)$ corresponds to a D5-brane. In (\ref{t})-(\ref{tp}) in this case $\ang_{p,q}=0$, so $t$ and $t'$ are real. We parameterize $\RR^+$ (perhaps we should better call it $\RR^-$) by $y<0$. The boundary condition at $y=0$ is the tilted Nahm pole. It prescribes the fields to approach at $y\rightarrow 0$ a model singularity
\beqn
A&=&\fr{e}{y}\sin\beta+\omega+\dots\,,\label{betanahm1}\\
\phi&=&-\fr{e}{y}\cos\beta+\dots\label{betanahm2}\,,
\eeqn
where $\beta$ is the angle determined by the fivebrane gauge field flux. As before, $\omega$ and $e$ are the Levi-Civita connection and the vielbein on $W$. It is straightforward to see that for these fields the singular terms in the Kapustin-Witten equations (\ref{KWeqs}) vanish, if $t$ is related to $\beta$ as in (\ref{t}), with $\ang_{p,q}=0$. With some work, one can also verify that this model singularity is invariant under the vector supersymmetries with parameter $t'$ (\ref{tp}), see appendix~\ref{5branesusy}. Interestingly, there are three possible values of the Nahm pole angle $\beta$ for each real value of the Kapustin-Witten parameter $t$.

Consider the theory in the half-space $W\times \RR^+$ with the tilted Nahm pole at $y=0$ and with the fields approaching a fixed flat bundle ${\bf a}$ at infinity. It requires some care to define the instanton number for a gauge field with a singularity (\ref{betanahm1}). Let us set $A'=A+f(y)\phi$, where $f(y)$ is a function with $f(y)=0$ for $y<-3\eps$ and $f(y)=\tan\beta$ for $y>-2\eps$. (By assumption, $\beta\ne\pm\pi/2$, since for those values the boundary condition is not expected to make sense.) The instanton number is set to be
\beq
s=\fr{1}{8\pi^2}\int_{\RR^+\times W} \tr(F'\wedge F')+\fr{1}{2\pi}\CS(\omega)\,,
\eeq
where $F'$ is the field strength for $A'$. Since $A'$ is non-singular, this instanton number is finite. Clearly, it is also gauge-invariant. Equivalently we can write it as
\beqn
s&=&\fr{1}{8\pi^2}\int_{y<-\eps} \tr(F\wedge F)\nonumber\\
&+&\fr{1}{8\pi^2}\int_{y=-\eps}\tr\left(\fr{2}{3}\tan^3\beta\,\phi\wedge\phi\wedge\phi+\tan^2\beta\,\phi\wedge\d_A\phi+2\tan\beta\,\phi\wedge F\right)\nonumber\\
&+&\fr{1}{2\pi}\CS(\omega)\,.
\eeqn
The counterterm in the second line subtracts the singularity. 

With the boundary condition (\ref{betanahm1})-(\ref{betanahm2}), the instanton number $s$ is $\Qb$-invariant and independent of the metric on $W$, but does depend on the framing. It takes values in 
\beq
s=\fr{1}{2\pi}\re\CS({\bf a})\quad{\rm mod}\,\,\ZZ\,,\label{smodz}
\eeq
where ${\bf a}$ is the flat connection at infinity.

The partition function on $W\times\RR^+$ is the so-called holomorphic or homological block, which we will also call a D5-block. It is equal to
\beq
Z^{{\rm D5},\beta}_{\bf a}(q)=\sum_s n_{{\bf a},s}^\beta q^{s}\,,\label{zd5}
\eeq
where ${q}=\exp(2\pi i\calK)$ and $n_{{\bf a},s}^\beta$ is the signed count of solutions\footnote{A mathematically-rigorous definition of this counting problem is not yet available due to problems with compactness of the moduli spaces \cite{Taubes}. Physics provides some examples of $q$-series that are good candidates for $Z^{\rm D5}_{\mathbf a}$, so one may hope that eventually it will be possible to make sense of the counting problem.} to the Kapustin-Witten equations with the Nahm pole of angle $\beta$ and the twisting parameter $t$ determined by $\beta$.

One can make several remarks on this formula. First, it is a Laurent series in ${q}$, up to an overall prefactor with a non-integer power of ${q}$, determined by (\ref{smodz}). An equivalent way of saying this is that the partition function $Z_{\bf a}^{\rm D5}$ is invariant under the transformation $T$ from the SL$(2,\ZZ)$ S-duality group, provided that $T$ is defined to act on the boundary conditions diagonally with eigenvalues
\beq
T_{{\bf a},{\bf a}}=\exp(-i\re\CS({\bf a}))\,.
\eeq
Second remark is that the counts of solutions $n_{\mathbf{a},s}^\beta$ are expected to be independent of $\beta$ away from the values of the parameter $t$ (which is related to $\beta$ by (\ref{t}) with $\ang_{p,q}=0$) when there are flows connecting different complex flat connections. Across these walls, the counts of solutions change according to the Stokes coefficients. The blocks thus depend on $\beta$ through their dependence on the chamber where they are computed. Another, related remark is that the parameters $q$ and $\beta$ are not quite independent. The quantum field theory instructs us to take $q=\exp(2\pi i\calK)$, where $\calK$ is given by (\ref{canonical}), which means that
\beq
\im\calK =\fr{4\pi}{g_\YM^2}\fr{t^2-1}{t^2+1}=-\fr{4\pi}{g_\YM^2}\sin 3\beta\,.
\eeq
Of course, the gauge coupling $g_\YM^2$ can be varied at will, but only with the constraint that it stays positive. Therefore, for $|q|>1$ the theory instructs us to restrict to chambers with $t^2<1$ or $\sin 3\beta<0$, and vice versa for $|q|<1$. This becomes important when the series in (\ref{zd5}) is infinite. A necessary condition for its convergence is that for $|q|>1$, there exists a bound on the instanton number $s$ from above, and for $|q|<1$ there exists a bound from below. One may hope that the corresponding bounds exist in chambers with $t^2<1$ and $t^2>1$ respectively, although this hasn't been proved. Interestingly, the holomorphic blocks for knot complements, as computed in the $\cN=2$ superconformal gauge theory \cite{HolomBlocks}, come in two versions, with the instanton number bounded on one or the other side.

It may be instructive to reflect on another possible use of the tilted Nahm pole boundary condition. So far we considered Teichm\"uller TQFT for real values of the coupling $b^2$, but one may expect that the theory exists for complex values as well, because Virasoro conformal blocks certainly do. Tiechm\"uller partition functions defined in \cite{AK} make sense for $b^2$ on the complex plane with a cut along $b^2<0$. 

An obvious generalization of our path-integral definition is to take the usual action (\ref{YMCS}) on the interval, but now with $b^2$ a complex number with some phase $\alpha$. One would guess that the twisting parameter $t$ now should be set to
\beq
t=-\tan\left(\fr{\pi}{4}+\fr{\alpha}{2}\right)\,,\label{tgrad}
\eeq
so that the Kapustin-Witten equations are \cite{WittenCS} the gradient flow equations for the real part of $ib^2$ times the Chern-Simons action. The boundary condition on the left end of the interval now is a tilted Nahm pole, with the value of $\beta$ chosen out of three possible values that are compatible with the given value of $t$. This is a nice $\Qb$-invariant setup, and the only question is whether the analog of Conjecture~1 holds, that is, whether the action is bounded from below. The answer to this question is likely to be negative. Indeed, consider a product three-manifold $W\simeq\RR\times C$. The tilted D5-brane then reduces to some brane in the Hitchin sigma-model. As follows from (\ref{betanahm1})-(\ref{betanahm2}), its support is the variety of opers, where the holomorphic structure of the bundle is defined by the operator $D_{\bar z}^w=\partial_\bz+A_\bz+w\phi_\bz$ with
\beq
w=-\tan\beta\,.
\eeq
This is related by a diffeomorphism to the usual variety of opers which has $w=i$.
Note that this statement does not make sense, if  $\beta=\pm\pi/2$, so that the Nahm pole is entirely in the gauge field. We expect that for these values of $\beta$, the tilted Nahm pole boundary condition is pathological.

 The support of the brane is holomorphic in complex structure $I_w$ and Lagrangian for the corresponding holomorphic symplectic form. Equivalently, it is Lagrangian for $\omega_J$ and
\beq
\omega_I\sin2\beta+\omega_K\cos 2\beta\,.
\eeq
On the other hand, symplectic form in the A-model determined by $t=-\tan(\pi/4-3\beta/2)$ is equal to
\beq
\omega_I\sin 3\beta+\omega_K\cos 3\beta\,.
\eeq
Unless $\beta=0$ or $\pi$, the support of the brane is not Lagrangian for the symplectic form! Since the action is $\Qb$-invariant by construction, it means that the brane supports a \emph{complex} Chan-Paton gauge field, whose curvature cancels the symplectic form, restricted to the brane.\footnote{We see in particular that the usual brane of opers, or any brane related to it by Hitchin diffeomorphism $\CC^*$, cannot be obtained by reduction from a brane with three-dimensional Lorentz symmetry.} There seems to be no way to make the action bounded from below with such a brane, at least if the other end of the string is on a coisotropic brane. 

\subsubsection{NS5-type boundary condition}
For an NS5-type 1/2-BPS boundary condition, we have $(p,q)=(1,0)$ and $\ang_{p,q}=\ang\equiv{\rm arg}(\tau)$. The action of the theory on $W\times\RR^+$ with this boundary condition put at $y=0$ is
\beq
i\calK\CS(\cA^u)+\{\Qb,\dots\}\,,\label{Su}
\eeq
where Chern-Simons functional is evaluated at $y=0$ and should be understood in the sense explained after eq.~(\ref{csa}). The complexified gauge field $\cA^u=A+u\phi$ is defined with parameter
\beq
u=\fr{tt'-1}{t+t'}=\fr{\sin\beta\cos\ang+i\cos 2\beta\sin\ang}{\cos\beta}\,.\label{uform}
\eeq
This formula for $u$ was obtained in \cite{5branes} for the boundary condition with $\beta=0$ and $t'=t$, and the generalization is derived in our appendix~\ref{5branesusy}. Note that $u$ becomes infinite and the action (\ref{Su}) makes no sense for $\beta=\pm\pi/2$, which is also when in the S-dual theory the tilted Nahm pole is unlikely to be a good boundary condition. Quite curiously, for $\beta=\pm\pi/4$ the imaginary part of $u$ vanishes. These values must be special for the NS5 boundary condition, but seem completely regular on the dual D5 side. We do not know, how to interpret this.

In topological theory, one usually looks at a different, ``topological'' NS5-type boundary condition, which is not 1/2-BPS, but has the advantage of possessing a clear interpretation. There, $u$ is set to $i$ and $t$ is chosen to be real and determined by the phase of the canonical parameter as in eq.~(\ref{tgrad}). It is easy to see that this topological boundary condition does not belong to the 1/2-BPS family.
The path-integral on a half-space in this setup, with the fields approaching a complex flat bundle $\mathbf{a}$ at infinity, is what in eq.~(\ref{NSblock}) we introduced as $Z^{\rm NS5,top}_{\mathbf a}(\calK)$. It is the Chern-Simons path-integral on a Lefschetz thimble, with some classical piece subtracted to make it independent of lifts of the flat bundle. We will call it a Lefschetz block.

The problem with the topological NS5-type boundary condition is that we do not have any useful description for its S-dual. To understand dualities, we have to work with 1/2-BPS boundary conditions. Hence we define $Z^{{\rm NS5},\beta}_{\mathbf a}(\calK)$ to be the partition function on $W\times\RR^+$ with the 1/2-BPS NS5-type boundary condition with angle $\beta$. We call this an NS5-block. The question is, what is this quantity? The action (\ref{Su}) suggests that it must be Chern-Simons partition function\footnote{Again, with some trivial classical piece subtracted.} on $W$ with some integration cycle which is determined by $\mathbf{a}$ and the couplings. The difference with the topological setup are the values of $t$ and $u$, given by (\ref{t}) and (\ref{uform}). One might assume that this difference is an inessential nuisance, and $Z^{\rm NS5,top}_{\mathbf a}$ and $Z^{{\rm NS5},\beta}_{\mathbf a}$ are just equal. The factor $u$ can be transformed into $\pm i$ by a $y$-dependent field redefinition, and the $\Qb$-exact terms perhaps can also be changed. One problem is that it would be hard to reconcile this with the Stokes phenomena on the two sides of S-duality. More practically, the problem is that, while deforming these terms, one has to ensure that the action stays bounded from below. Otherwise we may start with one admissible integration cycle and end with another. We propose that NS5-blocks and Lefschetz blocks are not identical, and the formulas (\ref{t}) and (\ref{uform}), though admittedly a little obscure at first sight, have physical meaning. It would be interesting to verify, whether or for what three-manifolds the action of the theory on $W\times\RR^+$ with these values of the parameters is bounded from below, so that the NS5-blocks are well-defined.

A basic fact that we know about the NS5-block is that it is related to D5-blocks by S-duality,
\beq
Z_{\bf a}^{{\rm NS5},\beta}(\calK)=\sum_{\mathbf b}S_{\mathbf{a},\mathbf{b}}Z^{{\rm D5},\beta}_{\mathbf b}(q)\,,
\eeq
where $S_{{\bf a},{\bf b}}$ are matrix elements of the action of S-duality on the integration cycles. In this formula, $q=\exp(-2\pi i/\calK)$. (It is the same $q$ as we defined in the S-dual theory, since $-1/\calK$ here is equal to $\calK$ there.) The S-duality matrix elements are expected to be some coupling-independent numbers. (For abelian flat connections, they were obtained in \cite{GPV1}.) We learn that $Z_{\mathbf a}^{{\rm NS5},\beta}$ has an expansion in (not necessarily integer) powers of $q$. It is locally independent of $\beta$, but jumps when $\beta$ crosses chamber boundaries. The gauge coupling $\tau$ in the theory is determined\footnote{It can be shown that $\tau$ lies on a semicircle in the upper half-plane, based at the points $0$ and $|\calK|^2/\re\calK$. The point on the semicircle is determined by $\beta$ and the phase $\alpha$, except when $\calK$ is real. In the latter case, $\tau$ can lie at any point of this semicircle \cite{Janus}.} by $\calK$ and $\beta$. 

The twisting parameter $t$, as given by (\ref{t}) for a supersymmetric NS5 boundary condition, generically is complex. It means that we cannot localize onto solutions of the BPS equations. (Indeed, the $\Qb$-exact terms are not a sum of positive squares.) Still, we can say something about the NS5-block by scaling both $\tau$ and $\calK$ by the same large factor. In this semiclassical limit one expects on general grounds that fields in the path-integral will tend to stay near the values, specified by the classical vacuum $\mathbf{a}$ at infinity.\footnote{Provided that  $\mathbf{a}$ is an isolated flat connection.} The $\Qb$-exact terms in the action (\ref{Su}) then vanish, since complex flat connections solve the KW equations for any $t$. Now, for our simple argument to work, these values of the fields better also be a critical point of the boundary Chern-Simons functional. Because $u\ne i$, this is not always true. Assume however that $\mathbf{a}$ is abelian. Then $u$ does not matter, and we observe that the supersymmetric NS5-block includes a contribution from the Lefschetz thimble, corresponding to $\mathbf{a}$. It may also have subleading contributions from lower-lying flat connections. For blocks labeled by non-abelian connections, we currently can say nothing about their decomposition in thimbles.

In fact, the formula (\ref{t}) does allow values of $t$ for which localization is possible, namely, $t=0$ or $t=\infty$. For definiteness, we consider $t=0$. To get the simplest interpretation in terms of Lefschetz thimbles, we also take $\calK\in i\RR^+$, which is when the KW equations with $t=0$ are gradient flow for the right functional. To have $t=0$, one can take $\beta=\pi/2$, $\pi/6$ or $5\pi/6$, which corresponds to $u=\infty$, $i/\sqrt{3}$ and $-i/\sqrt{3}$. The first choice is pathological, while the other two are equivalent up to an R-symmetry transformation, therefore we set $\beta=\pi/6$. The path-integral for $Z^{{\rm NS5},\pi/6}_{\mathbf{a}}$ becomes the integral over the usual Lefschetz thimble for $\calK\in i\RR^+$ of the Chern-Simons functional, in which the imaginary part of the gauge field has been rescaled by $1/\sqrt{3}$. This integral must be well-defined, if we believe that the NS5-blocks make sense. Without the rescaling, it is also well-defined and is equal to the Lefschetz block. Then it is natural to expect that $u$ can be deformed from $i/\sqrt{3}$ to $i$, with the path-integral being well-defined all the way. Therefore, in the special case of $\beta=\pi/6$ and $\calK\in i\RR^+$, NS5-blocks and Lefschetz blocks are the same thing. Actually, this statement needs a small correction. For $t=0$, the integral preserved by the flows is the same for any two unitary flat connections. Thus, there will be many flows between critical points. (These flows are what one studies in the usual Floer theory.) This makes some Lefschetz thimbles non-compact and Lefschetz blocks not well-defined.  However, for the NS5-block, taking $t\rightarrow 0$ should not be a singular limit. The simplest guess about how to make sense of it is to replace Lefschetz blocks by a symmetric combination just to the left and to the right of the Stokes line. Thus we may expect that for $\calK\in i\RR^+$
\beq
Z^{{\rm NS5},\pi/6}_{\mathbf a}(q)=\fr{1}{2}\left(Z^{\rm NS5,top}_{\mathbf a}(\ex^{i\eps}\calK)+Z^{\rm NS5,top}_{\mathbf a}(\ex^{-i\eps}\calK)\right)\,,\quad \eps\rightarrow 0.\label{qcomb}
\eeq
This formula nicely matches some known facts. In \cite{GMP} it was observed in examples that Chern-Simons integrals on Lefschetz thimbles in general do not possess presentation as series in $q$. However, a spectacular finding of that paper is that, if one forms the combination (\ref{qcomb}), it does have such a presentation, which can be then converted into a series in \emph{integer} powers of $q$ by making combinations with coefficients $S_{\mathbf {a,b}}$. (A string theory motivation and many examples of such series can be found in \cite{GPV1,GPV2}) It was also found that to reproduce the partition function of the ordinary, unitary Chern-Simons theory, for the examples considered in \cite{GPV1,GMP,GPV2} it is enough to consider (\ref{qcomb}) with abelian flat connections only. Our discussion seems to imply that NS5-blocks labeled by non-abelian flat connections also make sense, although for them it may be more difficult to find the S-duality transformations.

There certainly are lots of questions about the NS5-blocks that deserve better understanding. For the purposes of this paper, the takeaway of the discussion is that one should be careful with $\Qb$-exact terms, if one wishes to understand the integration cycles. S-duality works if one uses half-BPS boundary conditions, but it is unclear what is the dual of the topological NS5-type boundary condition. This will make it difficult to apply dualities to the brane setup for Teichm\"uller TQFT.

 The blocks $Z^{\rm NS5,top}_{\mathbf a}$ generally do not need to have presentation as power series in $q$. In particular, our conjectures of section~\ref{hyperb} do not contradict S-duality.

\subsection{Complex Chern-Simons theory from super Yang-Mills theory}\label{compsec}
\begin{figure}
 \begin{center}
   \includegraphics[width=6.5cm]{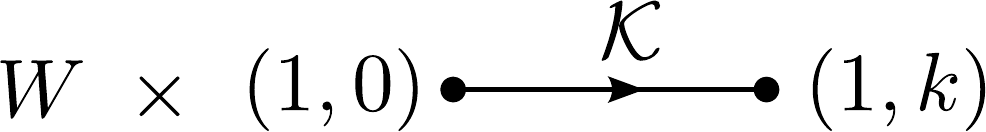}
 \end{center}
\caption{\small The topologically-twisted $\cN=4$ super Yang-Mills on an interval with an NS5 boundary condition on one side and a $(1,k)$ boundary condition on the other side is equivalent to a complex Chern-Simons theory with the integer level $k$ and the complex coupling related to the canonical parameter $\calK$.}
 \label{complcs}
\end{figure}
Let us put the twisted super Yang-Mills theory on an interval with an NS5-, or equivalently $(1,0)$-type boundary condition on one side and a $(1,k)$-type boundary condition on the other side, as shown on figure~\ref{complcs}. The $(1,k)$-type boundary condition is the same as the NS5-type, except that there is an extra Chern-Simons term with a coupling constant $k\in\ZZ$, supported at the boundary. The argument of the super Yang-Mills path-integral can be written as
\beq
\exp\left(-i\calK_\ell\CS(\cA_\ell)+i\calK_r\CS(\cA_r)+\{\Qb,\dots\}\right)\,,\label{ccs}
\eeq
where $\calK_r=\calK+k$, $\calK_\ell=\calK$, and $\calK$ is the canonical parameter. Another useful parameterization is
\beq
\calK_\ell=\fr{iv-k}{2}\,,\quad \calK_r=\fr{iv+k}{2}\,.\label{kv}
\eeq
The complex gauge fields in the action are
\beq
\cA_\ell=\left. A+u_\ell\phi\,\right|_{y=0}\,,\quad \cA_r=A+u_r\phi\,|_{y=y_0}\,,
\eeq
where the fields $A$ and $\phi$ are the restrictions of the corresponding bulk fields onto the boundary, and $u_\ell$ and $u_r$ are some complex parameters. We did not specify the boundary conditions precisely. They might be 1/2-BPS, in which case $u_{\ell}$ and $u_r$ are determined by the formulas (\ref{uform}), or it may be that 
the two parameters are set by hand to $\pm i$ or some other values. We only require that their imaginary parts be non-zero. Any such boundary coupling can be made $\Qb$-invariant by a suitable choice of the boundary conditions for the fermions.

We do not specify exactly the $\Qb$-exact terms in the action either. All that is needed from them is that the path-integral (\ref{ccs}) does not diverge, so the real part of the action in (\ref{ccs}) better be bounded from below. Whenever it is possible to make sense of the path-integral, it defines a partition function in some analytically-continued complex Chern-Simons theory with the couplings (\ref{kv}) and some integration cycle $\mathcal{C}(\calK)$ which inevitably has to depend on the canonical parameter, because of the Stokes phenomena. (For a given $\calK$ there may be more than one way to choose the $\Qb$-exact terms, which cannot be deformed into one another while keeping the action bounded from below. Then there will be several inequivalent versions of complex Chern-Simons theory with different integration cycles.) The integration cycle can be expanded as
\beq
\mathcal{C}(\calK)=\sum_{\substack{{\rm st.}\,\mathbf{a}_\ell\\{\rm st.}\, (\mathbf{a}_r,s_r)}}n_{\mathbb{a}_\ell,\mathbb{a}_r}^\calK\mathcal{C}^{\arg(-\calK)}_{(\mathbf{a}_\ell,s_0)}\times\mathcal{C}^{\arg(\calK+k)}_{(\mathbf{a}_r,s_r)}\,.\label{compcyc}
\eeq
Here we denote by $\mathcal{C}^\alpha_{(\mathbf{a},s)}$ the Lefschetz thimble for the critical point $(\mathbf{a},s)$, defined for the Chern-Simons action with the complex level of phase $\alpha$. That is, this integration cycle is obtained by the flow
\beq
\dot\cA=-i\ex^{-i\alpha}\star\bar{\mathcal{F}}\,,\label{alphaflow}
\eeq 
for the corresponding variable $\cA_\ell$ or $\cA_r$. (This flow equation is equivalent to the first two of the Kapustin-Witten equations with parameter $t$ given in terms of $\alpha$ by eq.~(\ref{tgrad}). Previously, we only needed Lefschetz thimbles for $\alpha=0$, that is, $t=-1$.)

The sum in (\ref{compcyc}) goes over lifts of the flat bundle $\mathbf{a}_r$, while the lift of $\mathbf{a}_\ell$ is some fixed $s_0$. The reason is that, as long as $k$ is an integer, the action (\ref{ccs}) is invariant under simultaneous large gauge transformations of the right and the left gauge fields, and therefore summing over both lifts would be redundant.

We expect that the sum includes only stable critical points. Unstable flat connections do not contribute as a consequence of the moment map equation $\d^*_A\phi=0$. Reducible connections should not contribute because of the vanishing phenomena related to the multiplet of the super Yang-Mills field $\sigma$, as was explained in section~\ref{lefsec} for Teichm\"uller TQFT. To make this argument precise, we would need to show that the expansion (\ref{compcyc}) can be interpreted as coming from stretching the interval $\mathcal{I}$. We cannot show this precisely without a good understanding of the $\Qb$-exact terms in the action, but the conclusion is very likely to be correct. In a special case, an alternative argument will be given later.

Let us make some general comments on complex Chern-Simons theories and their analytic continuation, following \cite{DimofteQuantum,Dimofte-k}. The space of states in such a theory is naturally a module for the algebra of quantized holonomies $\mathbf{A}_q\otimes \mathbf{A}_{\tilde q}$ of complex gauge fields $\cA_r$ and $\cA_\ell$. The parameters here are
\beq
q=\exp(2\pi i/\calK_r)\,,\quad \tilde{q}=\exp(-2\pi i/\calK_\ell)\,.
\eeq
This property seems to characterize the space of states uniquely. One can construct these modules, say, for real values of the coupling $v$, when the usual path-integral definition of complex Chern-Simons theory makes sense, and then continue the wavefunctions to complex $v$. In this sense, for any complex value of the coupling apart from the strong coupling singularities $v=\pm ik$, Chern-Simons theory exists and is unique. (Something special might also happen when $\calK$ is rational and $q$ and $\tilde q$ are roots of unity.) 

However, the space of states does not quite define a quantum field theory. For any orientation-reversing isomorphism of a pair of two-manifolds, we need a universal instruction on how to pair the states in the corresponding vector spaces. Since spaces of states in complex Chern-Simons theory are infinite-dimensional, one has to make sure that the corresponding integrals of products of wavefunctions do not diverge. An inner product on the states is a necessary part of a quantum field theory.

At the level of path-integrals, a quantum field theory is defined, once we have a universal rule to ascribe integration cycles (\ref{compcyc}) to three-manifolds, in a way consistent with factorization.

One obvious case is when $v\in\RR$ and the usual Chern-Simons path-integral makes sense. We will construct this theory in the super Yang-Mills language in a moment. We do not know, if this definition can be continued to complex values of $v$. We will also propose that there exists another version of complex Chern-Simons, for which we know the integration cycle for imaginary $v$ with $|v|>k$.

\subsubsection{The usual complex Chern-Simons theory}
Restrict for a moment to $\calK\in-\fr{k}{2}+i\RR$, so that the coupling constant $v$ in (\ref{kv}) is real. Also, set $u_r=i$ and $u_\ell=-i$. For the $\Qb$-exact terms, we take the usual squares of the Kapustin-Witten equations with the parameter $t$ determined as in (\ref{tgrad}) in terms of $\alpha=\arg(k+iv)$. Then it is easy to see that the action in (\ref{ccs}) is bounded from below, and therefore we have a consistent quantum field theory.

The interval $\mathcal{I}$ can be taken to be very small, in which case the fields have no time to fluctuate between $y=0$ and $y=y_0$, and in the path-integral (\ref{ccs}) we simply have $\cA_\ell=\bar\cA_r$. Upon reduction on the interval $\mathcal{I}$, we obtain the usual complex Chern-Simons theory with real coupling $v$.

On the other hand, take the interval $\mathcal{I}$ to be very long. Then the path-integral decomposes into a sum of products $Z^{\rm NS5,top}_{\bar{\mathbf a}}Z^{\rm NS5,top}_{{\mathbf a}}$, proving that the integration cycle is
\beq
\sum_{{\rm stable}\,\mathbf{a}}\mathcal{C}^{-\alpha}_{(\bar{\mathbf{a}},s_0)}\times\mathcal{C}^{\alpha}_{(\mathbf{a},s_0)}\,,\label{realcyc}
\eeq
where $\alpha=\arg(k+iv)$. Both left and right instanton numbers are fixed to some $s_0$, on which nothing depends. 

This formula is consistent with the Stokes phenomena. Indeed, looking at the flow equation (\ref{alphaflow}) one observes that if there is a flow at angle $\alpha$ from $\mathbf{a}$ to $\mathbf{b}$, then there is a flow at angle $-\alpha$ from $\bar{\mathbf{b}}$ to $\bar{\mathbf{a}}$. At the corresponding Stokes wall, the jumps of the summands for $\mathbf{a}$ and $\mathbf{b}$ in (\ref{realcyc}) cancel out, and the cycle stays invariant.\footnote{To be precise, we need to keep track of the instanton numbers. The flow may connect, say $(\mathbf{a},s_0)$ to $(\mathbf b,s_1)$. The argument still works, since we divide by simultaneous large gauge transformations on the left and on the right, and that identifies cycles $\mathcal{C}^\alpha_{({\mathbf{a}}_1,s_0)}\times\mathcal{C}^{\beta}_{(\mathbf{a}_2,s_0)}$ with different $s_0$. \label{insttt} } For consistency of this argument it is important that a reducible critical point cannot attach to an irreducible one.

Let us comment further on the fact that reducible connections do not contribute to (\ref{realcyc}). One way to see it is the argument about wavefunctions for the multiplet of the field $\sigma$, as we explained in section~\ref{lefsec}. More intrinsically, in complex Chern-Simons theory it can be understood as follows. 
In general, in expanding a path-integral near a reducible connection in gauge theory, one has to divide\footnote{Provided that the manifold is such that global gauge transformations are actually gauge, and not a part of the global symmetry.} by the volume of its isotropy subgroup. In complex Chern-Simons theory, isotropy subgroups are complex and their volumes are infinite. Equivalently, in a formulation of the theory where the complex part of the gauge group was gauge-fixed, these $1/\infty=0$ factors appear from zero modes of the ghosts. To obtain explicitly the complex Chern-Simons theory from our Yang-Mills setup, one makes a reduction on the interval $\mathcal{I}$. There is no complex gauge symmetry before the reduction, so it should automatically come as gauge-fixed after the reduction. One expects that the ghost multiplet would appear from the multiplet (\ref{Qsigma}), making contact between the two arguments for why reducible connections do not contribute.

One may think that throwing out reducible connections might be at tension with factorization in Chern-Simons theory. We had a very similar problem in eq.~(\ref{vacs}), where we were cutting the interval $\mathcal{I}$, and the solution was to keep track of the Hilbert space of the multiplet of $\sigma$. Similarly, one expects that to make factorization work in cutting the three-manifold $W$, one should be careful with the Hilbert spaces of the ghosts. Similar phenomena are important for supergroup Chern-Simons theory \cite{VMsuper}.

To define the quantum field theory away from $v\in\RR$, we would need to somehow pick the $\Qb$-exact terms in (\ref{ccs}) to keep the action bounded from below for generic complex $\calK$. We currently do not know, how to do this. 

For our following discussion it will be important that the analytic continuation of the partition functions of this ordinary complex Chern-Simons theory is expected to have at least three essential singularities in the plane of complex $v$. Two of them are the strong coupling singularities $v=\pm ik$, where the coefficient of one of the Chern-Simons terms in the action vanishes. The third is the weak coupling singularity $v\rightarrow\infty$. It is indeed an essential singularity, unless all complex flat connections on $W$ are actually unitary.

The integration cycle generically has non-trivial Stokes monodromies around each of the singularities. Clearly, it must be impossible to choose the $\Qb$-exact terms in the Yang-Mills action universally and consistently for all complex values of the coupling.

\subsubsection{An unusual complex Chern-Simons theory}\label{unus}
Let us answer the following question: does there exist an integration cycle for complex Chern-Simons theory, such that the partition function has no essential singularity in the weak coupling limit? (Everywhere by weak coupling limit we mean the limit of large coupling $v$. The integer level $k$ is kept fixed.)

There are plenty of integration cycles, such that the partition function is finite, say, for $v\rightarrow+\infty$. But the requirement that it does not blow up when $v$
approaches the infinity from any direction is very restrictive. As explained in section 3.2.1 of \cite{WittenCS}, it implies that the integration cycle contains only critical points with zero action, for any phase of $v$. It means that the integration cycle (\ref{compcyc}) has to be
\beq
|v|\rightarrow\infty:\quad \sum_{{\rm stable}\,\mathbf{a}}n_{\mathbf{a}}(\arg v)\,\mathcal{C}^{\arg(-iv)}_{(\mathbf{a},s_0)}\times\mathcal{C}^{\arg(iv)}_{(\mathbf{a},s_0)}\,,\label{cyc1}
\eeq
for some numbers $n_{\mathbf{a}}$ that possibly depend on $\arg(v)$. (We are assuming that the Chern-Simons action is different for all irreducible flat connections, so that no cancellations are possible.)

Suppose that at a Stokes wall at some $\arg(v)$, there is a flow between critical points $\mathbf{a}$ and $\mathbf{b}$, so that a Lefschetz thimble jumps,
\beq
\mathcal{C}^{\arg(-iv)}_{(\mathbf{a},s_0)}\rightarrow \mathcal{C}^{\arg(-iv)}_{(\mathbf{a},s_0)}+m\,\mathcal{C}^{\arg(-iv)}_{(\mathbf{b},s_1)}\,.
\eeq
The thimbles with angles differing by $\pi$ are dual, and therefore there is also a jump
\beq
\mathcal{C}^{\arg(iv)}_{(\mathbf{b},s_1)}\rightarrow \mathcal{C}^{\arg(iv)}_{(\mathbf{b},s_1)}-m\,\mathcal{C}^{\arg(iv)}_{(\mathbf{a},s_0)}\,.
\eeq
This creates a term\footnote{We used the identification, mentioned in  footnote~\ref{insttt}, to convert the cycle for $(\mathbf{b},s_0)$ in the sum (\ref{cyc1}) into a cycle for $(\mathbf{b},s_0)$ .} 
\beq
m(n_{\mathbf{a}}-n_{\mathbf{b}})\mathcal{C}^{\arg(-iv)}_{(\mathbf{b},s_1)}\times\mathcal{C}^{\arg(iv)}_{(\mathbf{a},s_0)}
\eeq
in the integration cycle, which violates the requirement of zero action. Therefore, for any value of $\arg(v)$, the coefficients $n_{\mathbf a}$ must be equal for all $\mathbf{a}$, and we may as well set them to one. The resulting integration cycle
\beq
\sum_{{\rm stable}\,\mathbf{a}}\,\mathcal{C}^{\arg(-\calK)}_{(\mathbf{a},s_0)}\times\mathcal{C}^{\arg(k+\calK)}_{(\mathbf{a},s_0)}\label{mycyc}
\eeq
is invariant under Stokes jumps around infinity in the $v$-plane. 

There is a gap in this argument. To equate all the coefficients $n_{\mathbf a}$, we need to assume that for each pair of the critical points there exists a Stokes line for some value of the phase $\arg(iv)$. This sounds like a plausible assumption generically, unless there is some special reason for the flows not to exist. Sometimes there is indeed a simple topological reason: flows cannot connect flat bundles of different topology. Thus in general, one may modify the expansion (\ref{mycyc}) by coefficients that depend on the topology of the bundle. Say, for gauge group SO$(3)$, they may depend on the Stiefel-Whitney class.

 Can we move away from infinity, while maintaining the cancellation of the Stokes jumps? The requirement for this is that $\arg(-\calK)=\pi+\arg(k+\calK)$, or 
\beq
\fr{\calK+k}{\calK}\in\RR^+\,.\label{ineq}
\eeq
This means first of all that when finite, $\calK$ should be real.\footnote{A slight subtlety is that for real $\calK$ there generically exist flows, connecting pairs of complex conjugate critical points. To define the integration cycle, we would need to displace infinitesimally from the Stokes wall. But since it is invariant under Stokes jumps, it does not actually change across the wall.} The inequality says that $\calK$ can be moved from infinity along the real line, until we hit a strong coupling singularity. We cannot continue the integration cycle (\ref{mycyc}) to the interval between the two singularities. 

At present, we do not know, how to choose the $\Qb$-exact terms in the super Yang-Mills action (\ref{ccs}), so as to construct complex Chern-Simons theory with this unusual integration cycle. We strongly believe that it should be possible, as follows from some duality arguments, to be presented in a moment, as well as a reduction from the six-dimensional $(2,0)$ theory, which is explained in section~\ref{redu}. If such construction is possible, it would mean that this Chern-Simons theory is an actual quantum field theory, and not just a set of partition functions.

Assuming optimistically that this is a quantum field theory, we may wonder, whether it can be unitary. Of course, this cannot be true for any complex coupling constants, but we can restrict to real $\calK$,  which is when our integration cycle makes sense.  A change of orientation of the manifold $W$ should act on the partition function by complex conjugation. First, it flips the sign of the Chern-Simons action. Since the coupling constant is real, for this to be equivalent to complex conjugation the gauge fields should also get conjugated. The integration cycle (\ref{mycyc}) is invariant under the orientation flip, but it is also real, and therefore conjugating the gauge fields makes no difference. Our Chern-Simons theory, if it exists, would be an example of the ``second unitary branch'' \cite{WittenCCS}.

\subsection{Dualities of Teichm\"uller TQFT}\label{tdualsec}
\begin{figure}
 \begin{center}
   \includegraphics[width=17cm]{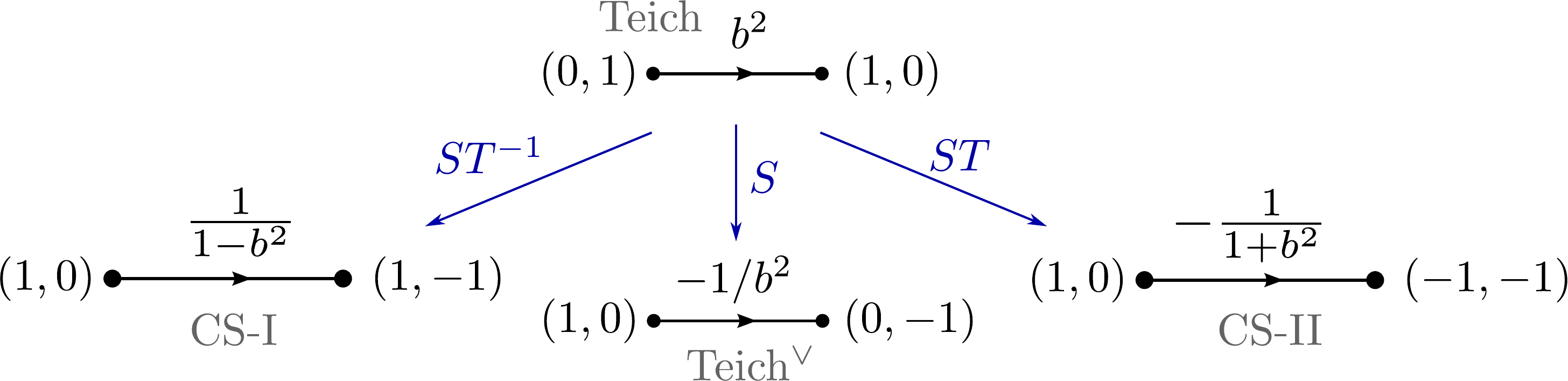}
 \end{center}
\caption{\small Brane construction for Teichm\"uller TQFT, and its duals. For each configuration, we show the boundary condition types $(p,q)$ and the value of the canonical parameter $\calK$.  Blue arrows are the S-duality transformations. }
 \label{TtoCS}
\end{figure}

Having a realization of Teichm\"uller TQFT in terms of the $\cN=4$ super Yang-Mills theory, it is natural to try to act on it with elements of the S-duality group. In doing so, we would like to keep the boundary conditions simple. By that we mean that they should be of the D5-, NS5- or $(\pm 1,k)$-type, which are the cases  that do not involve strongly coupled boundary theories. There are, up to trivial equivalences, only three elements of SL$(2,\ZZ)$ that transform our D5-NS5 system into something with simple boundary conditions. The resulting brane configurations are shown on figure~\ref{TtoCS}. These dualities have been used recently in \cite{Corners} in the study of vertex algebras at the corner of some slightly more general brane configurations.

Unfortunately, applying S-dualities to our brane setup is not as straightforward as one might hope. The reason was explained in section~\ref{secblocks}: the ``topological'' NS5-brane, used in the definition of Teichm\"uller TQFT, is not 1/2-BPS, and it is not clear, what are its duals. We are not being overly pedantic: as it was explained, ignoring these subtleties, say, for NS5-blocks and Lefschetz thimbles would lead to manifestly wrong predictions.

However, some observations suggest that in the case of Teichm\"uller TQFT, the dualities should work. First, the Liouville theory has symmetry $b\rightarrow b^{-1}$. In the mathematical construction of Teichm\"uller TQFT \cite{AK}, this symmetry is maintained. (In physics literature, this duality in Teichm\"uller TQFT was considered in \cite{DGS}.) Second, it is an experimental fact that the partition function of Teichm\"uller TQFT admits expansions into holomorphic blocks \cite{HolomBlocks,Dimofte-k,Garouf}
\beq
Z_{\rm Teichm}(b^2)=\sum_{{\rm stable}\,{\mathbf a}}B_{\mathbf a}(q)\tilde{B}_{\mathbf a}(\tilde q)\,,\label{holob}
\eeq
which are Laurent series in $q$ and $\tilde q$.

Based on these observations, we propose that it is possible to deform the super Yang-Mills configuration for Teichm\"uller TQFT to make the NS5 brane half-BPS. To do so, one first needs to rescale the field $\phi$ near the NS5-brane, in order to change the parameter $u$ in the complexified gauge field $\cA=A+u\phi$ from $u=i$ to $u=i\sin\ang$. Then one changes the $\Qb$-exact terms in the action, so that near the NS5-brane they correspond to the Kapustin-Witten parameter $t$ given by (\ref{t}) for the half-BPS $(1,0)$-brane. The result is a Janus configuration, where $t$ varies with $y$ from $t=-1$ at the D5-brane to $t=-|\tau|/\tau$ at the NS5-brane, which now is a supersymmetric $(1,0)$-brane with angle $\beta=0$. The gauge coupling $\tau$ also varies, so that the canonical parameter $\calK$, as defined in (\ref{canonical}), stays constant, which is needed for the Janus configuration to be $\Qb$-invariant. This is possible to achieve with real $g_\YM^2$ and $\theta_\YM$, as long as $t$ stays on the unit circle. We will call this super Yang-Mills setup the BPS setup for Teichm\"uller TQFT. It should be possible to make all these deformations, while keeping the action bounded from below. So far, we were not able to prove this directly.

The Janus configuration is not\footnote{Indeed, a half-BPS Janus configuration preserves the same supersymmetry as a $(p,q)$-brane \cite{Janus}. It is not possible to interpolate between two half-BPS branes of different types with a half-BPS Janus configuration.} half-BPS. However, there is an obvious proposal for how the S-dualities should act on it. One simply acts on the gauge coupling $\tau$ and the parameter $t$ at each value of $y$ according to the usual rules, valid for $y$-independent couplings. This should work at least when the couplings vary with $y$ slowly, and hopefully also more generally.\footnote{There may be subtleties in understanding duality transformation of $\Qb$-exact terms that involve $y$-derivatives of the couplings.}

The experimental facts listed above can now be easily understood. Under the basic S-duality, the BPS setup for Teichm\"uller TQFT is mapped into itself, hence $b^2\rightarrow b^{-2}$ indeed should be a symmetry of the theory. (To see this precisely, one has to combine the S-duality shown on figure~\ref{TtoCS} with a reflection in $y$.) 

 The decomposition (\ref{holob}) is also evident: by stretching the interval, we decompose the partition function into blocks, which for supersymmetric branes are related by S-duality\footnote{A D5-block has an expansion in integer powers of $q$, multiplied by a prefactor with a non-integer power. An NS5-block is a combination of such blocks with coefficients $S_{\mathbf{a},\mathbf{b}}$. Yet, in the cited papers it is found that both $B_{\mathbf a}(q)$ and $\tilde B_{\mathbf{a}}(\tilde q)$ are series in integer powers, up to an overall prefactor. The reason must be that for the class of examples considered in those papers, S-duality acts on flat connections diagonally. } to suitable versions of D5-blocks, and those are Laurent series in $q$ or $\tilde{q}$. 

Further support to the idea that Teichm\"uller TQFT can be engineered by the BPS setup comes from the 3d-3d correspondence. Teichm\"uller TQFT partition functions can be computed in 3d $\cN=2$ superconformal theories on squashed three-sphere backgrounds. As we explain in section~\ref{redu}, reducing the 6d $(2,0)$ theory on these backgrounds very naturally produces our BPS setup with half-BPS boundary conditions and a Janus configuration. Unfortunately, in this story there exists an echo of the problems with the actions that are potentially unbounded from below, which is the fact that the squashed three-sphere backgrounds are complex.

\begin{figure}
 \begin{center}
   \includegraphics[width=10cm]{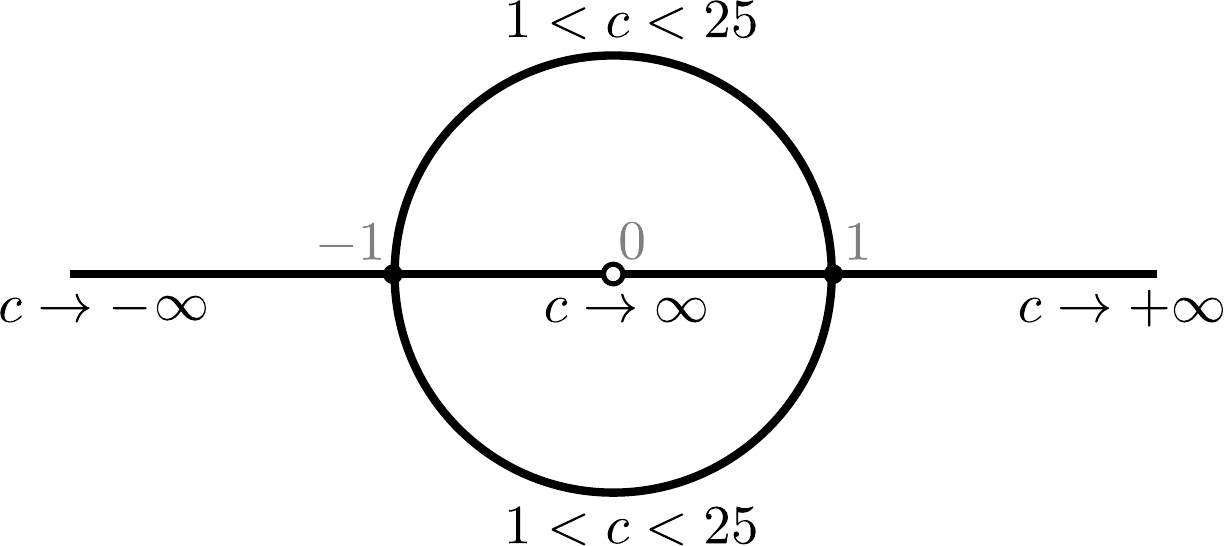}
 \end{center}
\caption{\small The complex $b^2$-plane with the value of the central charge $c=13+6(b^2+b^{-2})$ shown. It is real on the real line and on the unit circle. Teichm\"uller TQFT is defined for $b^2>0$, or $c>25$. On the unit circle, the usual real integration cycle makes sense for complex Chern-Simons theories with coupling constants (\ref{csi}) or (\ref{csii}).}
 \label{bplane}
\end{figure}
Now that we believe that S-dualities make sense, we can look at the two remaining arrows on figure~\ref{TtoCS}. But before that, we have to make a small comment. We will not be careful with the global forms of the gauge group. In the basic S-duality that acts by $b^2\rightarrow b^{-2}$, the gauge group SO$(3)$ changes into SU$(2)$. (If there is a discrete theta-angle turned on before the duality, then the gauge group remains SO$(3)$.) Whatever the gauge group in the dual theory is, the D5-type boundary condition fixes the gauge bundle to be topologically trivial, so the theories with different global forms of the gauge group should be essentially the same. Admittedly, a more careful argument is needed to fix factors like the volume of the center, and possible signs from discrete theta-angles. Unfortunately, we cannot consider these details here. In the following discussion, we will similarly not pay attention to the global form of the gauge group in the dual complex Chern-Simons theories. We hope to revisit these questions in the future.

First, let us focus on the brane configuration on the lower left of figure~\ref{TtoCS}. As was explained earlier in this section, it defines a complex Chern-Simons theory with the couplings
\beq
\calK_\ell=\fr{1}{1-b^2}\,,\quad \calK_r=\fr{b^2}{1-b^2}\,.
\eeq
Equivalently,
\beq
k=-1\,,\quad v=i\fr{b^2+1}{b^2-1}\,.\label{csi}
\eeq
We will call this theory CS-I. What is its integration cycle?
Teichm\"uller TQFT was defined for $b^2>0$, but it is a mathematical fact \cite{AK} that its partition functions can be smoothly continued to the complex $b^2$-plane with a cut along the negative real axis. (From the point of view of its path-integral definition, this is not at all surprising.) According to (\ref{csi}), the point $b^2=1$ should be the weak coupling singularity for CS-I, yet we see that the partition function is perfectly smooth at and around this point. This fixes CS-I to be the ``unusual'' theory of section~\ref{unus}. (If the gauge group is SO$(3)$, there are additional choices in fixing the integration cycle, as well as some interesting subtleties related to the fact that $k=1$ is not a properly quantized  coupling for a Chern-Simons term.)

On figure~\ref{bplane}, we have drawn the complex $b^2$-plane. Teichm\"uller TQFT with its usual integration cycle is defined on the ray $b^2>0$, or $c>25$. The limits $b^2\rightarrow+\infty$ and $b^2\rightarrow 0$, or equivalently $c\rightarrow+\infty$, are the weak/strong and strong/weak coupling limits for the two S-dual versions of Teichm\"uller TQFT. For CS-I, these two points are the strong coupling singularities. The theory with its integration cycle (\ref{mycyc}), according to the inequality (\ref{ineq}), is defined precisely for $b^2>0$. Its weak coupling point $b^2=1$ corresponds to $c=25$. Unitarity of Teichm\"uller TQFT for $b^2>0$ implies that CS-I is also unitary, which is consistent with the properties of its partition function, explained in section~\ref{unus}.

Now we turn to the lower-right corner of figure~\ref{TtoCS}. There, we have a Chern-Simons theory, to be called CS-II, with coupling constants
\beq
\calK_\ell=-\fr{1}{b^2+1}\,,\quad \calK_r=\fr{b^2}{b^2+1}\,,
\eeq
or equivalently
\beq
k=1\,,\quad v=-i\fr{b^2-1}{b^2+1}\,.\label{csii}
\eeq
This is the formula\footnote{To be precise, in \cite{CJ} a factor of $i$ seems to be missing in $v$.} obtained by Cordova and Jafferis \cite{CJ} by reduction of the 6d $(2,0)$ theory on a three-sphere. We see that there are in fact four different but dual Chern-Simons-like theories that can be obtained in this reduction, and the formula (\ref{csii}) is a consequence of S-duality. We will present a new simple method to do reduction from six dimensions in section~\ref{redu}.

CS-II has strong coupling singularities at $b^2=0$ and $b^2\rightarrow\infty$, and its weak coupling point would be at $b^2=-1$, or $c=1$. It must be some unitary complex Chern-Simons theory, defined for $b^2>0$, which on this ray is never weakly coupled. This makes it difficult to identify its integration cycle, and we can only speculate.

One possible guess is that it is the ``unusual'' Chern-Simons theory for $b^2<0$,  analytically continued to positive $b^2$. (Then one has to specify, whether it was continued through the upper or the lower half-plane.) The argument in favor of this guess is that, according to the definition of Teichm\"uller TQFT, it should be possible to continue its partition function to the left half-plane, and there seems to be no reason for it to be singular at $b^2=-1$. The argument against this guess is that it would imply, via dualities, that analytic continuation of Teichm\"uller TQFT to negative $b^2$ is again (the complex-conjugate of) Teichm\"uller TQFT. This is a highly non-trivial constraint on the Stokes behavior of the integration cycle $\mathcal{S}$, and we currently do not see any reason for it to be correct.

For $|b^2|=1$, the coupling constant $v$ in both CS-I and CS-II is real, in which case the usual complex Chern-Simons integration cycle makes sense. Is it possible that CS-II, continued to the unit circle in the $b^2$-plane, is the usual complex Chern-Simons theory? If so, it would mean that the analytically-continued Teichm\"uller partition function has an essential singularity at $b^2=-1$, of the form dictated by the integration cycle (\ref{realcyc}). It is hard to imagine how this could happen, but to make a definite conclusion, we would need to look more carefully at Teichm\"uller partition functions in examples.

There clearly still is a lot to understand in this story. Let us mention just one particular point. For $b^2$ living on the unit circle, it makes sense to choose the usual real integration cycle for complex Chern-Simons theories with the coupling constants (\ref{csi}) or (\ref{csii}). Most likely, such theories are not related to Teichm\"uller TQFT by any simple analytic continuation. Yet, their Hilbert spaces are naturally the spaces of Liouville conformal blocks, in the sense that each of them is a module for $\mathbf{A}_q\otimes\mathbf{A}_{\tilde{q}}$ with the dual values of the parameters $q=\exp(2\pi i/b^2)$ and $\tilde q=\exp(2\pi i b^2)$. The Hilbert space structures on these vector spaces however have no reason to be related to those in Teichm\"uller TQFT. By putting these Chern-Simons theories on an interval with suitable boundary conditions,\footnote{For Teichm\"uller TQFT, these boundary conditions were explored in \cite{Harlow} and are reviewed in our section~\ref{cocol}.} one may hope to obtain an analog of the Liouville theory which is well-defined in the region $1<c<25$ and is weakly coupled for $c\rightarrow 1$ or $c\rightarrow 25$.

\section{Reduction from six dimensions}\label{redu}
Chern-Simons theories with complex gauge groups can be obtained by reduction from the six-dimensional $(2,0)$ theory. This fact was first understood in the context of 3d-3d correspondence \cite{Yamazaki} (see also \cite{Dimofte-k}), and then was derived explicitly in \cite{CJ,YagiM5,YamazakiM5}. Here we would like to exhibit, how one can obtain the same results using the approach via $\cN=4$ super Yang-Mills theory, taken in the present paper, combined with some recent developments on rigid supersymmetric theories in curved backgrounds \cite{Komar3d1}. This will also clarify, how our story fits into the bigger context.

\subsection{The geometry}
To get Chern-Simons theory with a complex simply-laced gauge group G$_\CC$ on a three-manifold $W$, one starts with the corresponding ADE 6d $(2,0)$ theory on a product manifold $W\times L$, where $L$ is one of the lens spaces. The theory should be coupled to a suitable supergravity background, so as to preserve some supersymmetry. In the topological theory, the preserved supercharge would then become a BRST operator.
The 6d $(2,0)$ theory has SO$(6)$ Lorentz group and SO$(5)_{\rm R}$ R-symmetry group, and supercharges transforming in the ${\bf 4}_+\otimes {\bf 4}$ spinorial representation. We  first put the theory on $W$, making a topological twist by identifying a subgroup SO$(3)_W\subset\SO(6)$ of the Lorentz symmetry with a subgroup SO$(3)_{\rm X}\subset\SO(5)_{\rm R}$ of the R-symmetry, embedded in the obvious way. The twist preserves four supercharges, and we formally view the 6d theory on $W\times \RR^3$ as a 3d $\cN=2$ supersymmetric theory on $\RR^3$. (The preserved R-symmetry group, to be denoted $\SO(2)_{\rm Y}$ or U$(1)_Y$, is the commutant of $\SO(3)_{\rm X}$ in $\SO(5)_{\rm R}$.) If we were to take the volume of $W$ to zero, the compactified theory would flow to the three-dimensional $\cN=2$ superconformal theory $T[W]$ \cite{labelled, Dimofte-rev}, but we do not take this limit. The next step is to compactify the remaining three flat directions onto the lens space $L$.

A 3d $\cN=2$ supersymmetric theory cannot be twisted so as to preserve some supersymmetry on a general curved three-manifold. However, it can be put in a supersymmetric way on some three-manifolds. This is done by coupling to a particular $\cN=2$ supergravity background which is more complicated than just a flux of the R-symmetry gauge field.\footnote{The systematic exploration of rigid supersymmetric theories in curved space was initiated in \cite{SeibergFest}. For a recent review, see \cite{ThomasRev}.} For this background to exist, the three-manifold has to admit an extra geometric structure \cite{Komar3d1}, known as the transversely-holomorphic foliation (THF), compatible with the metric. A detailed exposition on THFs and rigid supersymmetry in three dimensions can be found in the original papers \cite{Komar3d1,Komar3d2}. A short summary is also given in our appendix~\ref{THFs}. Here we will need some basic statements.

A THF is a natural three-dimensional analog of an integrable complex structure.  It is defined by an equivalence class of coordinate atlases, where the local coordinates are a real\footnote{Not to be confused with the Kapustin-Witten parameter $t$.} $t$ and a complex $z$, and the allowed coordinate transformations are $t'=t+f(z,\bar z)$, $z'=g(z)$, with $g(z)$ holomorphic. These are called the adapted coordinates. A metric is called compatible with the THF, if the vector $\partial_t$ is unit and the metric is Hermitian when restricted onto planes, orthogonal to $\partial_t$. Having a THF on a three-manifold, very roughly speaking, allows to twist an $\cN=2$ supersymmetric theory by turning on a U$(1)$ R-symmetry flux that compensates for the spin-connection holonomies in the $\partial_z$-planes. This intuition is made precise in \cite{Komar3d3}.

For our application, the curved three-manifold is a lens space $L$. It can be presented as two solid tori, glued together by a PSL$(2,\ZZ)$ transformation. Equivalently, it is a $T^2$-bundle over an interval, where particular one-cycles of the torus shrink at the ends of the interval. We choose to work with a simple class of metrics on $L$ which have two Killing vectors along the fibers,
\beq
\d s^2=\d y^2+\fr{a}{\im\tau}\left((\d\varphi_1+\re\tau\,\d\varphi_2)^2+(\im\tau\,\d\varphi_2)^2\right)\,,\label{met}
\eeq
where $y$ parameterizes the interval ${\mathcal I}\simeq [0,\ell]$, and $\varphi_1$ and $\varphi_2$ are angular coordinates on the torus fiber, both with period $2\pi$. The torus area $4\pi^2a>0$ and the modular parameter $\tau$, $\im\tau>0$, are functions of $y$. Near $y=0$ or $y=\ell$, where some one-cycle shrinks, these functions should behave suitably to have a smooth metric. 

As explained in \cite{Komar3d1}, to define a THF, compatible with a given metric, it is sufficient to pick a real unit vector field $\xi$, satisfying a particular integrability condition. The adapted coordinates can then be introduced so that $\xi=\partial_t$. We will be interested in the case that the background preserves at least two supersymmetries $\zeta$ and $\tilde{\zeta}$ of opposite U$(1)_Y$-charge. Their anticommutator generates a translation along some, possibly complex, Killing vector $K$. For the metric (\ref{met}), Killing vectors, up to scaling, can be parameterized as
\beq
K=-\kappa\,\partial_{\varphi_1}+\partial_{\varphi_2}\,,\label{Killing}
\eeq
where $\kappa\in \CC\cup\{\infty\}$. (A minus sign was introduced for future convenience.) A particularly simple case is when the two supersymmetry generators $\zeta$ and $\tilde\zeta$ are Hermitian conjugates of each other. Then $\kappa\in\RR\cup\{\infty\}$, and (as shown in \cite{Komar3d1}) the unit vector $\xi$ which defines the THF is equal to $K/|K|$. The opposite is also true: if $K$ is some real nowhere vanishing Killing vector, then the unit vector field $\xi=K/|K|$ satisfies the integrability condition and defines a THF which preserves two conjugate supersymmetries. To avoid possible confusion, let us stress that the fibers of the THF are \emph{not} the tori $y={\rm const}$.

We remark that some combinations of $\partial_{\varphi_1}$ and $\partial_{\varphi_2}$ vanish at the ends of the interval. This excludes two (or one, if our lens space is $S^1\times S^2$) possible values of $\kappa$.

Importantly, the partition function (and supersymmetric observables) on a three-manifold depends holomorphically on the THF, but is independent of the details of the compatible metric \cite{Komar3d2}: the changes of the metric with fixed THF induce changes of the action, which are exact in preserved supersymmetries. In appendix~\ref{THFs} we apply this statement to our setup. We show that, at least for real $\kappa$, this fact implies that the partition function depends on $\kappa$, but is independent of $\tau(y)$ and $a(y)$, as long as they behave well near the ends of the interval. We will assume, though we did not prove it, that this logic extends to complex $\kappa$ as well: a general complex Killing vector (\ref{Killing}) defines a THF, preserving two supersymmetries of opposite chirality, and the supersymmetric partition function is a holomorphic function of $\kappa$, independent of $\tau(y)$ and $a(y)$.

Keeping $\kappa$ fixed, we are allowed to deform the metric so that the typical size of the torus fiber is much smaller than the size of the base interval ${\mathcal I}$, $a(y)\ll \ell^2$. The functions $\tau(y)$ and $a(y)$ can be taken to be slowly-varying, or even constant away from the ends of the interval. In this case the 6d $(2,0)$ theory can be reduced on the torus fiber, giving the topologically-twisted $\cN=4$ super Yang-Mills theory on ${\mathcal I}\times W$, with some boundary conditions at $y=0$ and $y=\ell$. We choose the reduction to go first on $\varphi_1$ and then on $\varphi_2$. Then the complexified four-dimensional gauge coupling is equal to $\tau$. It may be a non-trivial function of $y$, which case in gauge theory is known as a Janus configuration \cite{Janus}.

\subsection{Matching the parameters}
The six-dimensional configuration on $W\times L$ has reduced to the four-dimensional twisted super Yang-Mills theory on $W\times{\mathcal I}$. Let us understand this reduction in more detail. First, suppose that the gauge coupling $\tau$ is constant. As we recalled in section \ref{secblocks}, the topologically twisted Yang-Mills theory on $W\times{\mathcal I}$ has four scalar supercharges, which transform as two doublets of the unbroken R-symmetry group SO$(3)_Y$. In (\ref{V2}), we introduced the notation ${\bf 2}\otimes V_2$ for the space of these supercharges, with $V_2$ being the multiplicity vector space. The BRST supercharge $\Qb$ was chosen to be some vector in ${\bf 2}$ (declared to be of ghost charge plus one) times a vector in $V_2$, parameterized by the twisting parameter $t$. The action of the theory can be recast in the form (\ref{N4act}), which we repeat here for convenience,
\beq
I_{\YM}=\fr{i\calK}{4\pi}\int\tr(F\wedge F)+\{\Qb,\dots\}\,,\label{act22}
\eeq
where 
\beq
\calK=\re\tau+i\fr{t-t^{-1}}{t+t^{-1}}\im\tau\,.\label{calK2}
\eeq
Starting from this expression, it is easy to write a BRST-invariant action for a Janus configuration, that is, to make the coupling constant $\tau$ a function of the coordinate $y$ on the interval $\mathcal{I}$. One simply takes both $\tau$ and\footnote{To be precise, one has to take care of a small subtlety. The BRST transformations of the fields depend on $t$, and therefore will start to depend on $y$, which is bad. Before making $t$ $y$-dependent, one should redefine the fermions and rescale the supercharge so that the dependence on $t$ does not appear in the BRST transformations. This is possible, as long as $t\ne\pm i$.} $t$ to be $y$-dependent, so that their combination (\ref{calK2}) stays constant. The action is the same expression (\ref{act22}), where the gauge fermion, on which $\Qb$ is acting, is now defined with $y$-dependent parameters. Because the action preserves the SO$(3)_Y$ symmetry, the Janus configuration constructed in this way is invariant under two supercharges of opposite ghost number, which are $\Qb$ itself and its SO$(3)_Y$ transformation. This Janus configuration must be precisely what one obtains from six dimensions by reducing on a torus fiber in the lens space $L$. But how do we match the parameters?  The partition function of the (2,0) theory on $W\times L$ depends holomorphically on the THF parameter $\kappa$, while its dependence on the modular parameter $\tau(y)$ of the torus fiber is $\Qb$-exact. Similarly, the twisted super Yang-Mills theory partition function is independent of the coupling constant $\tau(y)$, as long as one keeps fixed $\calK$, on which it depends holomorphically.  We note that under large diffeomorphisms of the $(\varphi_1,\varphi_2)$ torus the THF parameter $\kappa$ transforms by a M\"obius transformation. In four dimensions, these diffeomorphisms become S-dualities, and the canonical parameter $\calK$ transforms under them in the same way \cite{KW}. Thus, we are led to conjecture that $\calK=\kappa$. Let us verify this explicitly. Again, we restrict to the case that $\kappa$ is real.
 
It is convenient to use string theory language. The $\cN=4$ super Yang-Mills theory can be obtained (for a unitary gauge group) on a stack of D3-branes wrapped on $W\times\mathcal{I}$ in the type IIB string theory. For a moment, forget about the boundary conditions and the variation of $\tau$. The space-time is $T^*W\times \RR\times \RR^3$ with coordinates $x^0,x^1,x^2$ along $W$, $x^4,x^5,x^6$ along the fibers of the cotangent bundle, $x^3\equiv y$ along $\RR$ and $x^7,x^8,x^9$ along the remaining $\RR^3$. The unbroken subgroup SO$(3)_Y$ of the R-symmetry group acts in the $789$ directions. Rotations in the tangent planes to $012$ and $456$ subspaces are the Lorentz group SO$(3)_W$ and the R-symmetry subgroup SO$(3)_X$, and the Lorentz group of the twisted theory is SO$(3)_W'\simeq{\rm diag}({\rm SO}(3)_W\times{\rm SO}(3)_X)$, as was already introduced in section \ref{secblocks}. In fact, it will be sufficient to take $W$ just the flat space. 

The space of preserved supercharges in the twisted theory is ${\bf 2}\otimes V_2$. Following \cite{GWbc}, we introduce operators $B_1=\Gamma_{3456}$ and $B_2=\Gamma_{3789}$, which commute with SO$(3)_W\times {\rm SO}(3)_X\times{\rm SO}(3)_Y$ and act in $V_2$. The BRST charge is proportional to a particular vector in $V_2$ which can be singled out by a suitable projection operator, assembled from $B_1$ and $B_2$. As shown in \cite{5branes} (see section 2), the correct condition is
\beq
\left(\fr{t+t^{-1}}{2}B_1+i\fr{t-t^{-1}}{2}B_2+1\right)\eps_t=0\,,\label{ssy2}
\eeq
where $\eps_t$ is the generator of the BRST symmetry and $t$ is the Kapustin-Witten twisting parameter.

Next we would like to find, which supersymmetry is preserved by the transversely-holomorphic foliation, and to match it to the formula above.
A direct and complete treatment of the problem would require lifting the 3d $\cN=2$ supergravity background of \cite{Komar3d1} to six dimensions and then finding, how it reduces to a background for $\cN=4$ super Yang-Mills theory in four dimensions. We will not attempt to do that, but will instead take a shortcut. Suppose that the metric on the lens space $L$ is such that the vector $\xi$ which defines the THF is covariantly constant on some three-dimensional submanifold $L^0\subset L$. Then the holonomy group on this submanifold naturally reduces to U$(1)$ which acts in the planes orthogonal to $\xi$. The coupling to the supergravity background on this submanifold then reduces \cite{Komar3d1} simply to a twist by the U$(1)_Y$ R-symmetry of the 3d $\cN=2$ supersymmetry, to compensate for the U$(1)$ holonomies of the Levi-Civita connection. It is then easy to understand, what this twist means in terms of the four-dimensional Yang-Mills theory.

Let the parameters $\tau(y)$ and $a(y)$ of the metric (\ref{met}) be constant on some subinterval $\mathcal{I}_0\subset\mathcal{I}$, and $L^0\subset L$ be the torus fibration over $\mathcal{I}_0$. Then vector $\xi$ is covariantly constant on $L^0$. As explained above, the supersymmetries preserved by the background\footnote{Of course, the metric on $L^0$ is just flat, and therefore on $L^0$ all supersymmetries are preserved, as long as one ignores the boundary terms. The equation (\ref{11dsusy}) selects those supersymmetries which can be extended to $L$ for a given THF.} are the ones that on $L^0$ are invariant under the twisted rotation group in the plane, orthogonal to $\xi$,
\beq
(\Gamma_{3{\bf v}}+\Gamma_{89})\eps=0\,,\label{11dsusy}
\eeq
where $\partial_3\equiv\partial_y$ and ${\bf v}$ are the two orthonormal vectors, orthogonal to $\xi$, and $\Gamma_{89}$ is the generator of U$(1)_Y$.

To understand what this equation means in four dimensions, we do a reduction and a T-duality from M-theory to type IIB string theory. We start with an M-theoretic setup  on $T^*W\times L\times \RR^2$, with a stack of M5-branes wrapped on $W\times L$. The group SO$(2)_Y$ acts on the $\RR^2$ which is the subspace $89$. The three-manifold $L^0\subset L$ for now is just $\mathcal{I}_0\times T^2$, with $\mathcal{I}_0$ along $x^3=y$ and $T^2$ in the $7,10$ subspace. The metric on $L^0$ is (\ref{met}), with constant $\tau$ and $a$. We choose orthonormal coordinates on $T^2$,
\beqn
x^7&=&\sqrt{a\,\im\tau}\,\varphi_2\,,\nnr
x^{10}&=&\sqrt{\fr{a}{\im\tau}}\,(\varphi_1+\re\tau\varphi_2)\,.
\eeqn
In the equation (\ref{11dsusy}), $\eps$ is a Dirac spinor in eleven dimensions, and vector ${\bf v}$ is
\beq
{\bf v}=a_1\partial_{10}+a_2\,\partial_7\,,
\eeq
with
\beqn
a_1^2+a_2^2&=&1\,,\nnr
a_2/a_1&=&\fr{\kappa-\re\tau}{\im\tau}\,.\label{a1a2}
\eeqn
We reduce on a circle $\partial_{\varphi_1}$, to obtain type IIA string theory in ten dimensions. The coordinate $x^{10}$, which is periodic with period proportional to the radius of the M-theory circle, is dropped upon reduction. The gamma-matrix $\Gamma_{10}$ becomes the 10d chirality operator $i\Gamma_{-1}$, and the spinor decomposes into $\eps=\eps_++\eps_-$, with $i\Gamma_{-1}\eps_\pm=\pm \eps_\pm$. The equation (\ref{11dsusy}) decomposes into two equivalent ones, of which we keep one,
\beq
\left(a_2\Gamma_{37}+\Gamma_{89}\right)\eps_+-a_1\Gamma_3\eps_-=0\,.\label{11dsusy3}
\eeq
Next, we do a T-duality on the circle $x^7$ to type IIB string theory. The type IIB supersymmetry generators are $\eps_1=\eps_+$ and $\eps_2=\Gamma_7\eps_-$. The branes that started their life as M-theory fivebranes are now D3-branes in directions $0123$. Using the D3-brane supersymmetry condition $i\Gamma_{0123}\eps_1=\eps_2$ and some gamma-matrix algebra, one can transform the equation (\ref{11dsusy3}) into
\beq
(a_1 B_1+a_2 B_2-1)\eps_1=0\,.\label{susyeq}
\eeq
Comparing with (\ref{ssy2}), we find that the supersymmetry in four dimensions is the same as the one preserved by the transversely-holomorphic foliation, provided that
\beq
a_2/a_1=i\,\fr{t-t^{-1}}{t+t^{-1}}\,.
\eeq
Therefore, from (\ref{calK2}) and (\ref{a1a2}),
\beq
\kappa=\calK\,,\nonumber
\eeq
as was claimed.

\subsection{The boundary condition}
So far we stayed away from the ends of the interval $\mathcal{I}$, where the torus fiber degenerates. Now let us understand the boundary conditions in the $\cN=4$ super Yang-Mills at the ends of the interval. As usual, we restrict to real $\kappa$. We focus on the left end at $y=0$ and assume for definiteness that the circle $\varphi_1$ shrinks there. As long as we keep away from the right end of $\mathcal{I}$, we can use the freedom to change $\tau(y)$ and $a(y)$ to set $\re\tau=\kappa$, $\im\tau=R/f(y)$ and $a=Rf(y)$. We choose $R$ to be a constant and $f(y)$ to be a function which vanishes as $y$ for $y\rightarrow 0$, is positive for $y>0$ and approaches a constant for large $y$. The metric becomes
\beq
\d s^2=\d y^2+f^2(y)\d\psi^2+R^2(\d\varphi_2)^2\,,
\eeq
where $\psi=\varphi_1+\kappa\varphi_2$. This is a twisted product of a cigar $C$ and an $S^1$, with the cigar rotating by $2\pi\kappa$ upon going around the $S^1$. The holonomy of this metric is contained in U$(1)$ that acts in the tangent planes to the cigar. The Killing vector $K=-\kappa\partial_{\varphi_1}+\partial_{\varphi_2}$ in the basis $(\partial_\psi,\partial_{\varphi_2},\partial_y)$ is simply $\partial_{\varphi_2}$, and the vector $\xi$ defining the THF is $R^{-1}\partial_{\varphi_2}$. Since $\xi$ is covariantly constant, the supergravity background corresponding to the THF reduces to the usual R-symmetry twist which compensates for the U$(1)$ Levi-Civita holonomy on the cigar. 

In M-theory, we can realize this setup by taking the spacetime to be $T^*W\times T^*C\times_\kappa S^1$, with the Taub-NUT metric on $T^*C$, and $\times_\kappa$ being a twisted product, where the Taub-NUT is rotated by its isometry by an angle $2\pi\kappa$ in going around the $S^1$. A number of M5-branes are wrapped on the supersymmetric cycle $W\times C\times_\kappa S^1$. After reducing on the Taub-NUT circle $\varphi_1$ and T-dualizing on $S^1$, the M5-branes become D3-branes ending on a D5-brane in the type IIB theory. Thus, the boundary condition at $y=0$ in the four-dimensional gauge theory is the one that comes from the usual 1/2-BPS D5-brane, that is, the (untilted) Nahm pole. Similarly, if a general $(p,q)$-cycle of the torus fiber shrinks at $y=0$, the boundary condition is of the general 1/2-BPS fivebrane type. 

If $\kappa$ is not real, then $\xi\ne K/|K|$ and our argument does not quite work. This is consistent with the four-dimensional gauge theory: the untilted D5-brane boundary condition, which has $t^2=1$, is compatible only\footnote{Assuming the gauge theory parameters $g_\YM^2$ and $\theta_\YM$ are real, which is the case in our geometric setup.} with real $\calK=\kappa$. For complex $\kappa$, the supergravity background near $y=0$ should be more complicated than just an R-symmetry twist on the cigar. Upon reduction and T-duality to the type IIB theory, one expects again to find a D5-brane, but now with a non-trivial flux of the fivebrane U$(1)$ gauge field. (Or, equivalently, a non-trivial background of the type IIB 2-form fields.) That would correspond to a non-trivial angle $\beta$ of the Nahm pole, in the language of section \ref{secblocks}. It would be interesting to generalize the present discussion to complex $\kappa$ explicitly. 

\subsection{Putting things together}
Let the Killing vector $\partial_{\varphi_2}$ vanish at $y=0$ and $\partial_{\varphi_2}+k\partial_{\varphi_1}$ vanish at $y=y_0$.  Topologically, this torus fibration over the interval is the $|k|$-th lens space\footnote{In our notations, $L_0$ is $S^1\times S^2$, while for $k>0$ $L_k\equiv L_{k,1}$, where $L_{p,q}$ is the factor of the three-sphere $|z_1|^2+|z_2|^2=1$ by $(z_1,z_2)\sim(\ex^{2\pi i/p}z_1,\ex^{2\pi i q/p}z_2)$. Reducing on lens spaces $L_{p,q}$ with $q\ne \pm 1$ would lead to theories with no weakly coupled description, which we do not consider.} $L_{|k|}$. The THF parameter $\kappa$ is a possibly infinite complex number, not equal to $0$ or $-k$, for the Killing vector $K$ to be everywhere non-vanishing. Upon reducing on the torus fiber, one gets a configuration in $\cN=4$ Yang-Mills theory on an interval $\mathcal{I}$.

It is fairly obvious that the boundary condition is of type $(\pm 1,0)$ on the left and $(\pm 1,\pm k)$ on the right of the interval. It takes a little more effort to fix the signs. Let us do this carefully, restricting as usual to real $\kappa$. A preliminary comment is that a change of coordinates 
\beq
\left(\bea{c}\varphi_1\\\varphi_2\eea\right)\rightarrow \left(\bea{cc}a&b\\c&d\eea\right)\left(\bea{c}\varphi_1\\\varphi_2\eea\right)\label{S1}
\eeq
is an S-duality transformation $\tau\rightarrow (a\tau-b)/(-c\tau+d)$, which acts on the fivebrane charges by
\beq
(p,q)\rightarrow (p,q)\left(\bea{cc}d&b\\c&a\eea\right)\,.\label{S2}
\eeq
Another preliminary comment is that the local holomorphic coordinate in the fibers of the THF, as shown in appendix~\ref{THFs}, is 
\beq
z=\varphi_1+\kappa\varphi_2+\int^y({i|K|}/{a})\d y'\,.\label{holomz}
\eeq

Near $y=0$, we go to coordinates $\varphi_1'=\varphi_2$ and $\varphi_2'=-\varphi_1$, in which the vanishing Killing vector is $\partial_{\varphi_1'}$. The coupling constant can be chosen to be $\re\tau'=\kappa'=-1/\kappa$, $\im\tau'=R/f(y)$, with $f(y)\approx y$ for small $y$. The space near $y=0$ is a twisted product of an $S^1$ with coordinate $\varphi_2'$ and a cigar with coordinates $y$ and $\psi'=\varphi_1'+\kappa'\varphi_2'=\varphi_1'-\kappa^{-1}\varphi_2'$. As discussed previously, in the duality frame associated to $(\varphi_1',\varphi_2')$, the fivebrane at $y=0$ is a D5. More precisely, it is a D5 or a $\bar{\rm D5}$, but the two choices of the boundary condition are exchanged by an R-symmetry transformation in the $\cN=4$ Yang-Mills, and thus we are free to choose the boundary condition to be precisely D5, or type $(0,1)$. According to (\ref{S1}) and (\ref{S2}), an S-duality back to the frame defined by $(\varphi_1,\varphi_2)$ produces an NS5, or a $(1,0)$-type boundary condition.
In a moment, we will need to know the THF holomorphic coordinate near $y=0$, so let us compute it. One easily finds that the norm of the Killing vector for small $y$ is $|K|=|\kappa|R$, and then, accordingly to (\ref{holomz}),
\beq
z=i|\kappa|(-i\psi'{\rm sign}(\kappa)+\log y)+{\rm const}\,.
\eeq
A good holomorphic coordinate $w'$ in the THF near $y=0$ is a multiple of $\exp(-i|\kappa|z)$,
\beq
w'=y\exp(-i\psi'{\rm sign}(\kappa))\,.\label{wp}
\eeq

\begin{figure}
	\begin{center}
		\includegraphics[width=13.5cm]{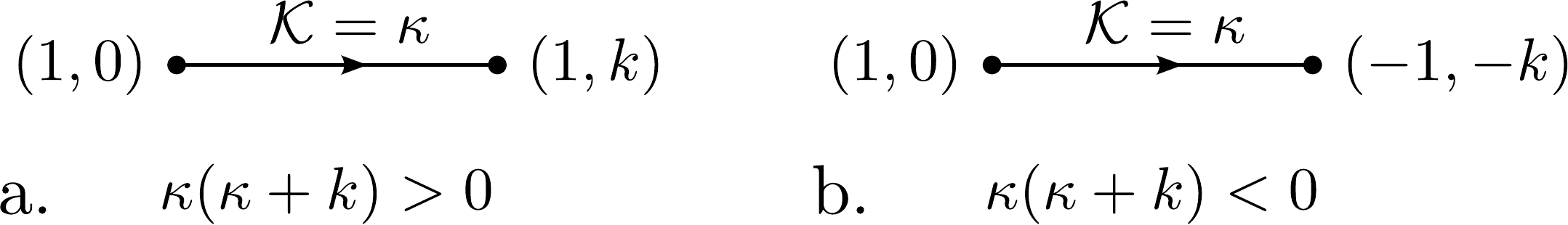}
	\end{center}
	\caption{\small These brane configurations are obtained from reducing on a lens space $L_k$ for real a THF parameter $\kappa$.}
	\label{complcs1}
\end{figure}
Now we turn to the neighborhood of $y=y_0$. Here we choose coordinates $\varphi_1''=\varphi_2$ and $\varphi_2''=k\varphi_2-\varphi_1$, so that the vanishing Killing vector becomes again $\partial_{\varphi_1''}$. The space near $y=y_0$ is again a twisted product of a circle and a cigar, with the coordinates on the cigar $y_0-y$ and $\psi''=\varphi_1''+\kappa''\varphi_2''=\varphi_1''-\fr{1}{\kappa+k}\varphi_2''$. We need to decide, whether this defines a D5- or a $\bar{\rm D5}$-type boundary condition. This is a physically meaningful question, since the symmetry that exchanges the two types has already been used to fix the boundary condition at $y=0$. Let us compute the THF holomorphic coordinate. For $y$ close to $y_0$, one finds that $|K|=|\kappa+k|R$, and then (\ref{holomz}) gives
\beq
z=-i|\kappa+k|(i\psi''{\rm sign}(\kappa+k)+\log(y_0-y))+{\rm const}\,,
\eeq
and the good holomorphic coordinate is $w''=\exp(iz/|\kappa+k|)$, or
\beq
w''=(y_0-y)\exp(i\psi''{\rm sign}(\kappa+k))\,.\label{wpp}
\eeq
Reducing on a cigar with holomorphic coordinate $y\exp(-i\psi)$ near $y=0$ preserves the same supersymmetry as reducing on a cigar with holomorphic coordinate $(y_0-y)\exp(i\psi)$ near $y=y_0$, because the two cigars can be glued together into a holomorphic $\mathbb{P}^1$ in ${\rm T}^*\mathbb{P}^1$. Comparing the THF holomorphic coordinates (\ref{wp}) and (\ref{wpp}), we conclude that in our setup, the boundary condition at $y=y_0$ is D5 for $\kappa(\kappa+k)>0$ and $\bar{\rm D5}$ for $\kappa(\kappa+k)<0$. This, of course, was in the duality frame $(\varphi_1'',\varphi_2'')$.

Making an S-duality transformation back to the original frame, we obtain the configurations shown on figure~\ref{complcs1}.
 
Let us restrict for a moment to the three-sphere, that is, $k=1$. Just to keep the discussion very explicit, let us start with the popular background of the SU$(2)\times{\rm U}(1)$-symmetric squashed sphere \cite{Imamura}. It has metric
\beq
\d s^2=h^2\left(\d\varphi_1-\sin^2\fr{\theta}{2}\d\varphi_2\right)^2+\d\theta^2+\sin^2\theta\,\d\varphi_2^2\,,\label{imamura}
\eeq
where $h>0$ controls the radius of the Hopf fiber. Here $\theta\in[0,\pi]$, and the angles $\varphi_{1,2}$ are $2\pi$-periodic. This metric falls into the class of torus fibrations (\ref{met}) with the identification $y=\theta$. The Killing vector $\partial_{\varphi_2}$ vanishes at $\theta=0$, and $\partial_{\varphi_2}+\partial_{\varphi_1}$ vanishes at $\theta=\pi$.
To fix the rigid supersymmetric background, one has to also specify the values of other bosonic fields of the 3d new minimal supergravity, and for this background we use the values presented in section 5.2 of \cite{Komar3d1}. In appendix~\ref{apex} we show that these values correspond to 
\beq
\kappa=-\fr{1}{1+b^2}\,,
\eeq
if one chooses the usual parameterization $h=b+b^{-1}$. (There is of course the second solution for $\kappa$ with $b$ and $b^{-1}$ exchanged, but the two are equivalent.)

For $b^2>0$, the inequality on the right of figure~\ref{complcs1} is satisfied. The corresponding brane configuration is precisely the one for CS-II on figure~\ref{TtoCS}, with the correct value of $\kappa$! We even correctly reproduce the brane types $(1,0)$ and $(-1,-1)$. Reducing on the circles $\varphi_1$ and $\varphi_2$ in different order would produce the other three dual brane configurations of figure~\ref{TtoCS}.

We conclude this section with some general remarks. For any integer $k$, the brane configurations of figure~\ref{complcs1} define complex Chern-Simons theories with levels $k$ and $v=-i(2\kappa+k)$, according to the equations of section~\ref{compsec}. However, lens space rigid supersymmetric backgrounds in general are not real, and we did not keep track of the $\Qb$-exact terms. As always in this paper, this raises the question of what are the integration cycles in these Chern-Simons theories. Does this question have physical meaning? May be, we can just choose Chern-Simons integration cycle at will and declare it to be the reduction from six dimensions? To answer these questions, one first of all has to look at lens space partition functions of the 3d $\cN=2$ superconformal theories T[W]. When they converge, they should define particular integration cycles for the Chern-Simons.

One may also desire that the Chern-Simons partition function does not have singularities, as long as the background is well-defined. As we have argued, the obvious bad values of $\kappa$ for our backgrounds are where the Killing vector $K$ becomes vanishing at one of the ends of the interval $\mathcal{I}$. These correspond precisely to the strong coupling singularities in Chern-Simons. Then it is natural to declare that our Chern-Simons theory has no weak coupling singularity, so it must be the ``unusual'' Chern-Simons theory of section~\ref{unus}. This argument allows to define a smooth partition function for $\kappa\in\mathbb{P}^1$ with a cut between the two strong coupling points.

Clearly, a more systematic understanding of integration cycles would be very desirable.

\section{Conformal blocks, corners and Liouville}\label{cocol}
From the point of view of a physicist, a conformal block is not an element of some abstract vector space, but rather a particular function of Liouville momenta and the complex structure. The formula for this function in terms of Teichm\"uller TQFT states was given in eq.~(\ref{conf1}), which we repeat here for convenience,
\beq
\Psi^{\sigmatt,{\bf l}}({\bf q})=\langle{\bf q}|\sigmatt,{\bf l}\rangle\,.\label{conf2}
\eeq
Here $\sigmatt$ is a pants decomposition of the Riemann surface $C$,  ${\mathbf l}\equiv\{l_1,\dots, l_{3g-3}\}$ is a set of Liouville momenta, and ${\mathbf q}\equiv\{q_1,\dots, q_{3g-3}\}$ are values of a set of holomorphic coordinates, parameterizing the complex structure. The state $|\sigmatt,{\bf l}\rangle$ belongs to the space of $({\mathcal B}_{\mathcal T},{\mathcal B}_{\rm c})$ strings, the state $\langle{\bf q}|$ belongs to the space of $({\mathcal B}_{\rm c},{\mathcal B}_{\mathcal T})$ strings, and the conformal block is given by the natural pairing.\footnote{We could have defined both states in the same Hilbert space and then used the Hermitian scalar product. The result would be essentially the same.}

\begin{figure}
	\begin{center}
		\includegraphics[width=15cm]{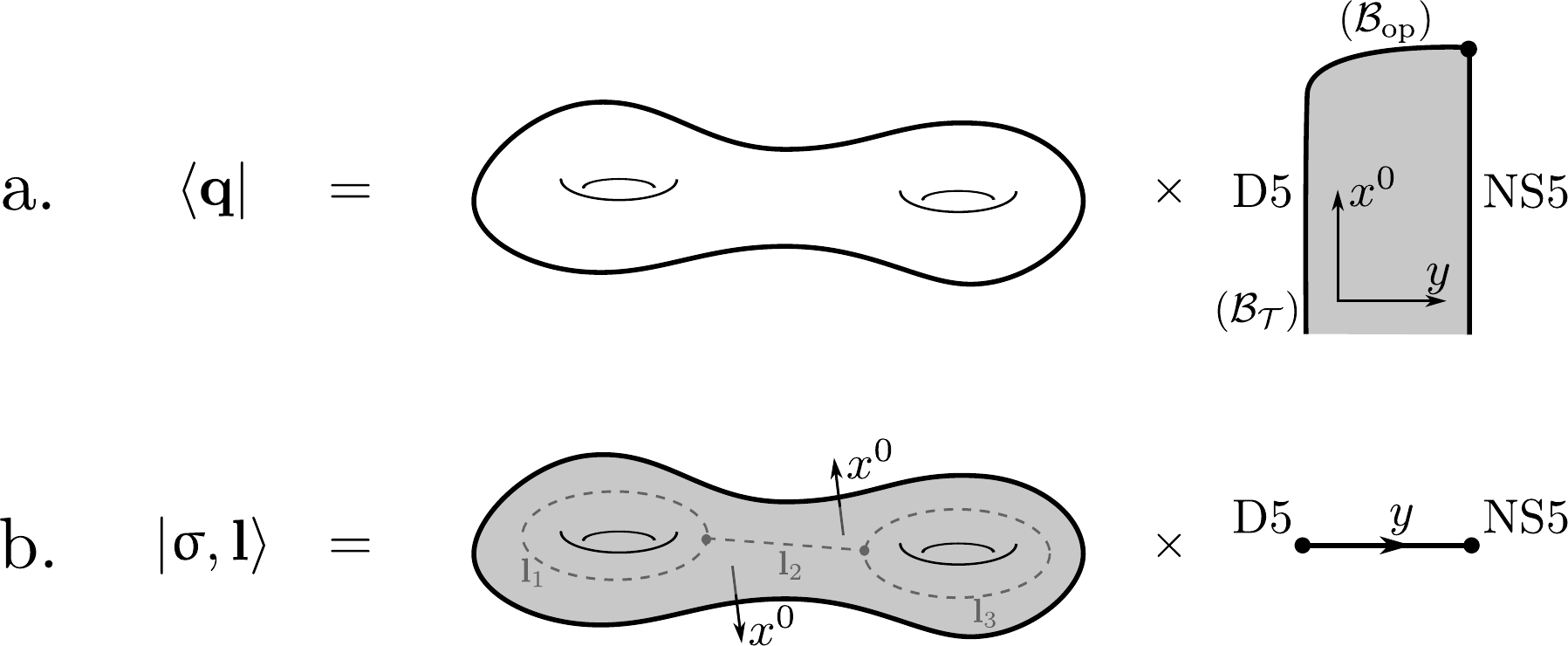}
	\end{center}
	\caption{\small {\large a.} The state $\langle\bf{q}|$ is generated on the boundary $C\times\mathcal{I}$ of $C$ times a brane corner. {\large b.} The state 
      $|\sigmatt,{\mathbf l}\rangle$ is created on the boundary $C\times\mathcal{I}$ of the solid handlebody $\bar C_{\sigmatt,{\mathbf l}}$ times the interval $\mathcal{I}$. The time direction $x^0$ is the outgoing normal vector at the boundary of the handlebody.  }
	\label{operstate}
\end{figure}
The question is, how to create the states $\langle{\mathbf{q}}|$ and $|\sigmatt,{\mathbf l}\rangle$ in our setup for Teichm\"uller TQFT. The configurations that produce these states are shown on figure~\ref{operstate}. For $|\sigmatt,{\mathbf l}\rangle$, we take a handlebody $\bar C_{\sigmatt,{\mathbf l}}$, with $\partial \bar C_{\sigmatt,{\mathbf l}}=C$, decorated by a network of monodromy defects, according to the pants decomposition $\sigmatt$, labeled by hyperbolic conjugacy classes, corresponding to the length parameters ${\mathbf l}$. The trivalent graph supporting the monodromy operators will be also called $\sigmatt$. The state produced by Teichm\"uller TQFT on the boundary $C$ of $\bar C_{\sigmatt,{\mathbf l}}$ is an eigenstate of a maximal set of commuting length operators with eigenvalues ${\mathbf l}$, and therefore is precisely $|\sigmatt,{\mathbf l}\rangle$. In the 4d Yang-Mills picture, this geometry should be tensored with the usual interval with D5 and NS5 boundary conditions. To construct the state $\langle{\mathbf q}|$, we put the 4d Yang-Mills theory on the geometry which is shown in figure~\ref{operstate}-a. It is the product of the Riemann surface $C$ and a brane corner. The definition of the D5 boundary condition depends on the metric on $C$, which determines some complex structure. The claim is that the brane corner configuration creates the state $\langle\mathbf{q}|$, with the complex structure ${\mathbf q}$ determined by the metric. Before proving this, we explain a subtlety involved in the definition of the brane corner.

In the super Yang-Mills configuration which defines Teichm\"uller TQFT with real coupling $b^2\in\RR^+$, at $y=0$ we used a particular sort of D5-type boundary condition which requires the super Yang-Mills adjoint one-form field $\phi$ to have a Nahm pole, while the gauge field $A$ is non-singular. In the Hitchin sigma-model, the support of the corresponding Lagrangian brane $\mathcal{B}_{\mathcal{T}}$ is the Hitchin section. This boundary condition preserves the topological supersymmetry $\Qb$ with the Kapustin-Witten parameter $t=-1$, which we used in the definition of the model. Of course, the same supersymmetry is preserved by the NS5-type boundary condition at $y=1$. However, it does not necessarily follow that it is possible to define a $\Qb$-invariant action with the two boundaries meeting at a corner. In fact, we will argue in a moment that it is not possible. 

The D5-type boundary conditions come in a family, parameterized by the field strength of the fivebrane gauge field. They all can be called tilted Nahm pole boundary conditions, because they require some components of the fields $A$ and $\phi$ to have the Nahm's singularity near the brane, and some combinations of the components of these fields to be non-singular. In section~\ref{d5sec}, we described a one-parametric subspace of such boundary conditions, singled out by the property of invariance under the 3d Lorentz symmetry group SO$(3)_W'$ of the twisted theory. For the brane corner configuration, it is enough that the two-dimensional Lorentz symmetry along $C$ is preserved. The boundary conditions with this property require that  the combination $A_\bz-w\phi_\bz$ of the fields, for some complex $w$, is non-singular near the brane, and other components have appropriate Nahm pole behavior. The support of the corresponding  brane in the Hitchin sigma-model is a Lagrangian submanifold, holomorphic in the complex structure $I_w$. For $w=0$, it is the Hitchin section, and the brane is $\mathcal{B}_{\mathcal T}$, which we used for Teichm\"uller TQFT. For $w=-i$, the brane is $\mathcal{B}_{\rm op}$ with the support on the variety of opers, while for general $w\ne 0$, the support is a submanifold, diffeomorphic to the variety of opers. We claim that it is possible to define a corner for our NS5-brane and a D5-brane which corresponds to $w=-i$, but not to $w=0$. Here are some supporting arguments, and the reader may choose the one they find the most convincing.
\begin{itemize}
	\item On the NS5-brane, we effectively have an analytically-continued Chern-Simons theory with the $\Qb$-invariant complexified gauge field $\cA=A+i\phi$. The corner with the D5-brane should define a meaningful boundary condition in this Chern-Simons theory. The tilted D5-brane with $w=-i$ corresponds to the oper boundary condition, which certainly does make sense. The D5 boundary condition with $w=0$ does not.
	\item Recently, brane corners like the one that we need here were studied in \cite{Corners}. There, they were obtained by twisting brane corner configurations in the physical theory. The supersymmetric NS5-D5 corners considered there have the tilted D5-brane, imposing the oper boundary condition in Chern-Simons theory.
	\item In the Hitchin sigma-model, the coisotropic brane $\mathcal{B}_c$ supports a Chan-Paton gauge field with non-trivial curvature.\footnote{To be precise, for topological reasons we had to move this curvature into the bulk, by turning on a $B$-field. This is unimportant for the present argument.} For the corner with a Lagrangian brane $\mathcal{B}_L$ to be $\Qb$-invariant, the curvature should vanish, when restricted to the support of $\mathcal{B}_L$, just like the symplectic form. Therefore, a corner configuration is possible, if the brane $\mathcal{B}_L$ is complex Lagrangian for the holomorphic form $\Omega_J$ which governs the physics of our coisotropic brane. The $(A,B,A)$ brane of opers $\mathcal{B}_{\rm op}$ has this property. The variety of opers is holomorphic Lagrangian in complex structure $J$.
\end{itemize}
Let us make a digression. A part of the important paper \cite{NW} of Nekrasov and Witten addresses the question of why the Liouville conformal blocks appear in the context of the AGT correspondence. In that paper, the 6d $(2,0)$ theory is put on a four-ball $B^4_{\eps_1,\eps_2}$ with an Omega-background. Upon reduction on a torus fiber, one obtains the Hitchin sigma-model with the brane corner configuration, as in figure~\ref{operstate}-a. The obvious brane setup to obtain the conformal blocks is the one that we used for quantization of the Teichm\"uller space, with the Lagrangian brane $\mathcal{B}_{\mathcal T}$ on the left and the coisotropic brane $\mathcal{B}_{\rm c}$, governed by $\Omega_J$, on the right. Yet in \cite{NW}, perhaps surprisingly, it was found that the Lagrangian brane obtained from the reduction is $\mathcal{B}_{\rm op}$, not $\mathcal{B}_{\mathcal T}$. In other words, this brane is Lagrangian not just for the symplectic form of the A-model, but for the holomorphic symplectic form $\Omega_J$, which appears in the definition of the coisotropic brane on the right. We have just explained that this must be a consequence of $\Qb$-invariance of the brane corner.

So, how do we define the state $\langle\mathbf{q}|$ in Teichm\"uller TQFT? We propose that there exists a brane corner configuration as in figure\ref{operstate}-a, where the NS5-brane on the right is the one that we used for Teichm\"uller TQFT, and the topological supercharge is the Kapustin-Witten supercharge $\Qb$ with $t=-1$, but the angle of the Nahm pole of the D5-brane on the left rotates with time $x^0$. At large time the Nahm pole becomes untilted, as needed for Teichm\"uller TQFT with 3d covariance.  At small time, the Nahm pole is tilted, so that the brane corner configuration is supersymmetric. In the language of the Hitchin sigma-model, the brane is holomorphic Lagrangian in complex structure $I_w$ with $w(x^0)\in i\RR$, interpolating between $w=0$, corresponding to $\mathcal{B}_{\mathcal{T}}$, and $w=-i$, corresponding to $\mathcal{B}_{\rm op}$. It is easy to see that for any of these values of $w$, the brane is supersymmetric in the $A$-model in complex structure $K$, and we propose that it can be also made supersymmetric, when $w$ varies with $x^0$. This would produce an isomorphism between the spaces of $({\mathcal B}_{\mathcal T},{\mathcal B}_{\rm c})$ strings and $({\mathcal B}_{\rm op},{\mathcal B}_{\rm c})$ strings, needed to complete the derivation of the AGT correspondence in the approach of \cite{NW}. We will not try to understand the details of the sigma-model definition of this rotating brane here, but will optimistically assume that it makes sense.

In section~\ref{redu} we analyzed, how the brane configuration needed for Teichm\"uller TQFT could arise from six dimensions. We started with a product geometry $W\times S^3$ with a particular three-dimensional supergravity background on the three-sphere, and reduced on a torus fiber. A similar story could be repeated starting from the six-dimensional geometry $C\times B^4$. On the four-ball $B^4$, we choose a supergravity background, such that a theory with 4d $\cN=2$ supersymmetry can be put on it, preserving some supercharges. Such backgrounds are reviewed in \cite{PestunRev}, the Omega-background being a particular example.\footnote{See also \cite{Komar3d2} for a similar story for 4d $\cN=1$ theories with unbroken U$(1)$ R-symmetry.} It is expected that the supersymmetric observables do not depend on much of the details of the background. A four-ball is a bundle of a two-torus over a corner geometry. It should be possible to change the metric on $B^4$, while affecting only the $\Qb$-exact terms, so as to make the torus fibers small. Upon reducing on them, one obtains the brane corner configuration of figure~\ref{operstate}-a. For the Omega-background, this has been done by Nekrasov and Witten \cite{NW}. We believe that there exist more general backgrounds which near the center of $B^4$ reduce to the usual omega-background, while near its boundary reduce to a pullback of the supersymmetric $S^3$ background that we have used in this paper. Moreover, these backgrounds should differ from the usual omega-background by $\Qb$-exact terms only. Upon reducing on a torus fiber, one would obtain our brane corner with a D5-brane with a rotating Nahm pole. It would be interesting to understand these backgrounds and the corresponding reduction more explicitly. (Note that a generic background in 4d upon reduction would lead to branes with rotating Nahm pole angle, just like in 3d a generic background lead to Janus configurations with spatially varying parameter $t$. This supports the view that such branes should make sense and be supersymmetric.)

After this long digression, we return to the state $\langle{\mathbf q}|$ and its relation to the brane corner. We recall some standard facts, closely following \cite{TeschnerLanglands}. The notion of an oper connection depends on a complex structure on the Riemann surface $C$. Fixing a particular complex structure defines a holomorphic Lagrangian subspace of opers in $\mathcal{M}_H$. Varying the complex structure, we get an affine bundle $\mathcal{P}$ of opers over $\mathcal{T}$ which is canonically isomorphic to $T^*\mathcal{T}$ as holomorphic affine bundles.\footnote{Since both the fibers of $T^*\mathcal{T}$ and of $\mathcal{P}$ are a torsor for the vector space of holomorphic quadratic differentials on $C$, in order to build the isomorphism, it is enough to construct a canonical holomorphic cross-section of $\mathcal{P}$. For that, see \cite{Kawai}. Note that this is not the cross-section of $\mathcal{P}$ given by the metric PSL$(2,\RR)$ opers. The latter is not holomorphic.} Moreover, the map from $\mathcal{P}$ to $\mathcal{M}_H$, given by computing the monodromy of an oper, is a local biholomorphism. In fact, upon identifying $\mathcal{P}$ and $T^*\mathcal{T}$, this map becomes a holomorphic symplectomorphism which pulls back the holomorphic symplectic form $\Omega_J$ on $\mathcal{M}_H$ to the standard symplectic form on $T^*\mathcal{T}$. Let ${\bf q}=\{q_a\}$ and ${\bf H}=\{H_b\}$ be the canonically conjugate holomorphic coordinates on the base and in the fiber of $T^*\mathcal{T}$. Upon quantization of the algebra of holomorphic functions on $\mathcal{M}_H$ with the symplectic form $b^2\Omega_J$, they become operators. The state $\langle{\mathbf q}|$ is an eigenstate of the operators $\hat q_a$. (As usual, a wavefunction in the ${\mathbf q}$-representation is given by the Hermitian product of a particular state with $\langle \mathbf{q}|$. The oper, or equivalently, the energy-momentum tensor $\hat T(z)$ acts on these wavefunctions by the familiar differential operator, given by substituting $H_b$ in the expansion of the oper in quadratic differentials by the operators $\partial_{q_b}$. The conformal block (\ref{conf2}) is an example of such wavefunction.) Now, the corner with an oper brane lands us in a fiber in $\mathcal{P}$ which lies above the complex structure $\mathbf{q}$ of $C$ which is used in the definition of the brane. When acting with the operators $\hat{q}_a$ on the state, produced by the quantum theory in the geometry of figure~\ref{operstate}-a, we can move these operators right into the corner, which makes it clear that they act by multiplication by $c$-numbers $\mathbf{q}$. The state produced by the corner must be the eigenstate~$\langle\mathbf{q}|$.

\begin{figure}
	\begin{center}
		\includegraphics[width=3.8cm]{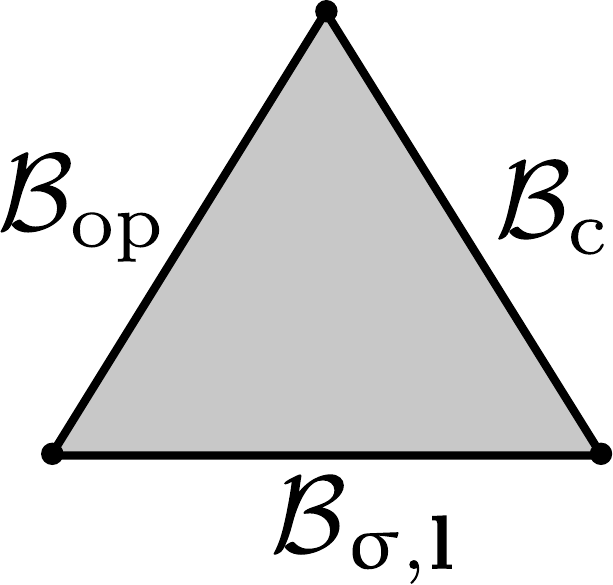}
	\end{center}
	\caption{\small A Liouville conformal block can be represented by a brane triangle.}
	\label{triangle}
\end{figure}
The geometry of figure~\ref{operstate} can be usefully deformed into a brane triangle,\footnote{The corners in this triangle are not some arbitrary states of the corresponding open strings. We have said some words about the sigma-model definition of the $(\mathcal{B}_c,\mathcal{B}_{\rm op})$ corner, and have nothing to say about the $(\mathcal{B}_{\sigmatt,{\mathbf l}},\mathcal{B}_{c})$ corner. One may hope that the Lagrangian branes $\mathcal{B}_{\rm op}$ and $\mathcal{B}_{\sigmatt,{\mathbf l}}$ intersect over a single point. If so, the state corresponding to the $(\mathcal{B}_{\rm op},\mathcal{B}_{\sigmatt,{\mathbf l}})$ corner is defined uniquely. } shown on figure~\ref{triangle}. Here $\mathcal{B}_c$ and $\mathcal{B}_{\rm op}$ are the familiar coisotropic brane and the brane of opers, while $\mathcal{B}_{\sigmatt,{\mathbf l}}$ is a Lagrangian brane, whose support are the complex flat connections that can be continued into the interior of the handlebody $\bar C_{\sigmatt,{\mathbf l}}$. This brane picture for conformal blocks has been recently proposed in \cite{coin}. Note that it makes sense for any non-zero complex $b^2$. Indeed, both Lagrangian branes are actually holomorphic Lagrangian for the symplectic form $\Omega_J$. Then it is possible to set the symplectic form and the B-field to be the real and the imaginary parts of $b^2\Omega_J$, while taking the Chan-Paton connections on all branes to be flat.

\begin{figure}
	\begin{center}
		\includegraphics[width=10cm]{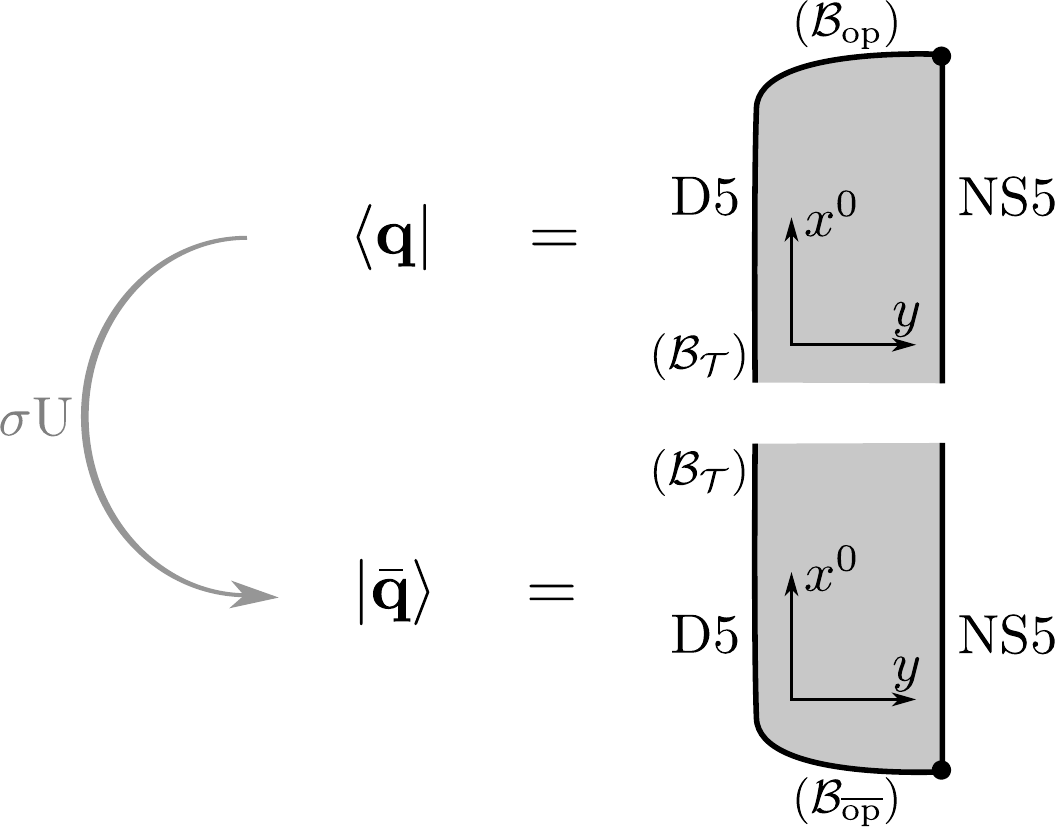}
	\end{center}
	\caption{\small Hermitian conjugation in topological theory is realizaed by operator $\sigma U$. It transforms the state $\langle\mathbf{q}|$ into $|\bar{\mathbf q}\rangle$, as shown. Gluing these two states together gives the Liouville partition function.}
	\label{lstates}
\end{figure}
We can also understand the Liouville partition function in the language of Teichm\"uller TQFT, making contact with some results in \cite{Harlow}. It is equal to the norm of the state $\langle \mathbf{q}|$, or
\beq
{Z}_{\rm Liouv}=\langle {\bf q}|\bar{\bf q}\rangle\,.
\eeq
To make sense of this formula, we need to define the conjugate state $|\bar{\bf q}\rangle=(\langle{\bf q}|)^\dagger$. In the branes and quantization setup, the appropriate antilinear map between the dual vector spaces $\mathcal{H}_{({\mathcal B}_{\mathcal T},{\mathcal B}_{\rm c})}$ and $\mathcal{H}_{({\mathcal B}_{\rm c},{\mathcal B}_{\mathcal T})}$  was described by Gukov and Witten \cite{BQ}. First, one uses the antilinear map, to be called $U$, that defines the scalar product in the physical sigma-model. That, however, maps the topological supercharge $\Qb$ into its conjugate. To bring it back, one also applies a symmetry that acts in the target space by an antiholomorphic\footnote{In the complex structure in which the A-model is defined.} involution, preserving the Lagrangian brane. In our setting, this can be taken to be the symmetry that flips the sign of the Higgs field $\phi_z$. It was called $\sigma$ in section~\ref{tquant}. If the state $\langle {\mathbf q}|$ is produced by the path-integral on a corner, shown on figure~\ref{operstate}-a, then the state, obtained by acting with the antilinear involution $U$, is given by the path-integral on the same geometry, but with the opposite choice of the orientation. Further applying  the involution $\sigma$ leads to the brane corner configuration, shown in the bottom of figure~\ref{lstates}. There, the D5-brane type interpolates from the untilted Nahm pole, corresponding to the Teichm\"uller brane $\mathcal{B}_{\mathcal T}$, to the tilted Nahm pole, corresponding to the brane of antiopers~$\mathcal{B}_{\bar{\rm op}}$. Our terminology here means the following. A complex connection is an oper, if it can locally be gauged into the form
\beq
\cA_z=\left(\bea{cc}0&1\\T&0\eea\right)\,,\quad \cA_\bz=0\,.
\eeq
By an antioper we mean a complex connection which can be locally gauged into the form
\beq
\cA_z=0\,,\quad \cA_\bz=\left(\bea{cc}0&\bar{T}\\1&0\eea\right)\,.
\eeq

Stacking the two corners on figure~\ref{lstates} together produces a geometry with two branes and two corners. It can be obtained from an $S^4$ geometry by reducing on a torus fiber. Note that the usual Pestun's $S^4$ background \cite{PestunOld} approaches an Omega-background at one pole, with the BPS configurations being instantons, and an anti-Omega-background at the other pole, with the BPS configurations being anti-instantons. According to Nekrasov and Witten \cite{NW}, reduction on a torus fiber should lead to a $(\mathcal{B}_{\rm op},\mathcal{B}_c)$ and a $(\mathcal{B}_{\bar{\rm op}},\mathcal{B}_c)$ corner, correspondingly, and this is what is seen in our picture. It is also clear that, whatever the details of the supergravity background and the reduction are, the result will include branes with the supersymmetry (equivalently, the Nahm pole angle) varying with time.

\subsection{Integration cycles and holography}
Let us look back at the setup of figure~\ref{operstate} for the conformal block. So far we mostly focused on what is happening at the NS5-brane, with the conclusion that we have an analytically-continued Chern-Simons theory on  the handlebody $\bar C_{\sigmatt,{\mathbf l}}$ with the oper boundary condition on $\partial \bar C_{\sigmatt,{\mathbf l}}$. What is the correct integration cycle in this theory? What integration cycles are possible in principle?

Critical points of the action are complex flat connections on the handlebody $\bar{C}_{\sigmatt,{\mathbf l}}$ which have an oper structure on $\partial \bar C_{\sigmatt,{\mathbf l}}$. In other words, they are points of intersection of the Lagrangian branes $\mathcal{B}_{\rm op}$ and $\mathcal{B}_{\sigmatt,\mathbf{l}}$. Explicitly, we need to solve a Riemann-Hilbert type problem and fix $3g-3$ accessory parameters of an energy-momentum tensor $T$ from the prescribed conjugacy classes of monodromies of the equation $(\partial_z^2+T)\psi=0$ along $3g-3$ cycles on $C$. Solutions to this problem are expected to exist generically. For example, in the simple case when $C$ is $\mathbb{P}^1$ with four punctures,\footnote{So far we did not consider punctures on $C$, but that restriction is inessential and was made just for notational simplicity.} the solution exists and is unique, see {e.g.} appendix~D of \cite{Harlow}. 

We remind that even if the flat connection is unique modulo gauge transformations, the number of possible integration cycles is infinite, because we do not divide by large gauge transformations. An elementary topologically non-trivial gauge transformation, corresponding to $\pi_3$ of the gauge group, multiplies the path-integral by $q=\exp(2\pi ib^2)$. In the presence of monodromy defects, the gauge group is reduced along these defects to its maximal torus. Then there are also topologically non-trivial gauge transformations, corresponding to $\pi_1$ of the maximal torus. The factors produced by these gauge transformations are $\exp(2\pi i b^2l_a)$, where $l_a$ are parameters of the monodromy operators, see {e.g.} \cite{Harlow}. Let $\mathcal{C}_0$ be the Lefschetz thimble, corresponding to some oper flat connection $\mathbf{a}_0$, taken in some particular gauge. Let $\mathcal{C}$ be some other integration cycle which is obtained from $\mathcal{C}_0$ by making large gauge transformations and taking sums with integer coefficients. From what we have just said, it is clear that the path-integral over $\mathcal{C}$ differs from the integral over $\mathcal{C}_0$ by multiplication by some Laurent series $\mathcal{P}$ in $\exp(2\pi i b^2)$ and $\exp(2\pi i b^2l_a)$ with integer coefficients.

As always in Teichm\"uller TQFT, the correct integration cycle should be obtained as the locus of endpoints of the Kapustin-Witten flows in $y$. In our figure~\ref{operstate}, the D5-brane is drawn in a suggestive way: its horizontal part is meant to impose the oper boundary condition for the complex gauge field $\cA$ along $\partial \bar C_{\sigmatt,{\mathbf l}}$, while its vertical part imposes the Nahm pole initial condition for the Kapustin-Witten flows.\footnote{We could perhaps deform this D5-brane into a brane corner with a vertical $\mathcal{B}_{\mathcal T}$ and a horizontal $\mathcal{B}_{\rm oper}$. The supports of these two Lagrangian branes intersect at a single point, so there is only one possible boundary changing operator.} It is 
natural to speculate that the conjectures of section~\ref{hyperb} still apply in this setting. Let us see, how this could possibly work. First we explain, what is the analog of the conjugate geometric flat connection here.

Let $\d s^2_C=\ex^{2\varphi}\,\d z\,\d\bz$ be a metric on $C$, Hermitian in the complex structure $\mathbf{q}$. Suppose we want to construct a complete hyperbolic metric on the handlebody $\bar{C}_{\sigmatt,{\mathbf l}}$ which near the boundary is locally asymptotic to the hyperbolic space ${\rm H}_3$ and conformal to $\d s^2_C$. In other words, asymptotically, the metric should look like
\beq
\d s^2=\fr{\d x^2+\d s_C^2}{x^2}+\dots\,,
\eeq
where $x$ is the radial coordinate which goes to zero at the conformal boundary. (On figure~\ref{operstate} it was called $x^0$.) The dots stay for less singular terms. With a convenient choice of the vielbein, the leading terms in the vielbein and the spin-connection are
\beq
e=\fr{i}{2x}\left(\bea{cc} \d x & \ex^{\varphi}\,\d z\\ \ex^{\varphi}\,\d\bar z&-\d x\eea\right)+\dots\,,\quad \omega=\fr{1}{2x}\left(\bea{cc}0&\ex^{\varphi}\,\d z\\-\ex^{\varphi}\,\d \bz&0\eea\right)+\dots\,.
\eeq
(Recall that $e$ and $\omega$ are locally one-forms, valued in $\mathfrak{su}(2)$, that is, in anti-Hermitian matrices.) The complexified  gauge field $\cA=\omega-ie$ is then
\beq
\cA_z=\fr{\ex^{\varphi}}{x}\left(\bea{cc}0&1\\0&0\eea\right)+\dots\,,\quad
\cA_\bz={\rm O}(1)\,,\quad
\cA_x=\fr{1}{2x}\left(\bea{cc}1&0\\0&-1\eea\right)+\dots\,.\label{tilnam}
\eeq
To solve the Einstein equations, we are supposed to find a flat gauge field with these asymptotics, modulo complex gauge transformations that are trivial at $x\rightarrow 0$. The behavior (\ref{tilnam}) is precisely the tilted Nahm pole, and the solution \cite{GWJones} to the problem is precisely that the flat connection is an oper. (This is a well-known fact in the 3d gravity context, see {e.g.} \cite{Carlip}.)

Inside $\bar{C}_{\sigmatt,{\mathbf l}}$, there is a network of defects along the graph $\sigmatt$, and we want our geometric connection $\cA=\omega-ie$ to have the corresponding monodromies. When the parameters $\mathbf{l}$ are real, the conjugacy classes of the monodromies are hyperbolic. The graph $\sigmatt$ is the world-history of a number of colliding BTZ black holes inside $\bar C_{\sigmatt,{\mathbf l}}$. To make sense of this statement and to obtain a complete hyperbolic metric, we would need to cut out a neighborhood of $\sigmatt$ from $\bar C_{\sigmatt,{\mathbf l}}$ and to glue in some appropriate smooth manifold to represent the black hole geometry. It is not clear what is the right way to do this, so let us instead continue the Liouville conformal block to purely imaginary values of the Liouville momenta $\mathbf{l}$, and moreover, restrict them to be rational. Then the monodromy defects along $\sigmatt$ define orbifold singularities of the metric.

Suppose that $\bar C_{\sigmatt,{\mathbf l}}$ has a complete hyperbolic metric, as described: its conformal structure at infinity agrees with the complex structure $\mathbf{q}$, and in the bulk it has orbifold singularities along the graph $\sigmatt$, with the deficit angles determined by $\mathbf{l}$. From what we have just said, it is clear that the corresponding flat connection $\omega-ie$ belongs to the intersection of $\mathcal{B}_{\rm op}$ and $\mathcal{B}_{\sigmatt,\mathbf{l}}$. The converse need not be true: a point in the intersection $\mathcal{B}_{\rm op}\cap\mathcal{B}_{\sigmatt,\mathbf{l}}$ may not define a complete hyperbolic structure on $\bar C_{\sigmatt,{\mathbf l}}$. Explicitly, any oper flat connection can be gauged into the form
\beq
\cA=\left(\bea{cc}-\fr{1}{2}\d\rho & \ex^{\rho}\,\d z\vspace{2mm}\\ T\ex^{-\rho}\,\d z&\fr{1}{2}\d\rho\eea\right)\,,
\eeq
where $T$ is a holomorphic stress tensor, and $\rho=-\log x+\varphi$. (Note that $\rho$ has to transform non-trivially between the patches on $C$, while $x$ does not.) This defines a metric
\beq
\d s^2=\d \rho^2+|\ex^{\rho}\d z+\bar T\ex^{-\rho}\d\bz|^2\,,
\eeq
which is a non-degenerate hyperbolic metric for
\beq
x<{\rm min}_C\left(\ex^\varphi|T|^{-1/2}\right)\,.
\eeq
If $C$ has no punctures, this metric makes sense in a finite neighborhood of the boundary of the handlebody, yet it is not obvious that it can be continued as a complete non-degenerate hyperbolic metric onto the whole  $\bar C_{\sigmatt,{\mathbf l}}$. 

In fact, a theorem due to Sullivan \cite{Sul,McM} implies that for ``nice'' hyperbolic three-manifolds, the hyperbolic structure is unique, once the conformal structure at infinity is fixed. This theorem is an extension of Mostow rigidity to non-compact hyperbolic three-manifolds. For an example of its application in physics, see \cite{Bir}. To honestly apply it to our case, we would need to check one technical condition (that our tentative hyperbolic three-manifolds are geometrically finite), and to understand, how to deal with the orbifold singularities. We will not try to do this, but will just optimistically hope that the theorem does apply. Thus, by our natural assumptions, the intersection $\mathcal{B}_{\rm op}\cap\mathcal{B}_{\sigmatt,\mathbf{l}}$ contains one and precisely one point that defines a complete hyperbolic structure on $\bar C_{\sigmatt,{\mathbf l}}$. The corresponding flat connection will be called the conjugate geometric flat connection $\cA_{\bar{\rm geom}}$. Of course, there may be any number of other flat connections in $\mathcal{B}_{\rm op}\cap\mathcal{B}_{\sigmatt,\mathbf{l}}$.

Conjecture~3 of section~\ref{hyperb}, applied to this situation,\footnote{Explicitly, we use the hyperbolic metric on the handlebody to define the $\cN=4$ super Yang-Mills path-integral on $\bar C_{\sigmatt,{\mathbf l}}\times \mathcal{I}$. Two differences with the setup of figure~\ref{operstate}~---~as obtained by reduction from six dimensions,~---~is that the boundary of $\bar C_{\sigmatt,{\mathbf l}}$ is now at infinite distance, and the Gukov-Witten surface operators for the monodromy defects on $\sigmatt\times\mathcal{I}$ are replaced by an orbifold singularity. This should not affect the path-integral of the topological theory. Also, in section~\ref{hyperb} it was important that the reducible connections do not appear, but our argument used the fact that the volume of the three-manifold was finite. In the present section, reducible connections do not appear because of the oper boundary condition.} says that the dominant contribution to the Teichm\"uller TQFT path-integral for the Liouville conformal block comes from the Lefschetz thimble for the conjugate geometric flat connection. Thus, the setup of figure~\ref{operstate} can be loosely understood as a chiral gravitational path-integral on $\bar C_{\sigmatt,{\mathbf l}}$ with an asymptotically AdS metric. We obtain something that is well-known in the context of holography.

There exists a vast literature on holographic computations of semiclassical Virasoro conformal blocks, see {e.g.} \cite{Hijano}, \cite{Perlmutter} and references therein. (Note that in the semi-classical limit, the fact that the gravity on ${\rm AdS}_3$ computes Virasoro conformal blocks is essentially a consequence of the Ward identities \cite{Hijano}.) It mainly focuses on the case of $\mathbb{P}^1$ with four punctures, at most two of which carry operators that are heavy in the semi-classical limit $c\rightarrow \infty$. There is a beautiful story on how to incorporate light operators by geodesic Witten diagrams and how to actually compute the expansions of the blocks and to compare them to the known CFT results. We have nothing to say about this. Our goal was to explain that this holographic picture can be naturally obtained in Teichm\"uller TQFT, which in particular implies that it should work not only semi-classically, but also quantum-mechanically.

We made an assumption about the existence of an appropriate hyperbolic metric on $\bar C_{\sigmatt,{\mathbf l}}$. This was necessary both for our argument to work, and for the holographic picture to make sense. It would be interesting to construct this hyperbolic metric explicitly, by identifying the corresponding subgroup of the isometries of ${\rm H}_3$. Also, in eq.~(\ref{bound}) of section~\ref{hyperb}, we stated a bound on the Chern-Simons action in terms of the hyperbolic volume. It was saying essentially that the conjugate geometric flat connection was the most low-lying critical point. If an analog of this statement holds as well for suitably regularized volumes in the non-compact setting of the present section, then the integration cycle for the path-integral for the conformal block is precisely the Lefschetz thimble for the hyperbolic flat connection on $\bar C_{\sigmatt,{\mathbf l}}$. In this case, the degrees of freedom in the bulk of the holographic setup are truly gravitational. If the statement above is not true, the conformal block may receive subleading contributions from lower-lying flat connections, which have no geometric meaning.

On a further philosophical note, it may be questionable if the holographic setup for the conformal blocks deserves to be called ``true holography''. Indeed, 
the hallmark of holography is that on one side of the duality, we have gravity. In the present context, even if local degrees of freedom in the bulk are components of an invertible metric, the theory there is a QFT and not quantum gravity. We do not sum over possible topologies, but instead explicitly prescribe what should be happening in the bulk according to the parameters $(\sigmatt,{\mathbf l})$ of the conformal block.

We have to admit that in our derivation, we were not careful about the overall normalization of the states $\langle {\bf q}|$ and $|\sigmatt,{\bf l}\rangle$. The function $\Psi^{\sigmatt,{\bf l}}({\bf q})$ could differ from the conformal block in the standard normalization by a $b^2$-dependent prefactor, which would make pointless much of the above discussion of integration cycles. {E.g.}, for the four-point conformal block on $\mathbb{P}^1$, there exists up to gauge transformations only one complex flat connection, and possible integration cycles differ just by Laurent series $\mathcal{P}$, which could be missed with improper normalization.  However, the brane picture of figure~\ref{operstate} could have been derived from six dimensions, and therefore, via the AGT correspondence, our conformal blocks are expected to be normalized in the standard way.

Finally, we point out that K\"ahler quantization basis of states $\langle {\bf q}|$ is useful not only for conformal blocks. For any three-manifold $W$ with boundary, we can choose conformal structures on each component of $\partial W$ and use the oper boundary conditions to write the state $Z_{\rm Teicm}(W)$ in this basis. The advantage of doing so is that, if $W$ is hyperbolic, the Teichm\"uller TQFT integration cycle is simple. From the point of view of hyperbolic geometry, it is very natural to consider hyperbolic three-manifolds with fixed conformal structures at the boundary. It is well-known in mathematics, how to glue them together to produce new hyperbolic manifolds. In this sense, the basis $\langle\bf{q}|$ is very natural geometrically. We hope to discuss elsewhere, how to glue Teichm\"uller wavefunctions in this basis. 

\subsection{The Liouville partition function}
According to figure~\ref{lstates}, the Liouville partition function on a Riemann surface $C$ is computed by analytically-continued Chern-Simons theory on $C\times\mathcal{I}'$, with the oper and the antioper boundary conditions at the two ends of the interval $\mathcal{I}'$. This picture was first obtained in the paper \cite{Harlow}, which was actually the motivation for our discussion.

Let us look at the possible integration cycles. The flat connections on $C\times \mathcal{I}'$, satisfying the boundary conditions, are points of the intersection of the Lagrangian branes $\mathcal{B}_{\rm op}$ and $\mathcal{B}_{\bar{\rm op}}$. One such point is the real, or ``geometric'' oper which defines the hyperbolic structure on $C$, compatible with the chosen complex structure. Explicitly, if $\d s^2_C=\ex^{2\varphi}\d z\d \bar z$ is the hyperbolic metric on $C$, so that $\partial\bar\partial\varphi=\ex^{2\varphi}$, then the real oper is
\beq
\cA^{(0)}=\left(\bea{cc}ia&\ex^\varphi\d z\\\ex^\varphi\d\bz&-ia \eea\right)\,,\label{geomop}
\eeq
where $a=-\fr{i}{2}(\partial\varphi\d z-\bar\partial\varphi\d\bz)$ is the spin connection. The reader may check that (\ref{geomop}) can be gauged into the oper and the antioper form. The real oper is the only oper with PSL$(2,\RR)$ holonomies, or equivalently, the only point in $\mathcal{B}_{\rm op}$, invariant under the automorphism $\sigma$. (Therefore, it is also the only point of intersection of $\mathcal{B}_{\rm op}$ and $\mathcal{B}_{\mathcal{T}}$. It is the point in $\mathcal{B}_{\mathcal{T}}$ that lies over the origin of the base of the Hitchin fibration.) We do not know if the intersection of the branes $\mathcal{B}_{{\rm op}}$ and  $\mathcal{B}_{\bar{\rm op}}$ contains any other points. If it does, those would come in pairs, exchanged by the automorphism $\sigma$, since the only $\sigma$-invariant point is the real oper. Perhaps, this fact could be used to prove that there are no other points, but we will not assume that.

The real oper defines a complete hyperbolic metric on $C\times\mathcal{I}'$. Explicitly, the flat connection is obtained from (\ref{geomop}) by a gauge transformation ${\rm diag}(\ex^{\rho/2},\, \ex^{-\rho/2})$, 
and the metric is
\beq
\d s^2=\d\rho^2+2\cosh^2\hspace{-0.8mm}\rho\,\d s_C^2\,,\label{hypint}
\eeq
with the two asymptotic ends at $\rho\rightarrow\pm\infty$. A more geometric way to view this is the following. Let $\Gamma\subset{\rm PSL}(2,\RR)$ be the Fuchsian subgroup by which we quotient the upper half-plane to obtain $C$ with its chosen complex structure. The hyperbolic three-space ${\rm H}_3$ is topologically a ball, and its boundary is an $S^2$. $\Gamma$ is naturally a subgroup of the isometry group PSL$(2,\CC)$ of ${\rm H}_3$. The quotient $H_3/\Gamma$ is precisely $C\times{\mathcal I}'$ with the hyperbolic metric (\ref{hypint}). In particular, $\Gamma$ acts properly discontinuously on the boundary $S^2$ of $H_3$ away from a great circle. The quotients of the two hemispheres in $S^2$ become the two asymptotic ends of $C\times\mathcal{I}'$. The theorem due to Sullivan, mentioned in the previous subsection, implies that this hyperbolic structure on $C\times{\mathcal I}'$ is unique \cite{McM}. In this particular geometry, it is a special case of the Bers simultaneous uniformization theorem \cite{Bers}.
Thus, the real oper is the only point in $\mathcal{B}_{{\rm op}}\cap\mathcal{B}_{\bar{\rm op}}$ that defines a complete hyperbolic metric on $C\times \mathcal{I}'$.

As figure~\ref{lstates} suggests, the correct integration cycle for the Teichm\"uller TQFT on $C\times\mathcal{I}'$ should be obtained by solving the Kapustin-Witten equations on $C\times\mathcal{I}'\times\mathcal{I}$ with the Nahm pole initial condition. Assuming that Conjecture~3 of section~\ref{hyperb} holds in this setting, the dominant contribution to the functional integral comes from the Lefschetz thimble for the conjugate geometric flat connection, that is, for the real oper (\ref{geomop}). If $\mathcal{B}_{{\rm op}}\cap\mathcal{B}_{\bar{\rm op}}$ contains any lower-lying critical points, there can be additional contributions, and if not, then the geometric Lefschetz thimble is all that there is.

Did we learn anything new about the Liouville partition function? Certainly not. Just from the definition of the usual Liouville path-integral, it is obvious that the dominant contribution in it comes from the real critical point, that is, from the Lefschetz thimble for the real oper (\ref{geomop}). What we did obtain is a non-trivial consistency check of the conjectures of section~\ref{hyperb}. Note that this is true, even if the flat connection on $C\times\mathcal{I}'$ with the oper and the antioper boundary conditions is unique. Indeed, even in this case, there is an infinite number of possible integration cycles for Chern-Simons due to the fact that we do not divide by  large gauge transformations. A generic integration cycle would differ from the correct one my multiplying the partition function by a Laurent series $\mathcal{P}$ in $q=\exp(2\pi ib^2)$. Our conjectures assure that such a prefactor does not appear. We remind the reader that all this applies to the regime, when the Liouville coupling constant $b$ is real. When it is not, our conjectures say nothing about the integration cycle. A beautiful analysis  in \cite{Harlow} has shown that a prefactor of the form $\mathcal{P}$ does appear, when $b$ becomes imaginary.

The setup for computing the Liouville partition function that we have just considered is holographic in spirit, but we should again stress that the bulk theory is a QFT rather than gravity, even though its dynamical fields have a metric interpretation. Moreover, instead of a non-chiral theory in a geometry with one asymptotic region, we have a chiral theory in a geometry with two asymptotic regions. This is quite unusual.

Note that the existence and uniqueness of a hyperbolic structure on $C\times\mathcal{I}$ continue to hold, if we independently vary the complex structures at the two asymptotic ends. This is the content of the simultaneous uniformization theorem. Then there is a natural continuation of the integration cycle to the situation when $\mathbf{q}$ and $\bar{\mathbf q}$ are independent variables, so that the Liouville partition function becomes a function on $\mathcal{T}\times\bar{\mathcal T}$.
\beq
Z_{\rm Liouv}(\mathbf{q}_1,\bar{\mathbf{q}}_2)=\langle {\mathbf{q}}_1|\bar{\mathbf q}_2\rangle\,.
\eeq
That such a continuation should be possible is clear from the decomposition into conformal blocks. In the  4d $\cN=2$ class-$\mathcal{S}$ gauge theory, related to this setup via the AGT correspondence, this quantity must be the $S^4$-partition function with a $1/2$-BPS Janus interface inserted near the equator. 

\section{Remarks on PSL$(2,\RR)$ Chern-Simons theory and other matters}\label{SecRemarks}
Here we present an informal discussion on the relations between Teichm\"uller TQFT, PSL$(2,\RR)$ Chern-Simons theory, as well as on some other matters. This section is not meant to contain any essentially new results, but may help in putting our story in broader context.

\subsection{PSL$(2,\RR)$ Chern-Simons theory}
Since ancient times \cite{VerlindeVerlinde,Verlinde}, Teichm\"uller TQFT is viewed as a sort of PSL$(2,\RR)$ Chern-Simons theory. We recall the reason for this, and highlight important differences between the two theories.
As usual, let $C$ be an oriented Riemann surface of genus $g\ge 2$. The classical phase space of PSL$(2,\RR)$ Chern-Simons theory on $C$ is the moduli space of flat PSL$(2,\RR)$ connections. It decomposes as
\beq
{\mathcal M}^{\rm \RR}\simeq\bigcup_{d=-2g+2}^{2g-2}{\mathcal M}^{\RR}_d\,,\label{decomp}
\eeq
according to the Euler number $d$ of the flat vector bundle. The component $\mathcal{M}_{2g-2}^\RR$ is isomorphic to the Teichm\"uller space.\footnote{Similarly, $\mathcal{M}_{-2g+2}$ is isomorphic to the conjugate Teichm\"uller space, that is, parametrizes complex structures which induce the orientation on $C$, opposite to the chosen.} It consists of flat PSL$(2,\RR)$ bundles that come from uniformization of $C$. In PSL$(2,\RR)$ Chern-Simons theory, we are supposed to quantize $\mathcal{M}^\RR$ with the symplectic form that descends from
\beq
-\fr{k}{4\pi}\int_C\tr\left(\delta \cA\wedge\delta\cA\right)\,,
\eeq
where $\cA$ is the PSL$(2,\RR)$ gauge field and $k$ is the Chern-Simons level. The subspaces in (\ref{decomp}) with $|d|<2g-2$ are topologically non-trivial, and the restriction of the symplectic form to them is non-trivial in the cohomology. This is one way to see that some appropriate multiple of the level $k$ has to be an integer, for quantization to make sense. This is the most obvious difference with Teichm\"uller TQFT: there, one quantizes only the topologically-trivial component $\mathcal{M}_{2g-2}^\RR$, and the level $b^2$ need not be an integer.

The Hilbert space of Chern-Simons theory with phase space (\ref{decomp}) is a direct sum
\beq
\mathcal{H}_{\rm CS}\simeq \bigoplus_{d=-2g+2}^{2g-2}\mathcal{H}_d\,.\label{Hd}
\eeq
The Hilbert space of Teichm\"uler TQFT is the component $\mathcal{H}_{2g-2}$.
The mapping class group (MCG) of $C$ acts on $\mathcal{M}^\RR$ by symplectomorphisms which preserve the Euler number of the bundle, and therefore the components $\mathcal{M}^\RR_d$ of the moduli space. Then each of the vector spaces $\mathcal{H}_d$ produced by quantization should carry a MCG representation. In other words, assigning vector spaces $\mathcal{H}_d$ to Riemann surfaces defines what is known as a modular functor.

As long as we are concerned with Hilbert spaces only, we can consider PSL$(2,\RR)$ Chern-Simons theory on a geometry, say, $\RR^+\times C$, and place a suitable boundary condition at the origin of $\RR^+$ that would pick flat bundles on $C$ with the maximal Euler number. In this way we can project to the component $\mathcal{H}_{2g-2}$ of the Hilbert space which we would see in Teichm\"uller TQFT. (The boundary condition breaks all gauge symmetry at the origin of $\RR^+$. It is easy to see that this makes it impossible to make large PSL$(2,\RR)$ gauge transformations on $\RR^+\times C$. Therefore, we can also continue the level $k$ away from integers, as is appropriate for Teichm\"uller TQFT.) This is what was done in \cite{VerlindeVerlinde,Verlinde}.
But there is more to a 3d TQFT than its Hilbert spaces and MCG representations. Its amplitudes should be defined for any oriented three-manifold with boundary, and should respect factorization under arbitrary cuttings of the manifold. The PSL$(2,\RR)$ Chern-Simons theory has the usual path-integral definition with 3d covariance, but there seems to be no natural way to project it down to Teichm\"uller TQFT. By projecting we mean doing something with the path-integral, so that in cutting along an arbitrary embedded Riemann surface, the intermediate state would live in the component $\mathcal{H}_{2g-2}$. (One obvious thing that can be done with Chern-Simons theory is choosing a different global form of the gauge group, that is SL$(2,\RR)$ or further covers. It is easy to see that this wouldn't help.) Thus, PSL$(2,\RR)$ Chern-Simons theory and Teichm\"uller TQFT are two different quantum field theories in three dimensions, even though they have a relation at the level of Hilbert spaces. To further highlight the difference, we point out that, for a hyperbolic three-manifold, the Teichm\"uller partition function is dominated by the PSL$(2,\CC)$ flat connection $\cA_{\bar{\rm geom}}$, which is not even a PSL$(2,\RR)$ connection. (If it were possible to conjugate it into a PSL$(2,\RR)$ subgroup, the hyperbolic volume would be zero.) 

\subsection{Observables}
Let us next comment on the algebra of observables in Teichm\"uller TQFT and  Chern-Simons theory. We go back to the geometry $\RR\times C$ and see, what operators act on the Hilbert space. In PSL$(2,\RR)$ Chern-Simons theory, for any non-self-intersecting cycle $\gamma$ on $C$ we can consider a Wilson loop labeled by a representation of the gauge group. The algebra of Wilson operators corresponding to the finite-dimensional representations is what we called the operator algebra ${\mathbf A}_q$ in section~\ref{tquant}. It is unlikely that the Hilbert space $\mathcal{H}_{\rm CS}$ is an irreducible module of this algebra. One would rather expect that Wilson loops for some classes of infinite-dimensional unitary representations must be included to make the operator algebra act irreducibly. For Teichm\"uller TQFT, the action of $\mathbf{A}_q$ on its Hilbert space $\mathcal{H}_{2g-2}$ is reducible as well. To understand what operators have to be added in this case, one can again look at the moduli space of flat connections on $C$. For simplicity, for now think of SL$(2,\RR)$ instead of PSL$(2,\RR)$, and consider a Wilson line $W_\gamma$ in the fundamental representation. Let $U$ be the two-by-two matrix of the SL$(2,\RR)$ holonomy around the cycle $\gamma$. Then
\beq
W_\gamma=\tr\, U_\gamma=2\cosh \fr{l_\gamma}{2}\,,
\eeq
where $\exp(\pm l_\gamma/2)$ are the eigenvalues of the matrix $U_\gamma$. If $U_\gamma$ belongs to a hyperbolic conjugacy class, which is the case in particular if our flat connection lies in the Teichm\"uller component $\mathcal{M}_d^\RR$, then $W_\gamma\ge 2$ and $l_\gamma$ is a real positive number, equal to the length of the closed geodesic, homotopic to $\gamma$. 
If the conjugacy class of $U_\gamma$ is elliptic, $l_\gamma$ is an imaginary number, which moreover is not uniquely defined. In Teichm\"uller theory, we restrict to the Teichm\"uller component $\mathcal{M}_d^\RR$ of the moduli space, and $l_\gamma$ is a good observable. Upon quantization, it becomes an operator $\hat l_\gamma$ acting on the Hilbert space $\mathcal{H}_{2g-2}$. In PSL$(2,\RR)$ Chern-Simons theory, where we keep all components of the moduli space of flat connections, the length $l_\gamma$ is not well-defined, and there is no corresponding operator in the quantum theory. We only have the Wilson lines $\hat W_\gamma$, which generate the algebra $\mathbf{A}_q$ acting on $\mathcal{H}_{\rm CS}$. (Of course, there are also Wilson lines for infinite dimensional representations.) With length operators included, the algebra acting on $\mathcal{H}_{2g-2}$ is larger\footnote{We are being sloppy about the issues related to infinite dimensionality of the spaces of states. For example, if we declare the Hilbert space to be $L^2(\RR^{3g-3})$, so that the wavefunctions are functions of $3g-3$ length variables, then exponential operators like $\hat W_\gamma$ take us out of such Hilbert space. We will not attempt to make our statements precise.}

It is instructive to recall some further details on this enhancement of the operator algebra, following {e.g.} \cite{DimofteQuantum}. Let us go to the Fock-Goncharov coordinates on the moduli space of flat connections. This set of coordinates is defined for a choice of an ideal triangulation of the surface $C$. (This requires the surface to have punctures, which we so far did not consider.) There is a coordinate function $z_a$ for each edge $a$ of the triangulation, and there are some relations imposed, see {e.g.} \cite{FG,GMN,DimofteQuantum}. Note that if we restrict to the Teichm\"uller component in the moduli space, the logarithms $Z_a=\log z_a$ of the coordinates are equal to some linear combinations of lengths of geodesics, connecting the punctures \cite{FG}. In Fock-Goncharov coordinates, the symplectic form on the moduli space takes a simple form. Upon quantization, the coordinates become operators $\hat{z}_a$ with the commutation relation
\beq
\hat z_a\,\hat z_b=q^{\langle a,b\rangle}\,\hat z_b\,\hat z_a\,,
\eeq
where  $\langle a,b\rangle$ is the signed count of faces that the edges $a$ and $b$ share, and $q=\exp(2\pi i/b^2)$, as usual. The operator algebra with generators $\hat z_a$ is the same as the algebra of Wilson loop operators, and was previously denoted by ${\mathbf A}_q$. Again, it does not contain the length operators $\hat{Z}_a$. However, we can define another algebra ${\mathbf A}_{\tilde q}$, where $\tilde q=\exp(2\pi i b^2)$, with generators $\hat z_a'=\exp(b^2\hat Z_a)$. One easily sees that it commutes with ${\mathbf A}_{ q}$. Moreover, the algebra generated by $\hat z_a$ and $\hat z_a'$ together is equivalent \cite{Faddeev} to the algebra generated by the  length operators $\hat{Z}_a$, after suitable completion. Thus, having the length operators in Teichm\"uller TQFT means that the algebra $\mathbf{A}_q$ of Wilson loops is effectively doubled. In section~\ref{tquant} we recalled how this doubling occurs from the point of view of the Hitchin sigma-model, following \cite{NW}. 

One may ask similar questions about observables on a general three-manifold $W$. Suppose that $W$ is hyperbolic. Then semiclassically, in Teichm\"uller TQFT we are expanding near the (conjugate) geometric flat connection, and it makes sense to compute expectation values of length operators. If we work in the 4d Yang-Mills picture, the natural observables are two sets of line operators, the usual supersymmetric Wilson lines, supported at the NS5-type boundary, as well as 't~Hooft operators, supported at the D5-type boundary. It would be interesting to see if, after localizing on the BPS equations, one can relate these two sets of line operators and length observables. 

\subsection{Modular functors}
There is one more point to be made regarding the construction of Teichm\"uller TQFT. Our path-integral definition via the 4d $\cN=4$ super Yang-Mills theory may be the most direct physical way of seeing that Teichm\"uller TQFT exists as a 3d quantum field theory. But there is another way that starts from two dimensions. A 3d TQFT with line operators defines what is called a $\mathcal{C}$-extended modular functor, where $\mathcal{C}$ is the category of line operators. An extended modular functor associates vector spaces with MCG actions to two-manifolds with punctures. Importantly, these data have prescribed behavior under cutting the two-manifolds along closed non-intersecting curves. If $C$ is a punctured Riemann surface and $\gamma$ is a simple closed curve on $C$, we can cut along $\gamma$ to produce a (possibly disconnected) Riemann surface $C'$ with two more punctures. Then there exists a canonical isomorphism of Hilbert spaces
\beq
\mathcal{H}_C\simeq\bigoplus_R \mathcal{H}_{C',R,R^\vee}\,,\label{fact}
\eeq
respecting the MCG action. Here the sum goes over the objects $R$ of $\mathcal{C}$, and $\mathcal{H}_{C',R,R^\vee}$ is the Hilbert space for ${C}'$ with the two punctures labeled by $R$ and its dual. This axiom is a manifestation of locality and allows to construct Hilbert spaces with MCG actions by gluing the Riemann surface from pairs of pants. (For details and precise definitions, see {e.g.} \cite{BK}.) It is fairly obvious that a TQFT produces an extended modular functor. What one may find more surprising is that having an extended modular functor is enough to reconstruct the 3d TQFT\footnote{An extended modular functor is equivalent to a modular tensor category \cite{MS1,BK}, which in turn produces a TQFT by Reshetikhin-Turaev construction. For details, other approaches and references, see \cite{BK}.} 
Quantizing the moduli spaces $\mathcal{M}_{2g-2}$ produces a modular functor which can be made into an extended modular functor, as we recall in a moment. (A rigorous construction of an analog of an extended modular functor for Teichm\"uller TQFT can be found in \cite{TeschnerMod}. Some references on the category $\mathcal{C}$ that appears in this context can be found in section~6.1 of \cite{TeschnerQRev}.) From this data, it should be possible to reconstruct a TQFT. This gives an alternative view on why Teichm\"uller TQFT exists and how to define it. An important caveat in this argumentation is that the machinery of modular functors and TQFTs, as described in \cite{BK}, applies to the situation with finite dimensional Hilbert spaces, while in our case the spaces of states are infinite dimensional. Extending these results to our situation is a non-trivial problem.  Let us also mention that instead of a TQFT, we should more precisely call our theory a restricted TQFT, since the partition function is finite only for some subclass of closed three-manifolds. (As we proposed in section~\ref{TeichCS}, what characterizes this nice subclass for Teichm\"uller TQFT may be a particular property of the Kapustin-Witten equations.)

Consider the Teichm\"uller space $\mathcal{T}_C$ for a (possibly punctured) Riemann surface $C$, and let as before $C'$ be a surface, obtained from $C$ by cutting along a simple curve $\gamma$. Let the PSL$(2,\RR)$ conjugacy classes at the two extra punctures on $C'$ be hyperbolic, corresponding to some fixed length $l$. The Teichm\"uller space for $C'$ with these punctures is isomorphic to the Hamiltonian reduction of $\mathcal{T}_C$ in which we quotient by the flow, generated by the length function $l_\gamma$, and impose the moment map constraint $l_\gamma=l$. (The isomorphism is obvious if one uses the Fenchel-Nielsen coordinates.) This statement is the classical precursor of the factorization axiom (\ref{fact}) and the reason, why the modular functor obtained by quantizing Teichm\"uller spaces can be promoted to an extended modular functor. Since $l$ is real, the index $R$ in (\ref{fact}) runs over the real numbers and is nothing but the Liouville momentum. Note that we had to allow punctures labeled by hyperbolic conjugacy classes. In Chern-Simons theory, one expects that they correspond to Wilson lines in representations, obtained by quantizing the hyperbolic coadjoint orbits. These representations are the ones belonging to the principal series.\footnote{For PSL$(2,\RR)$, there is one principal series of representations, parameterized by a real number. Had we worked with SL$(2,\RR)$ instead of PSL$(2,\RR)$, we would have to make a choice of a spin structure on the cutting circle $\gamma$. For the emerging puncture, this $\ZZ_2$ label would correspond to choosing between the two families of principal series representations that exist for SL$(2,\RR)$.} 
We note finally that a similar factorization argument for the full moduli space of PSL$(2,\RR)$ flat connections would show that the emerging punctures are labeled by arbitrary PSL$(2,\RR)$ conjugacy classes, not only the hyperbolic ones. Thus, in the full PSL$(2,\RR)$ Chern-Simons theory, the category $\mathcal{C}$ has more objects than in Teichm\"uller TQFT. One has to allow Wilson lines for representations not belonging to the principal series.\footnote{It would be interesting to find a relation between the grading (\ref{Hd}) and the classes of representations, used to label the legs of a conformal block.}

\section{Outlook}\label{out}
There are a lot of important questions that we were not able to answer in this paper. Let us list some of them, as well as some general open directions.

We worked with gauge group of rank one. A generalization to other gauge groups should be straightforward, and then one could check the perturbative expansions of section~\ref{pertexp} in this more general case. The global aspects like discrete theta-angle related to the topology of the gauge group were largely ignored in our paper. It would be interesting to keep track of them more carefully.

We formulated three conjectures about the Kapustin-Witten equations. In the absence of general proofs, it would be interesting at least to further test them. Since we understand well time-independent solutions of the Kapustin-Witten equations on $\RR\times C\times\RR^+$, one could try to use this knowledge to test our conjectures for hyperbolic three-manifolds that are mapping tori of pseudo-Anosov maps.

In section~\ref{dual}, we pointed out that the relation between supersymmetric blocks and Lefschetz thimble integrals is unclear. It would be interesting to understand it explicitly and generally by methods of $\cN=4$ super Yang-Mills. This question is a part of the more general problem of writing bounded actions with 1/2-BPS NS5-brane boundary condition. If we learn how to do that, we would also better understand the definition of the ``unusual'' complex Chern-Simons theory of section~\ref{unus}, as well as the integration cycles for the duals of Teichm\"uller TQFT.

Simpler versions of these questions can be asked purely in two dimensions. We have a well-defined setup for quantization of the Teichm\"uller space in the Hitchin sigma-model. Appropriate elements of the T-duality group should relate it to brane configurations that describe the Hilbert spaces of Chern-Simons theories CS-I and CS-II of section~\ref{tdualsec}. What are these configurations, and what do we learn from them about the Hilbert spaces of these Chern-Simons theories?

Can we apply our dualities to the holographic setup for the Liouville conformal blocks? Is it possible to see any traces of semiclassical complex Chern-Simons theories in the behavior of the conformal blocks near $c=1$ or $c=25$?

In section~\ref{redu}, our proofs and derivations are literally valid only for real values of the canonical parameter $\kappa$. It would be interesting to explicitly extend this story to complex $\kappa$. One would expect this generalization to be rather straightforward.

In this paper, we mostly focused on Chern-Simons theories at integer level $k=1$. The case of level $k=0$, related to supersymmetric indices of theories ${\rm T}[W]$, is also interesting and was explored in the literature \cite{indices}. It is possible that some of our questions may be answered more easily in that setting. Also, note that these indices have some nice structural properties, see \cite{DimofteRev2} and references therein. One may wonder if these properties can be observed from the counting problem for the Kapustin-Witten equations on $W\times\mathcal{I}$ with two Nahm poles.

\begin{figure}
	\begin{center}
		\includegraphics[width=6cm]{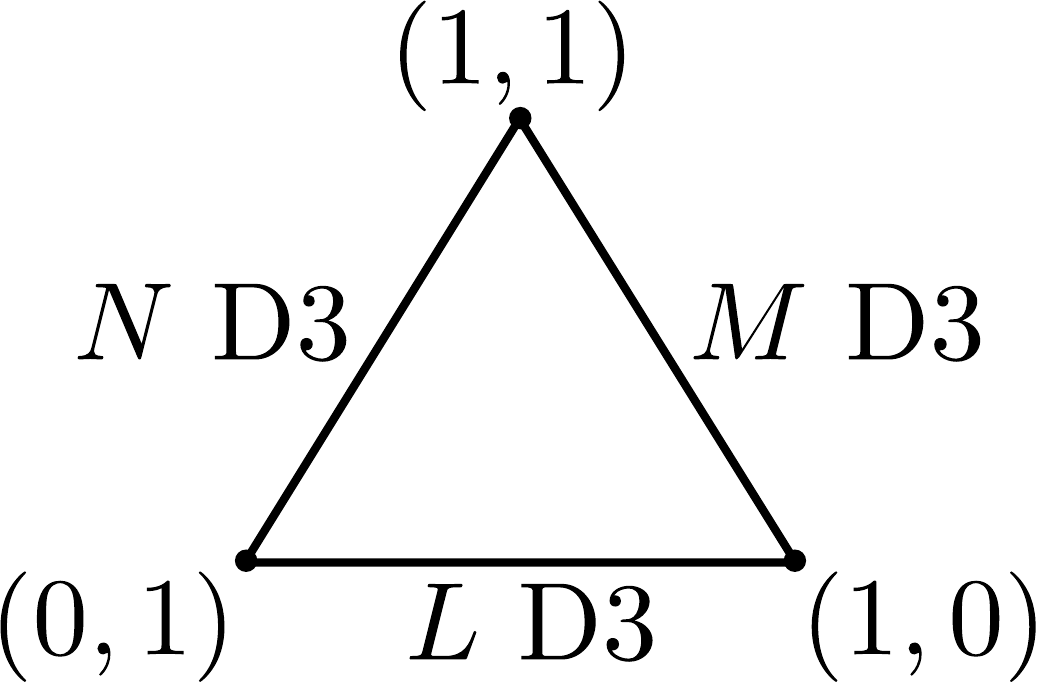}
	\end{center}
	\caption{\small A D5-, an NS5- and a $(1,1)$-brane joined by $N$, $M$, and $L$ D3-branes. This brane configuration produces an $\cN=4$ Yang-Mills theory with three half-BPS defects, which upon topological twisting is supposed to give a 3d TQFT. Teichm\"uller TQFT is the special case with $N=M=0$ and $L=2$.}
	\label{vertt}
\end{figure}
Recently, the elements of the SL$(2,\ZZ)$ S-duality group that we used in section~\ref{dual} were applied in \cite{Corners} to produce dualities of chiral algebras. The brane corners studied in that paper are a generalization of our brane corner of section~\ref{cocol}. They would arise if we wished to study the state $\langle{\bf q}|$ in a Chern-Simons theory engineered by a brane triangle of figure~\ref{vertt}. It would be interesting to better understand these Chern-Simons theories, if they are well-defined.

\appendix
\section{KW equations on a hyperbolic three-manifold}\label{hypap}
\subsection{Details on the symmetric ansatz}\label{apsyman}
Here we prove\footnote{The following proof was supplied by A.~DeBill.} that Conjecture~3 holds for the maximally symmetric ansatz (\ref{syman}). The flow equations are
\beqn
f'&=&-1-g^2+f^2\,,\nnr
g'&=&-2fg\,.
\eeqn
The model solution is $f(y)\equiv f_0(y)=-\coth y$ and $g(y)=0$. Other solutions with the right behavior for $y\rightarrow 0$ form a one-parametric family, labeled by the value of the integral of motion
\beq
I\equiv-4\pi\re(\CS(\cA))=\left(1+\fr{1}{3}g^2-f^2\right)g\,.
\eeq
(Here we dropped a constant $\CS(\omega)$.) The solution with $I=-c$ has an expansion near $y=0$
\beq
g=c y^2+\dots\,,\quad f(y)=f_0(y)-\fr{1}{7}c^2y^5+\dots\,,\label{modexp}
\eeq
thus, for small $y$, $f-f_0\le 0$. At the two critical points we have $I=0$. All trajectories that we are interested in lie to the left of the separatrix $f=-\sqrt{1+g^2/3}$. In particular, they all have $f<-1$. Also,
\beq
\left.\begin{array}{c}f'\le f^2-1\\ f'_0=f_0^2-1\end{array}\right\}\,\Rightarrow\, (f-f_0)'\le (f-f_0)(f+f_0)\,.
\eeq
Set $f-f_0=\exp\left(\int_1^y\d y'(f+f_0)\right)\tilde{f}(y)$. Then $\tilde{f}'\le 0$. Since $\tilde{f}\le 0$ for small $y$, it is also true for all $y>0$, and therefore $f\le f_0$.

Now,
\beqn
h'&=&-(f^2-1-g^2)^2-4f^2g^2=-1-(f^2+g^2)^2+2f^2-2g^2\le -(f^2-1)^2\,,\nnr
h_0'&=&-(f_0^2-1)^2\,.
\eeqn
Since $f\le f_0<-1$, we have 
\beq
h'\le h_0'\label{hineq}
\eeq
for all $y>0$. From the perturbative expansion it follows that for small $y$ it is true that $h\le h_0$, with equality for the model solution only. From (\ref{hineq}) it follows that the same is true for all $y>0$.

\subsection{Details on the perturbative expansion}\label{AppEx}
This computation largely follows \cite{Hen} and \cite{MazzeoWitten}. One defines a useful operator $L$: $\Omega^1_{\rm adj}\rightarrow\Omega^1_{\rm adj}$,
\beq
Lx=\star(e\wedge x+x\wedge e)\,.
\eeq
It acts in each of the subspaces $V_-$, $V_0$, $V_+$ diagonally with eigenvalues $2$, $1$ and $-1$, respectively. We also need a differential operator $\star\d_\omega$, which acts from the space $\Omega^1_{\rm adj}$ to itself. It has the property
\beq
\star \d_\omega:\quad  V_-\rightarrow V_0\,,\quad V_0\rightarrow V_-\oplus V_0\oplus V_+\,,\quad V_+\rightarrow V_0\oplus V_+\,.\label{stard}
\eeq
We will also need the fact that for any adjoint-valued one-forms $a$ and $b$,
\beq
\tr(e\wedge a\wedge b+e\wedge b\wedge a)=\fr{1}{2}\tr(a\wedge\star(Lb)+b\wedge\star(La))\,.\label{eab}
\eeq
The Kapustin-Witten equations in the gauge $A_y=0$ and with our ansatz (\ref{ansatz}) are
\beqn
\partial_y\varphi+\coth y\,L\varphi&=&-\star\d_\omega a+\d_\omega\phi_y+[a,\phi_y]+\star(\varphi\wedge\varphi-a\wedge a)\,,\nnr
\partial_y a-\coth y\,La-\coth y[e,\phi_y]&=&-\star\d_\omega\varphi+[\phi_y,\varphi]-\star(a\wedge\varphi+\varphi\wedge a)\,,\nnr
-\coth yPa+\partial_y\phi_y&=&\d_\omega^*\varphi-\star(a\wedge\star\varphi-\varphi\wedge \star a)\,,
\eeqn
where $Pa\equiv \star(a\wedge\star e-e\wedge\star a)=a_\mu^ae^\mu_b\eps_{abc}t^c$ is a projector onto $V_0$. Note that $[e,Pa_0]=2a_0$. 
The equations have a $\ZZ_2$ symmetry $y\rightarrow -y$, $a\rightarrow a$, $\varphi\rightarrow -\varphi$, $\phi_y\rightarrow\phi_y$. Only modes invariant under this symmetry survive \cite{MazzeoWitten}, thus, we can make an ansatz
\beq
\varphi=\sum_{k=1}^\infty \varphi_k y^{2k-1}\,,\quad a=\sum_{k=1}^\infty a_k y^{2k}\,,\quad \phi_y=\sum_{k=1}^\infty \phi_{y,k}y^{2k}\,.\label{alexp}
\eeq
(On a general three-manifold, the expansion will also contain logarithms. They are absent for an Einstein manifold.) The first two KW equations become
\beqn
(2k-1+L)\varphi_{k}&=&-L\,\sum_{\substack{p+q=k\\ p,q\ge 1}}b_p\varphi_q-\star\d_\omega a_{k-1}+\d_\omega\phi_{y,k-1}\nnr
&-&\sum_{\substack{p+q=k-1\\p,q\ge 1}}[\phi_{y,p},a_q]+\star\left(\sum_{\substack{p+q=k\\p,q\ge 1}}\varphi_p\wedge\varphi_q-\sum_{\substack{p+q=k-1\\p,q\ge 1}}a_p\wedge a_q\right)\,,\label{exec1}\\
(2k-L)a_k-[e,\phi_{y,k}]&=&L \sum_{\substack{p+q=k\\p,q\ge 1}}b_p a_q-\star\d_\omega\varphi_k+\sum_{\substack{p+q=k\\p,q\ge 1}}[e,b_p\phi_{y,q}]\nnr
&+&\sum_{\substack{p+q=k\\p,q\ge 1}}[\phi_{y,p},\varphi_q]-\star\left(\sum_{\substack{p+q=k\\p,q\ge 1}}(a_p\wedge\varphi_q+\varphi_q\wedge a_p)\right)\,,\label{exec2}
\eeqn
Here we defined the numbers $b_p$ as $\coth y=\sum_{p=0}^\infty b_p y^{2p-1}$. The third equation $\partial_y\phi_y-\d_A^\star\phi=0$ at order $y^{2k-1}$, $k\ge 1$, is\footnote{The order $y^{0}$ or $k=1/2$ of this equation is also important. It was used to exclude the possible order $y$ term in $a$.}
\beqn
-Pa_k+2k\phi_{y,k}&=&\sum_{\substack{p+q=k\\p,q\ge 1}}b_pPa_q+\d^\star_\omega\varphi_k\nnr
&-&\sum_{\substack{p+q=k\\p,q\ge 1}}\star(a_p\wedge\star\varphi_q-\varphi_q\wedge\star a_p)\,.\label{exec3}
\eeqn
To the first non-trivial order, the solution to the KW equations is
\beqn
\varphi_1&=& c_+          \,,\nnr
a_1&=&c_-+c_0-(\star\d_\omega c_+)_0-\fr{1}{3}(\star\d_\omega c_+)_+           \,,\nnr
\phi_{y,1}&=&\fr{1}{2}Pc_0          \,,
\eeqn
where $c_-$, $c_0$ and $c_+$ are unconstrained zero-modes. All higher order terms in the expansions (\ref{alexp}) are fixed in terms of these zero-modes by the equations.

 The functional that we are trying to bound can be expanded as
\beq
\fr{1}{2}(h(y)-h_0(y))=\sum_{k=1}^\infty y^{2k-1}\delta_k\,,
\eeq
where
\beqn
\delta_k&=&-\fr{1}{2}\sum_{\substack{p+q+r=k\\p\ge 0,\,q,r\ge 1}}b_p\,(a_q,La_r)+\fr{1}{2}\sum_{\substack{p+q+r=k+1\\p\ge 0,\,q,r\ge 1}}b_p\,(\varphi_q,L\varphi_r)-\sum_{\substack{p+q=k+1\\p\ge 0,\,q\ge 1}}d_p\,(e,\varphi_q)\nnr
&+&\sum_{\substack{p+q=k\\p,q\ge 1}}(\varphi_p,\star\d_\omega a_q)+\fr{1}{3}\sum_{\substack{p+q+r=k+1\\p,q,r\ge 1}}\int\tr\left(\varphi_p\wedge\varphi_q\wedge\varphi_r\right)-\sum_{\substack{p+q+r=k\\p,q,r\ge 1}}\int\tr(\varphi_p\wedge a_q\wedge a_r)\,.\nonumber
\eeqn
Here we defined $d_p$ from $y^2\sinh^{-2}y=\sum_{p=0}^\infty d_p y^{2p}$. Note that in principle the expansion of $h(y)-h_0(y)$ could start with $y^{-1}$, but the corresponding term is proportional to $(e,\varphi_1)$, which is zero. To find $\delta_1$ and $\delta_2$, we need $\varphi_1$, $a_1$, $\varphi_2$ and $(\varphi_3)_-$. The relevant equations are
\beqn
(3+L)\varphi_2&=&\fr{1}{3}c_+-\star\d_\omega a_1+\d_\omega\phi_{y,1}+\star(c_+\wedge c_+)\,,\nnr
(5+L)\varphi_3&=&-\fr{1}{45}c_+-\fr{1}{3}L\varphi_2-\star\d_\omega a_{2}+\d_\omega\phi_{y,2}-[\phi_{y,1},a_1]+\star\left(c_+\wedge\varphi_2+\varphi_2\wedge c_+-a_1\wedge a_1\right)\,.\nonumber
\eeqn

The leading non-trivial order $\delta_1$ gets contributions
\beqn
\fr{1}{2}b_0(\varphi_1,L\varphi_1)&=&-\fr{1}{2}|c_+|^2\,,\nnr
-d_0(e,\varphi_2)&=&-\fr{1}{5}(e,\star(c_+\wedge c_+))=\fr{1}{10}|c_+|^2\,,
\eeqn
and thus we obtain (\ref{delta1}). The next order $\delta_2$ gets the following contributions,
\beqn
-\fr{1}{2}b_0(a_1,La_1)&=&-|c_-|^2-\fr{1}{2}|c_0|^2-\fr{1}{2}|(\star\d_\omega c_+)_0|^2+(c_0,\star\d_\omega c_+)+\fr{1}{18}|(\star\d_\omega c_+)_+|^2       \,,\nnr
\fr{1}{2}b_0((\varphi_1,L\varphi_2)+(\varphi_2,L\varphi_1))&=&-(c_+,\varphi_2)\nnr
&=&-\fr{1}{6}|c_+|^2-\fr{1}{2}|(\star\d_\omega c_+)_0|^2-\fr{1}{6}|(\star\d_\omega c_+)_+|^2+\fr{1}{2}\int\tr(c_+^3)   \,,\nnr
\fr{1}{2}b_1(\varphi_1,L\varphi_1)&=&-\fr{1}{6}|c_+|^2      \,,\nnr
-d_1(e,\varphi_2)&=&-\fr{1}{30}|c_+|^2        \,,\nnr
-d_0(e,\varphi_3)&=&\fr{2}{21}(e,\varphi_2)+\fr{1}{7}(c_+,\varphi_2)+\fr{1}{14}(a_1,La_1)\nnr
&=&\fr{1}{70}|c_+|^2+\fr{1}{7}|c_-|^2+\fr{3}{14}|c_0|^2\nnr
&+&\fr{1}{7}|(\star\d_\omega c_+)_0|^2+\fr{1}{63}|(\star\d_\omega c_+)_+|^2-\fr{2}{7}(c_0,\star\d_\omega c_+)-\fr{1}{14}\int\tr(c_+^3)      \,,\nnr
(\varphi_1,\star\d_\omega a_1)&=&-|(\star\d_\omega c_+)_0|^2-\fr{1}{3}|(\star\d_\omega c_+)_+|^2+(c_0,\star\d_\omega c_+)     \,,\nnr
\fr{1}{3}\int\tr(\varphi_1^3)&=&\fr{1}{3}\int\tr(c_+^3)\,.
\eeqn
Altogether, we obtain,
\beqn
\delta_2&=&-\fr{6}{7}|c_-|^2-\fr{37}{105}|c_+|^2-\fr{2}{7}|c_0|^2\nnr
&-&\fr{3}{7}|(\star\d_\omega c_+)_+|^2-\fr{13}{7}|(\star\d_\omega c_+)_0|^2+\fr{12}{7}(c_0,(\star\d_\omega c_+)_0)\nnr
&+&\fr{16}{21}\int\tr (c_+^3)\,.
\eeqn
If one sets $c_0=0$, which is equivalent to $\phi_y=0$, one gets (\ref{delta2}). When $c_0\ne 0$, we expect that the conjecture is not true. Let us set $c_-=0$ and $c_0=3(\star\d_\omega c_+)_0$. Then $\delta_1=0$ and
\beqn
\delta_2&=&\fr{5}{7}|(\star\d_\omega c_+)_0|^2-\fr{3}{7}|(\star\d_\omega c_+)_+|^2\nnr
&-&\fr{37}{105}|c_+|^2+\fr{16}{21}\int\tr(c_+^3)\,.
\eeqn
We can take $c_+$ to be small, but quickly varying. Then the second line can be neglected. It looks plausible that $c_+$ can be chosen to make the first line positive, although we do not know, how to prove this.

For completeness, let us evaluate the integral of motion $I=-4\pi\re\CS(\cA)$. We find
\beq
I-I_0=-2(e,c_-)\,,
\eeq
where $I_0=-4\pi\CS(\omega)$ is the integral on the model solution. Since $I$ must be $y$-independent, there are no higher order corrections. For the order $y^2$ it is easy to check this explicitly.

\section{Transversely holomorphic foliations}\label{THFs}
First we give a very brief review of transversely holomorphic foliations and their deformations, following the papers \cite{Komar3d1,Komar3d2}, where the reader can find all the details. 

In three dimensions, the analog of a complex structure operator is a tensor $\Phi$, which is a section of $TL\otimes T^*L$ satisfying
\beq
\Phi^2=-1+\xi\otimes\eta\,,
\eeq
where $\xi$ is a nowhere vanishing vector field and $\eta$ is a one-form with the property $\eta(\xi)=1$. This data defines a THF, if it satisfies an extra integrability condition, an analog of vanishing of the Nijenhuis tensor. A metric $g$ is called compatible, if 
\beq
g(\Phi X,\Phi Y)=g(X,Y)-\eta(X)\eta(Y)\,,
\eeq
which is a natural analog of the Hermiticity condition in even dimensions. Given a compatible metric and a nowhere vanishing vector field $\xi$, the rest of the data can be restored,
\beqn
\eta_\mu&=&g_{\mu\nu}\xi^\nu\,,\nnr
\Phi^\mu_\nu&=&\sqrt{g}\,g^{\mu\sigma}\eps_{\sigma\nu\rho} \xi^\rho\,,
\eeqn
up to an ambiguity $\Phi\rightarrow -\Phi$. (As a consequence of the first equation, we will often not distinguish $\eta$ and $\xi$.)

The integrability condition ensures that there exist so-called adapted coordinates $t$, $z$ and $\bar z$, in which $\xi=\partial_t$, and $\d z$ is a holomorphic one-form, in the sense that $\Phi^T\d z = i\d z$. One has $\eta=\d t+h\d z+\bar h\d\bar z$, and for the compatible metric 
\beq
\d s^2=\eta^2+g\d z\d\bar z\,,
\eeq
where $h$, $\bar h$ and $g$ in general are functions of all the coordinates.

 Complex differential one-forms can be split according to the eigenvalues of $\Phi$,
\beq
\omega=\omega_+\d z+\omega_-\d \bar z+\omega_0\eta\,,
\eeq
with $\Phi^T\d z=i\d z$, $\Phi^T\d \bar z=-i\d \bar z$ and $\Phi^T\eta=0$. We write this decomposition as $T_\CC^*L\simeq T^*_+L\oplus T^*_-L\oplus T^*_0L$. It is convenient to treat $\d\bar z$ and $\eta$ together, and correspondingly to denote by $\Omega^{p,q}(L)$ the space of sections of $\oplus_{m+n=q}(\wedge^p T^*_+L)(\wedge^m T^*_-L)(\wedge^n T^*_0L)$.

For  vectors, the analogous decomposition $T_\CC L\simeq T^+L\oplus T^-L\oplus T^0L$ is
\beq
v=v^+(\partial_z-h\partial_t)+v^-(\partial_{\bar z}-\bar h\partial_t)+v^0\xi\,,
\eeq
since $\Phi(\partial_z-h\partial_t)=i(\partial_z-h\partial_t)$, $\Phi(\partial_{\bar z}-\bar h\partial_t)=-i(\partial_{\bar z}-\bar h\partial_t)$ and $\Phi\xi=0$. Vectors in $T^+L$, just like differential forms in $T^*_+L$, transform with a holomorphic factor under allowed changes of the adapted coordinates.
 
For a three-manifold with a THF, there exists a Dolbeault-like operator $\tilde\partial$. On $\Omega^{p,q}(L)$ it is defined by projecting the image of the de Rahm operator $\d$ onto $\Omega^{p,q+1}(L)$. The Dolbeault operator squares to zero, and the corresponding cohomology can be decomposed with respect to the $(p,q)$-grading. On functions, $\tilde \partial f=0$ if and only if $f$ is a holomorphic function of the adapted coordinate $z$, independent of $t$. This allows to define cohomology on differential forms, valued in a holomorphic line bundle, for example, in $T^+L$.

Suppose one makes a deformation of the THF and the compatible metric. Out of variations $\delta\xi$ and $\delta\Phi$ one can construct a $T^+L$-valued $(0,1)$-form 
\beq
\Theta
=(\partial_z-h\partial_t)\otimes(\Theta_0^+\eta+\Theta_-^+\d\bar z)\label{Theta}
\eeq
with
\beqn
\Theta^+_0&=&-2i\delta\xi^+\,,\nnr
\Theta^+_-&=&\delta\Phi^+_-\,.
\eeqn
Here various components are defined in the basis of the undeformed THF. In \cite{Komar3d2} the following has been shown: for an integrable deformation, $\Theta$ has to be $\tilde\partial$-closed, and all the components of the variations $\delta\xi$ and $\delta\Phi$ can be found from the ones that appear in (\ref{Theta});  $\tilde\partial$-exact tensors $\Theta$ correspond to trivial deformations, induced by diffeomorphisms. It has been also shown that the supersymmetric partition function of the 3d $\cN=2$ theory on the three-manifold $L$ depends on the geometry only through the THF, that is, through the $\tilde\partial$-cohomology class of $\Theta$. 

\subsection{THF for a torus fibration}
Let us apply this knowledge to our case (\ref{met}), (\ref{Killing}). We have
\beq
\xi=s^{-1}K\,,
\eeq
where
\beqn
K&=&-\kappa\partial_{\varphi_1}+\partial_{\varphi_2}\,,\nnr
s^2\equiv|K|^2&=&\fr{a}{\im\tau}(\kappa^2-2\kappa\,\re\tau+|\tau|^2)\,.
\eeqn
One also finds
\beqn
\eta&=&\fr{a}{s\im\tau}((-\kappa+\re\tau)\d\varphi_1+(-\kappa\,\re\tau+|\tau|^2)\d\varphi_2)\,,\label{eta}\\
\Phi&=&\fr{1}{s\im\tau}((\kappa\,\re\tau-|\tau|^2)\partial_{\varphi_1}+(-\kappa+\re\tau)\partial_{\varphi_2})\otimes\d y+\fr{a}{s}\partial_y\otimes(\d\varphi_1+\kappa\d\varphi_2)\,.
\eeqn

We can choose the holomorphic adapted coordinate to be 
\beq
z=\varphi_1+\kappa\varphi_2+i\int^y \fr{s(y')}{a(y')}\d y'
\eeq
The basis holomorphic one-form and the basis holomorphic vector field are
\beqn
\d z&=&\d\varphi_1+\kappa\d \varphi_2+\fr{i s}{a}\d y\,,\label{dz}\\
\partial_z-h\partial_t&=&\fr{a}{2s^2\,\im\tau}\left((-\kappa \re\tau+|\tau|^2)\partial_{\varphi_1}+(\kappa-\re\tau)\partial_{\varphi_2}\right)-\fr{ia}{2s}\partial_y\label{vz}\,.
\eeqn
With suitable periodic identifications, the adapted coordinates are good everywhere on $L$, away from the ends of the interval $\mathcal{I}$. Various line bundles in the decomposition of $T_\CC L$ and $T_\CC^* L$ have trivial gluing functions.

Using (\ref{eta}), (\ref{dz}) and (\ref{vz}), one finds the deformation tensor $\Theta$ in terms of the variations of those components of $\xi$ and $\Phi$ which can be non-zero for our family of THFs,
\beqn
\Theta^+_0&=&-2i(\delta\xi^1+\kappa\delta\xi^2)\,,\nnr
\Theta^+_-&=&\fr{ia}{2s}(\delta\Phi^1_3+\kappa\delta\Phi^2_3)+\fr{i}{2s\im\tau}((-\kappa\re\tau+|\tau|^2)\delta\Phi^3_1+(\kappa-\re\tau)\delta\Phi^3_2)\,,
\eeqn
where numerical indices on the components are in the coordinates $x^1=\varphi_1$, $x^2=\varphi_2$, $x^3=y$.

For a variation of $\tau(y)$, $a(y)$ and $\kappa$, the tensor $\Theta$ is automatically $\tilde\partial$-closed, since the deformation of the THF is integrable. Then locally $\Theta$ is $\tilde\partial$-exact, as follows from the analog of the Poincar\'e lemma. We look for a locally-defined vector $v=v^+(\partial_z-h\partial_t)$ in $T^+L$, such that $\Theta=\tilde\partial v$. (Any two such vectors differ by a shift of $v^+$ by a holomorphic function.) The deformation is trivial, if and only if such a $v$ also exists globally. 

First, suppose that $\tau(y)$ and $a(y)$ are varied, while $\kappa$ is kept fixed. Then we find
\beq
\Theta^+_0=0\,,\quad \Theta^+_-=-\fr{ia}{s}\delta\left(\fr{s}{a}\right)\,.
\eeq
(Here $\delta$ means variation.) The equations $(\tilde\partial v^+)_0=\Theta^+_0$ and $(\tilde\partial v^+)_-=\Theta^+_-$ become
\beqn
s^{-1}(-\kappa\partial_{\varphi_1}+\partial_{\varphi_2})v^+&=&0\,,\nnr
\fr{a}{2s^2\im\tau}((-\kappa\re\tau+|\tau|^2)\partial_{\varphi_1}+(\kappa-\re\tau)\partial_{\varphi_2})v^++\fr{ia}{2s}\partial_yv^+&=&-\fr{ia}{s}\delta\left(\fr{s}{a}\right)\,.
\eeqn
A solution is
\beq
v^+=-2\int^y\delta\left(\fr{s(y')}{a(y')}\right)\d y'\,,\label{sol1}
\eeq
which is good at least away from the ends of the interval $\mathcal{I}$. Let us show that it is well-defined everywhere. We look at the end of the interval at $y\rightarrow 0$, and with no loss of generality assume that the shrinking circle is $\varphi_1$. (In this case $\kappa$ is allowed to be any finite number.) For the metric to be smooth, we need
\beq
\im\tau\simeq R/y+\dots\,,\quad \re\tau\simeq{\rm const}+\dots\,,\quad a\simeq Ry+\dots\,,\label{asympt}
\eeq
where dots stay for terms of lower order in $y$, and $R$ is the radius of the circle $\varphi_2$ at $y=0$. First, assume that the variation $\delta(s/a)$ vanishes at $y=0$ (or at least is integrable), so that $v^z$ at $y=0$ is a finite number. Then the vector $v$ is well-defined globally, if the basis vector $(\partial_z-h\partial_t)$ is well-defined at $y=0$. From (\ref{vz}) and (\ref{asympt}), for $y\rightarrow 0$
\beq
\partial_z-h\partial_t\simeq \fr{1}{2}(\partial_{\varphi_1}-iy\partial_y)\,,
\eeq
which at $y=0$ is indeed well-defined, going to zero in orthonormal coordinates. The assumption about the behavior of $\delta(s/a)$ near $y=0$ is always true. Indeed, one can verify that 
\beq
\fr{s}{a}\simeq \fr{1}{y}+\dots\,,
\eeq
where the omitted terms are non-singular at $y=0$, and in fact go to zero. Then the variation of $s/a$ is vanishing at $y=0$, and is certainly integrable. We have proved that the THF is independent of the functions $\tau(y)$ and $a(y)$.

Now assume that $\tau(y)$ and $a(y)$ are fixed, and $\kappa$ is varied. Then we find
\beq
\Theta^+_0=2is^{-1}\delta\kappa\,,\quad \Theta^+_-=0\,,
\eeq
and the equations for $v^+$ become
\beqn
(-\kappa\partial_{\varphi_1}+\partial_{\varphi_2})v^+&=&2i\delta\kappa\,,\nnr
((-\kappa\,\re\tau+|\tau|^2)\partial_{\varphi_1}+(\kappa-\re\tau)\partial_{\varphi_2})v^++{is\,\im\tau}\,\partial_yv^+&=&0\,.
\eeqn
The first equation gives
\beq
v^+=2i\delta\kappa(\varphi_2+f(\varphi_1+\kappa\varphi_2,y))\,,
\eeq
where $f$ should satisfy
\beq
f(x+2\pi,y)=f(x,y)\,,\quad f(x+2\pi\kappa,y)=f(x,y)-2\pi\,,
\eeq
which is impossible for a continuous function.
Thus, $v$ doesn't exist globally, and $\kappa$ parameterizes non-trivial deformations of the THF.

\subsection{An example}\label{apex}
Here we find the Killing vector parameter $\kappa$ for the squashed three-sphere background (\ref{imamura}). The background supergravity fields include the $B$-field, an $R$-symmetry gauge field $A^{(R)}$, and another gauge field $C$ that couples to the central charge. For the squashed sphere that we are considering, explicit values of these fields can be found in section~5.2 of \cite{Komar3d1}. For our purposes it is sufficient to know that in this background $V^\mu$, which is the dual of the field strength for $C$, is a Killing vector, proportional to $\partial_{\varphi_1}$.

Given the supergravity fields, we would like to know which THF they correspond to. In general, the formula for $V^\mu$ is \cite{Komar3d1}
\beq
V^\mu=\fr{1}{\sqrt{g}}\eps^{\mu\nu\rho}\partial_\nu\xi_\rho+\lambda\xi^\mu\,,\label{Vofxi}
\eeq
where $\lambda$ is some function that satisfies $K^\mu\partial_\mu\lambda=0$. 
As before, $K=-\kappa\partial_{\varphi_1}+\partial_{\varphi_2}$ and we use notation $s\equiv|K|$. The equation for $\lambda$ implies that it is a function of $\theta$ only. We assume that $\kappa$ is real, in which case $\xi=K/s$. 
We substitute this $\xi$ into (\ref{Vofxi}). The condition that $V\propto\partial_{\varphi_1}$ fixes $\lambda$, and we get
\beq
V=-\fr{\partial_{\theta}s}{h\sin\theta}\partial_{\varphi_1}\,.
\eeq
For $V$ to be a Killing vector, the coefficient in front of $\partial_{\varphi_1}$ must be constant. After a small computation, this translates into an equation for $\kappa$ in terms of the parameter $h$ of the metric,
\beq
\kappa=-\fr{1}{2}\pm\sqrt{\fr{1}{4}-h^{-2}}\,.
\eeq
One usually takes parameterization $h=b+b^{-1}$. Then either $\kappa=-\fr{1}{1+b^2}$ or $\kappa=-\fr{1}{1+b^{-2}}$.

\section{Details on supersymmetries}\label{5branesusy}
\subsection{Half-BPS supersymmetry subalgebras}
In ten-dimensional notations, the supersymmetry transformations are
\beq
\delta A_I=i\eps^T B\Gamma_I\Psi\,,\quad \delta\Psi=-\fr{i}{2}F_{IJ}\Gamma_{IJ}\eps\,.
\eeq
Here $B=\Gamma_0\dots \Gamma_8$ is a symmetric matrix with the property $\Gamma_I^T B = B \Gamma_I$. (We use the basis where all gamma-matrices are hermitian, the even ones are real and symmetric, the odd ones are imaginary and antisymmetric.) The 16-component spinor $\Psi$ (and similarly $\eps$) is of positive chirality, $i\Gamma_{0..9}\Psi =\Psi$.

The boundary condition or the Janus configuration breaks the bosonic symmetry to SO$(3)^3\equiv {\rm SO}(3)_W\times{\rm SO}(3)_X\times{\rm SO}(3)_Y$, or further to a subgroup. One introduces operators 
\beq
B_0=i\Gamma_{456789}\,,\quad B_1=\Gamma_{3456}\,,\quad B_2=\Gamma_{3789}\,,
\eeq
which commute with SO$(3)^3$ and satisfy the algebra of sigma matrices. When acting on positive chirality 10d spinors, we have $B_0=\Gamma_{0123}$.

Under SO$(3)^3$, the spinors transform as $({\bf 2},{\bf 2},{\bf 2})\times V_2$, where $V_2$ is a two-dimensional multiplicity space, acted on by the matrices $B_{0,1,2}$. It is convenient to choose a basis of vectors in $V_2$, and decompose all spinors accordingly. One natural choice for the basis would be the eigenvectors of the 4d chirality operator $B_0$. But instead, following \cite{GWbc}, we choose the basis to be the eigenvectors of $B_2$. The advantage of this choice is that $B_2$ commutes with the group SO$(6)$, acting in directions $012456$, and we can work covariantly with respect to SO$(6)\times{\rm SO}(3)_Y$. Correspondingly we decompose
\beq
\Psi=\Psi_++\Psi_-\,,\quad B_2\Psi_\pm=\pm\Psi_\pm\,,
\eeq
and similarly for the supersymmetry generator $\eps$. Note that on chiral 10d spinors, $B_2=-i\Gamma_{012456}$, and therefore $\Psi_+$ and $\Psi_-$ transform in ${\bf 4}\times {\bf 2}$ and $\bf{\bar 4}\times {\bf 2}$ of SO$(6)\times{\rm SO}(3)_Y$, with appropriate definition of ${\bf 4}$ and $ {\bf \bar4}$. If we restrict to SO$(3)^3$ transformations, we can write $\Psi_+$ as $\Psi_+^{\alpha A\dA}v_+$, and similarly $\Psi_-$ as $\Psi_-^{\alpha A\dA}v_-$, where $v_\pm$ are the $\pm 1$ eigenvectors of $B_2$, which we choose to be related by $v_-=B_1 v_+$.

The preserved supersymmetry parameters should satisfy
\beq
(\tilde \eps,\eps)\equiv \tilde{\eps}^T B\Gamma_3 \eps=0\label{bg3}
\eeq
for any $\tilde\eps$ and $\eps$, in order for the supersymmetry anticommutators not to generate translations in the third direction. As found in \cite{GWbc}, a middle-dimensional subspace of spinors that satisfy this condition can be parameterized as
\beq
\eps_-=\slq\eps_+\,,\label{subsp}
\eeq
where 
\beq
\slq\equiv\fr{1}{3!}q_{pqr}\Gamma_{3pqr}\,,
\eeq
$p,q,r$ run over $012456$, and $q_{pqr}$ is an arbitrary self-dual\footnote{By self-dual here we mean $i\star_6 q =q$, where $\star_6$ is the Hodge star in six Euclidean dimensions.} three-form. (It's anti-self-dual part would cancel out, anyway.) Subspaces (\ref{subsp}) give a dense set of solutions to (\ref{bg3}), missing some points ``at infinity'', which can be covered by a similar condition, but expressing $\eps_+$ via $\eps_-$. The preserved supersymmetry transformations act as
\beqn
\delta Y_\da&=&-(\eps_+,\sigma_\da \left(\Psi_-+\slq\Psi_+\right))    \,,\nnr
\delta A_p&=&i(\eps_+,\Gamma_{3p}\Psi_+-\slq\Gamma_{3p}\Psi_-)          \,,\nnr
\delta \Psi_+&=&\left(-D_3Y_\da\sigma_\da+\fr{1}{2}\eps_{\da\db\dot c}[Y_\da,Y_\db]\sigma_{\dot c}-\fr{i}{2}F_{pq}\Gamma_{pq}-(D_pY_\da\sigma_\da+iF_{3p})\Gamma_{3p}\slq\right)\eps_+   \,,\nnr
\delta \Psi_-&=& \left((D_pY_\da\sigma_\da-iF_{3p})\Gamma_{3p}+(D_3Y_\da\sigma_\da+\fr{1}{2}\eps_{\da\db\dot c}[Y_\da,Y_\db]\sigma_{\dot c}-\fr{i}{2}F_{pq}\Gamma_{pq})\slq\right)\eps_+       \,.\label{decompsusy}
\eeqn

If we work explicitly in SO$(3)^3$ indices, we can write the symmetric bilinear form in (\ref{bg3}), up to a factor, as
\beq
\eps_{\alpha\beta}\eps_{AB}\eps_{\dA\dB}(\tilde\eps_+^{\alpha A\dA}\eps_-^{\beta B\dB}-\tilde\eps_-^{\alpha A\dA}\eps_+^{\beta B\dB})\,.
\eeq
An SO$(3)_Y$-invariant subspace on which this form vanishes can be characterized by
\beq
\eps_-^{\alpha A\dA}=\eps^{\alpha\alpha'}\eps^{AA'}m_{\alpha' A'\beta B }\eps_+^{\beta B\dA}\,,\label{m1}
\eeq
where $m_{\alpha A\beta B}$ is any matrix with the symmetry $m_{\alpha A\beta B}=m_{\beta B\alpha A}$. It can be written as
\beq
m_{\alpha A\beta B}=m_0\eps_{\alpha\beta}\eps_{AB} +m_{ia} \sigma^i_{\alpha\beta}\sigma^a_{AB}\,,\label{m2}
\eeq
with arbitrary $m_0$ and $m_{ia}$. It is straightforward to relate (\ref{m1}), (\ref{m2}) to (\ref{subsp}). Explicitly,
\beqn
q&=&-\fr{1}{2}m_0(i\d x^0\wedge \d x^1\wedge \d x^2-\d x^4\wedge\d x^5\wedge\d x^6)\nnr
&-&\fr{i}{2}m_{ia}\left(\fr{1}{2}\eps_{abc}\d x^i\wedge\d x^b\wedge\d x^c +\fr{i}{2} \eps_{ijk}\d x^j\wedge\d x^k\wedge\d x^a\right)\,.\label{qm}
\eeqn
In deriving this, it is helpful to know that
\beqn
&&\Gamma_{ij}=i\eps_{ijk}\sigma^k\,,\,i\ne j\,,\quad \Gamma_{ab}=i\eps_{abc}\sigma^c\,,\, a\ne b\,,\nnr
&&\Gamma_{3i}=i B_0\sigma_i\,,\quad \Gamma_{3a}=-i B_1\sigma_a\,,\quad \Gamma_{3\da}=-i B_2\sigma_\da\,,
\eeqn
when acting on spinors of positive 10d chirality.

In the special case that SO$(3)_W'$ invariance is preserved, we have\footnote{The Kronecker symbol here identifies $0,1,2$ with $4,5,6$.} $m_{ia}=m_1\delta_{ia}$ for some constant $m_1$. Then the preserved supersymmetries are
\beqn
\eps_{\rm scalar}^{\alpha A\dA}&=&((m_0-3m_1)v_-+v_+)\eps^{\alpha A}u^\dA\,,\nnr
\eps_{\rm vec\,i}^{\alpha A\dA}&=&((m_0+m_1)v_-+v_+)\sigma_i^{\alpha A}u^\dA\,.
\eeqn
From the identity (\ref{ssy2}) we find
\beq
m_0-3m_1=-i\fr{t-i}{t+i}\,,\quad m_0+m_1=-i\fr{t'-i}{t'+i}\,.
\eeq
In particular, the parameterization (\ref{subsp}), (\ref{m1}) breaks down on the subspaces $t=-i$ and $t'=-i$.

\subsection{Fivebrane supersymmetries}
A $(p,q)$-fivebrane along $012456$ in the type IIB string theory preserves the supersymmetries\footnote{As always in this paper, we work in Euclidean signature.}
\beq
-\eps_1\sin\fr{\ang_{p,q}}{2}+\eps_2\cos\fr{\ang_{p,q}}{2}=ih\Gamma
_{012456}\left(\eps_1\cos\fr{\ang_{p,q}}{2}+\eps_2\sin\fr{\ang_{p,q}}{2}\right)\,,\label{pq5susy}
\eeq
where we defined
\beqn
\ang_{p,q}&=&\arg(p\tau+q)\,,\nnr
h&=&\exp\left(\fr{1}{2}b_{pq}\Gamma^{pq}\right)\,,
\eeqn
the two-form $b_{pq}$ being related to the fivebrane gauge field background (and the type IIB two-form backgrounds).

The D3-brane located along directions 0123 preserves supersymmetries with $\eps_2=i\Gamma_{0123}\eps_1$. Using this, the condition (\ref{pq5susy}) can be written as
\beq
(B_1h-1)\left(\cos\fr{\ang_{p,q}}{2}+iB_0\sin\fr{\ang_{p,q}}{2}\right)\eps=0\,,\label{pq5susy2}
\eeq
where $\eps\equiv \eps_1$. This is equivalent to the condition (\ref{subsp}) with
\beq
\slq=\left(\cos\fr{\ang_{p,q}}{2}+h^{-1}\sin\fr{\ang_{p,q}}{2}\right)^{-1}B_1\left(h\cos\fr{\ang_{p,q}}{2}-\sin\fr{\ang_{p,q}}{2}\right)\,,
\eeq
provided that $B_1 h B_1=h^{-1}$, which is a necessary condition for preserving one-half of the supersymmetries.

If we wish to preserve the SO$(3)_W'$ symmetry, we have to take
\beq
h=\exp\left(\beta(\Gamma_{04}+\Gamma_{15}+\Gamma_{26})\right)\,,\label{symh}
\eeq
with some angle $\beta$. 

When acting on a spinor of positive ten-dimensional chirality, the following operators are equal,
\beq
\Gamma_{04}+\Gamma_{15}+\Gamma_{26}\simeq iB_2(\Gamma_{1256}+\Gamma_{0246}+\Gamma_{0145})\,.
\eeq
Using that $\Gamma_{1256}=\fr{1}{2}(\Gamma_{12}+\Gamma_{56})^2+1$,
together with similar equations for $\Gamma_{0246}$ and $\Gamma_{0145}$, we find
\beq
\Gamma_{04}+\Gamma_{15}+\Gamma_{26}\simeq i\Gamma_{3789}(3-2\hat J^2)\,,
\eeq
where $\hat J^2$ is the Casimir of the twisted Lorentz group SO$(3)_W'$, normalized to be equal to $j(j+1)$ in a spin $j$ representation. The supersymmetry condition (\ref{pq5susy2}) can then be rearranged into
\beq
\left(B_1(\cos\ang+i\sin\tilde\beta\sin\ang)+B_2(\sin\ang-i\sin\tilde\beta\cos\ang)-\cos\tilde\beta\right)\eps=0\,,
\eeq
where $\tilde\beta=\beta(3-2\hat J^2)$. Comparing to (\ref{ssy2}), we find that the physical fivebrane boundary condition preserves supersymmetries with
\beqn
t&=&-\ex^{-i\ang_{p,q}}\tan\left(\fr{\pi}{4}-\fr{3\beta}{2}\right)\,,\nnr
t'&=&-\ex^{-i\ang_{p,q}}\tan\left(\fr{\pi}{4}+\fr{\beta}{2}\right)\,,\label{ttprimeap}
\eeqn
as claimed in the main text.

\subsection{Simple half-BPS boundary conditions}
The boundary condition necessarily requires $\delta Y^{\dot a}=0$. Taking the variation of this condition by the preserved supersymmetries gives the boundary condition for the fermions,
\beq
\Psi_-+\slq\Psi_+=0\,.\label{fermbc}
\eeq
Taking a supersymmetry variation of this equation should give the Neumann boundary conditions for the bosonic fields. Explicitly,
\beq
\left(F_{3p}(\Gamma_{3p}+\slq\Gamma_{3p}\slq)+\fr{1}{2}F_{pq}(\Gamma_{pq}\slq+\slq\Gamma_{pq})\right)\eps_+=0\,.\label{bpsbc1}
\eeq
\subsubsection{D5}
For the D5-brane, we have $\slq=B_1h$. Then (\ref{bpsbc1}) becomes
\beq
\left(F_{3p}(h^{1/2}\Gamma_{3p}h^{-1/2}+B_1h^{1/2}\Gamma_{3p}h^{-1/2}B_1)+\fr{1}{2}F_{pq}(h^{1/2}\Gamma_{pq}h^{-1/2}B_1+B_1h^{1/2}\Gamma_{pq}h^{-1/2})\right)h^{1/2}\eps_+=0\,,\nonumber
\eeq
which means that turning on $h$ rotates the Nahm pole,
\beq
\Gamma_pA_p=\fr{t_a}{y}\,h^{-1/2}\Gamma_a h^{1/2}\,,
\eeq
as already explained in \cite{GWbc}. For $h$ given in (\ref{symh}), this gives the tilted Nahm pole
\beq
A_i=-\fr{t_i}{y}\sin\beta+\dots\,,\quad \phi_a=\fr{t_a}{y}\cos\beta+\dots\,.
\eeq

\subsubsection{NS5}
Equation (\ref{bpsbc1}) can be transformed using the following identities
\beq
F_{3p}\slq\Gamma_{3p}\slq\eps_+=2F_{3s}q_{p'q's}q_{p'q'p}\Gamma_{3p}\eps_+\,,\quad \fr{1}{2}F_{pq}(\Gamma_{pq}\slq+\slq\Gamma_{pq})\eps_+=-2q_{rsp}F_{rs}\Gamma_{3p}\eps_+\,.
\eeq
(It is sufficient to verify them for $q$ of the form (\ref{qm}) with $m_{ia}=0$.) The resulting boundary condition is
\beq
\left(\delta_{ps}+2q_{p'q's}q_{p'q'p}\right)F_{3s}-2q_{prs}F_{rs}=0\,.\label{bosneum}
\eeq

The bulk action of the theory in ten-dimensional notations is
\beq
I_{\rm bulk}=-\fr{1}{g_\YM^2}\int\tr\left(\fr{1}{2}F_{IJ}^2+\Psi^T B\Gamma_ID_I\Psi\right)\,,\label{ibulk}
\eeq
while the boundary action is\footnote{Note that $I_{\rm bdry}$ could possibly contain a fermion bilinear. It can be seen that this bilinear should vanish for the fermionic boundary condition (\ref{fermbc}) to follow from the boundary variation of the action.}
\beq
I_{\rm bdry}=\fr{8}{g_\YM^2}\left(\fr{1}{3!}z_{pqr}\Omega_{pqr}\right)\,,\label{ibdry}
\eeq
where $\Omega_{pqr}$ is an antisymmetric form 
\beq
\Omega_{pqr}=\int\tr(A_p[A_q,A_r])+{\rm derivatives}\,,
\eeq
so that
\beq
\delta\left(\fr{1}{3!}z_{pqr}\Omega_{pqr}\right)=\fr{1}{2}z_{pqr}\int\tr\left(\delta A_pF_{qr}\right)\,.
\eeq
Explicitly,
\beqn
&&\Omega_{012}=\fr{1}{2}\int\tr\left(A\d A+\fr{2}{3}A^3\right)\,,\quad\Omega_{aij}=\int\tr\left(\phi_aF_{ij}\right)\,,\nnr
&&\Omega_{aib}=\int\tr\left(\phi_aD_i\phi_b\right)\,,\quad \Omega_{456}=\int\tr\left(\phi_4[\phi_5,\phi_6]\right)\,.
\eeqn
To reproduce the bosonic Neumann boundary condition (\ref{bosneum}) by setting to zero the boundary variations of the fields, we have to fix
\beq
z_{pqr}=(\delta_{ps}+2q_{p'q'p}q_{p'q's})^{-1}q_{sqr}\,.\label{indry2}
\eeq
Despite its appearance, this three-tensor is fully antisymmetric. (Again, it is sufficient to verify this for $q$ of the form (\ref{qm}) with $m_{ia}=0$.) For small $q$, we have $z\approx q$, as was argued in \cite{GWbc} on symmetry grounds.

Under a supersymmetry transformation, the bulk part (\ref{ibulk}) of the action produces a boundary variation
\beq
\delta I_{\rm bulk}=-\fr{i}{g_\YM^2}\int_{y=0}\tr\left(F_{JK}\eps^T B\Gamma_K\left(\fr{1}{2}\Gamma_{3J}+\delta_{3J}\right)\Psi\right)\,.
\eeq
Using this expression and the boundary condition (\ref{bosneum}), it is easy to verify that the combined action $I_{\rm bulk}+I_{\rm bdry}$ is supersymmetric.

Now let us turn to the special case of the SO$(3)_W'$-invariant boundary conditions. From the topological field theory point of view it is clear that there should exist a complex number $u$ such that the complexified gauge field $A_i+u\phi_i$ is invariant under the scalar supersymmetries, when restricted to the boundary. Taking a supersymmetry variation (\ref{decompsusy}) for $\eps_+\sim \eps^{\alpha A}$, one readily finds that
\beq
u=\fr{tt'-1}{t+t'}=\fr{\sin\beta\cos\ang+i\cos 2\beta\sin\ang}{\cos\beta}\,,
\eeq
where we also used the natural parameterization (\ref{ttprimeap}) for $t$ and $t'$, with $\ang_{p,q}\equiv \ang$.

In the topologically twisted theory with an NS5-brane boundary condition, one expects to have the action
\beq
-i\calK\CS(A+u\phi)+\{\Qb,\dots\}\,,
\eeq
where the $\Qb$-exact terms are the usual squares of the Kapustin-Witten equations. Let us rewrite the bosonic part of this action as
\beq
I_{\rm bulk}|_{\rm bos}+\fr{i\theta_\YM}{8\pi^2}\int\tr(F\wedge F)+\fr{1}{g_\YM^2}\int_W\tr\left(2z_1\,\phi\wedge F+z_2\,\phi\wedge\d_A\phi+\fr{2}{3}z_3\,\phi^3\right)\,,\label{ns5zzz}
\eeq
where
\beqn
z_1&=&\fr{2}{t+t^{-1}}-\fr{i\calK g_\YM^2}{4\pi}u\,,\label{ztop1}\\
z_2&=&\fr{t-t^{-1}}{t+t^{-1}}-\fr{i\calK g_\YM^2}{4\pi}u^2\,,\label{ztop2}\\
z_3&=&-\fr{2}{t+t^{-1}}-\fr{i\calK g_\YM^2}{4\pi}u^3\,,\label{ztop3}
\eeqn
On the other hand, from (\ref{ibdry}) and (\ref{indry2}) one can compute
\beqn
\fr{\theta_\YM g^2_\YM}{8\pi^2}&=&4iz_{012}=i\fr{t+3t'+(3t+t')t'^2}{t+3t'-(3t+t')t'^2}=\cot\ang\,,\nnr
z_1&=&4z_{401}=\fr{2t'(t-t')}{t+3t'-(3t+t')t'^2}=-i\fr{\tan\beta}{\sin\ang}   \,,\nnr
z_2&=&4z_{045}=-\fr{(t-t')(t'^2-1)}{t+3t'+(3t+t')t'^2}=-\fr{\sin\beta(\sin\ang+i\sin\beta\cos\ang)}{\sin\ang\cos^2\beta}     \,,\nnr
z_3&=&4z_{456}=\fr{2(tt'^3-1)}{t+3t'-(3t+t')t'^2}=\fr{\left(\fr{3}{2}\cos^2\beta-1\right)\sin2\ang-i\sin^3\beta\cos2\ang}{\sin\ang\cos^3\beta}\,.\nonumber
\eeqn
It is straightforward to check that $z_{1,2,3}$ here match (\ref{ztop1})-(\ref{ztop3}).

\acknowledgments{I thank T.~Dimofte, S.~Gukov, C.~Manolescu and N.~Nekrasov for useful discussions. I am grateful to R.~Mazzeo for helpful discussions and hospitality at Stanford. I would like to thank E.~Witten for helpful discussions, encouragement and important comments on the draft. Some results of this paper were presented in the spring of 2017 at seminars at Simons Center for Geometry and Physics and Caltech. I thank people in the audiences for their insightful questions and comments. This work was performed in part at the Aspen Center for Physics, which is supported by National Science Foundation grant PHY-1066293. }

\enddocument